\documentclass[a4paper,11pt]{article}
\pdfoutput=1 

\usepackage{jheppub} 

\usepackage[T1]{fontenc} 

\usepackage[utf8]{inputenc}
\usepackage{amssymb}
\usepackage{amsmath}
\usepackage{amsfonts}
\usepackage{graphicx}
\usepackage{color}
\usepackage{xspace}
\usepackage{comment}
\usepackage{hyperref}
\usepackage[normalem]{ulem}
\usepackage[section]{placeins}
\usepackage{afterpage}
\usepackage{slashed}
\usepackage{enumerate}
\usepackage{enumitem}
\usepackage{caption}
\usepackage{subfigure}
\usepackage{float}
\usepackage{ulem}
\usepackage{verbatim}
\usepackage{multirow,rotating}
\usepackage[dvipsnames]{xcolor}

\definecolor{dcolour}{rgb}{.5, .5, .5}
\def\gsim{\raise0.3ex\hbox{$\;>$\kern-0.75em\raise-1.1ex\hbox{$\sim\;$}}}
\def\lsim{\raise0.3ex\hbox{$\;<$\kern-0.75em\raise-1.1ex\hbox{$\sim\;$}}}
\def\gsim{\raise0.3ex\hbox{$\;>$\kern-0.75em\raise-1.1ex\hbox{$\sim\;$}}}
\def\lsim{\raise0.3ex\hbox{$\;<$\kern-0.75em\raise-1.1ex\hbox{$\sim\;$}}}

\newcommand{\ba}[1]{\begin{eqnarray} \label{(#1)}}
\newcommand{\ea}{\end{eqnarray}}

\newcommand{\iab}{\rm ab^{-1}}

\newcommand{\mltp}{{\mkern-2mu\times\mkern-2mu}}

\newcommand{\met}{\slashed{E}_T}
\newcommand{\pmiss}{\vec{\slashed{p}}_T}

\title{\boldmath Search for heavy Majorana neutrinos in the $\tau$ final state at proton-electron colliders}

\author[a,b]{Haiyong Gu}
\author[a]{, Ying-nan Mao}
\author[c]{, Hao Sun}
\author[a,1]{and Kechen Wang\note{Corresponding author.}}

\affiliation[a]{Department of Physics, School of Science, Wuhan University of Technology,\\430070 Wuhan, Hubei, China}
\affiliation[b]{School of Information Science and Technology, Xiamen University Tan Kah Kee College, \\363105 Zhangzhou, Fujian,
China}
\affiliation[c]{Institute of Theoretical Physics, School of Physics, Dalian University of Technology,\\116024 Dalian, Liaoning, China}

\emailAdd{haiyong@xujc.com}
\emailAdd{ynmao@whut.edu.cn}
\emailAdd{haosun@dlut.edu.cn}
\emailAdd{kechen.wang@whut.edu.cn}

\abstract{
We utilize the lepton number violation signal process $p\, e^- \to \tau^+ jjj$ to search for heavy Majorana neutrinos at future proton-electron colliders. 
The LHeC (FCC-eh) is considered to run with an electron beam energy of 60 GeV, a proton beam energy of 7 (50) TeV and an integrated luminosity of 1 (3)  ab$^{-1}$,
and the electron beam is considered to be unpolarized.
We apply detector configurations and simulate signal and related 
standard model
background events for both hadronic $\tau_h$ and leptonic $\tau_\ell$ final states,
$\ell$ being a muon.
After preselection, multivariate analyses are performed to reject the background. 
The strategy to 
reconstruct
the heavy neutrino mass is developed and distributions of reconstructed mass are presented. 
Discovery sensitivities
on parameter $|V_{\tau N}|^2 |V_{eN}|^2 / ( |V_{\tau N}|^2 + |V_{eN}|^2 )$ for the heavy neutrino mass between 
10 and 3000 GeV
are predicted. 
At the 2-$\sigma$ significance, the best discovery sensitivity
is $\sim 1.2 \times10^{-5}  \,\,(5.0 \times 10^{-6})$ at the LHeC (FCC-eh) when $m_N \sim 100$ GeV for the hadronic $\tau_h$ final state. 
Sensitivities for the leptonic $\tau_\ell$ final state are found to be similar to those for the hadronic $\tau_h$ final state for most of the parameter space investigated.
We also derive the limits on mixing parameters from electroweak precision data (EWPD) and DELPHI experiment. Assuming $|V_{\tau N}|^2 = |V_{eN}|^2 = |V_{\ell N}|^2$, sensitivity bounds from the LHeC and FCC-eh experiments are found to be stronger than those from EWPD when $m_N \lesssim  900$ GeV, and also stronger than those from DELPHI when $m_N \gtrsim 70$ GeV. 
Constraints are also interpreted and compared in the $|V_{\tau N}|^2$ vs. $|V_{e N}|^2$ plane. Compared with current limits from EWPD, DELPHI, and LHC experiments, future $pe$ experiments can probe large additional regions in the parameter space formed by $|V_{\tau N}|^2$ and $|V_{e N}|^2$, and thus 
significantly enhance the discovery potential for a large portion of the $|V_{\tau N}|^2$ vs. $|V_{e N}|^2$ plane.
}

\begin{document}\maketitle
\flushbottom

\section{Introduction}
\label{sec:intro}

The experiments of neutrino oscillation~\cite{Super-Kamiokande:1998kpq,MINOS:2006foh,MINOS:2011amj,PhysRevLett.108.131801,Ling:2013fta,Kim:2013sza} show that at least two of three active neutrinos are massive. However, in the standard model (SM), neutrinos have only left-handed components and no right-handed components, so they cannot form a Dirac mass term, and the neutrino mass is strictly equal to zero. Therefore, the standard model needs to be slightly expanded. 
One important method is
the seesaw mechanism~\cite{FRITZSCH1975256,Minkowski:1977sc,Yanagida:1979as,Sawada:1979dis,Mohapatra:1979ia,Glashow:1979nm,GellMann:1980vs,Keung:1983uu,Foot:1988aq,Mohapatra:1986aw,MAGG198061}, which introduces right-handed Majorana neutrinos $N_{R}$ and couple them with SM neutrinos to produce small neutrino masses $M_{\nu^{\prime}} \sim M_D^2/M_R$, where $M_R$ is the Majorana mass of $N_{R}$ and $M_D$ is the Dirac mass and proportional to the Yukawa coupling between the standard model neutrino and $N_{R}$. When $M_R\gg M_D$, active neutrino acquires a small mass.

Search for the heavy neutrinos is crucial to verify the seesaw mechanism. Because the production cross section, decay width, and lifetime of $N$ are determined by its mass $m_N$ and the parameter $|V_{\ell N}|^2$ which is related to the matrix element describing the mixing of $N$ with the SM neutrino of flavor $\ell = e, \mu, \tau$, limits for such searches are usually given in the plane of the mixing parameter $|V_{\ell N}|^2$ vs. $m_N$. 

At colliders, they can be searched from the decays of Higgs bosons~\cite{Gao:2021one, Gao:2019tio}, $W$-bosons~\cite{Antusch:2018bgr, Dib:2017iva, Dib:2017vux, Dib:2016wge} and $Z$-bosons~\cite{Wang:2019xvx}.
Summaries of collider searches of heavy neutrinos can be found in 
Refs.~\cite{Atre:2009rg,Deppisch:2015qwa,Das:2015toa,Das:2016hof,Cai:2017mow,Das:2017rsu,Bolton:2019pcu,Ding:2019tqq,Shen:2022ffi}
and references therein.
Recent experimental studies on heavy neutrino searches can be found in 
Refs.~\cite{L3:1999ymc, L3:2001zfe, CMS:2018iaf, CMS:2018jxx,  CMS:2021lzm, ATLAS:2019kpx, LHCb:2020wxx, NA62:2020mcv, Belle:2013ytx, T2K:2019jwa, CMS:2022rqc},
and are reviewed in Ref.~\cite{Gu:2022muc}.

Compared with plentiful studies focusing on the mixing parameters $|V_{e N}|^2$ and $|V_{\mu N}|^2$,
because of the challenges in detecting the final state taus, 
the mixing parameter $|V_{\tau N}|^2$ 
is more difficult to be probed, 
making it not well studied at current experiments.
For heavy neutrinos with mass above 10 GeV, the main experimental constraints on the mixing parameter $|V_{\tau N}|^2$ are set by the DELPHI ~\cite{DELPHI:1996qcc}, and can be derived from the electroweak precision data (EWPD)~\cite{delAguila:2008pw,Akhmedov:2013hec,Basso:2013jka,deBlas:2013gla,Antusch:2015mia,Chrzaszcz:2019inj,Cheung:2020buy} and rare decays of 
$Z$-boson
~\cite{DELPHI:1996qcc}.

However, in phenomenology, there do exist some studies to probe the mixing parameter $|V_{\tau N}|^2$ in different heavy neutrino mass ranges.
Among them, 
Refs.~\cite{Bondarenko:2018ptm,Cvetic:2019shl,Dib:2019tuj,Zhou:2021ylt} study heavy neutrinos with $m_N \sim (0.1-5)$ GeV;
Refs. \cite{Abada:2018sfh,Cottin:2018nms} 
for $m_N \sim (1-20)$ GeV; 
Ref. \cite{Cheung:2020buy} for $m_N \sim (25-150)$ GeV; 
and 
Refs. \cite{Florez:2017xhf,Pascoli:2018rsg,Pascoli:2018heg}
for $m_N > 150$ GeV.
In the recent work~\cite{Bai:2022lbv}, some of our authors derive current constraints on $|V_{\tau N}|^2$ from the rare $Z$-boson decay and electroweak precision data (EWPD), and forecast sensitivities on $|V_{\tau N}|^2$ via the signal $p p \to \tau^{\pm} \tau^{\pm} j j$ at future proton-proton colliders.

Ref.~\cite{Azuelos:2021ese} and references therein have reviewed BSM physics searches at future electron-proton colliders, the Large Hadron electron Collider (LHeC)~\cite{Klein:2009qt,LHeCStudyGroup:2012zhm,Bruening:2013bga,Klein:2016uwv,LHeC:2020van,Holzer:2021dxw}
and the electron-hadron mode of the Future Circular Collider (FCC-eh)~\cite{Zimmermann:2014qxa,Klein:2016uwv,TOMAS2016149,FCC:2018byv,Holzer:2021dxw}.
The $\tau$ final state at $pe$ colliders is also studied in Ref.~\cite{Antusch:2020fyz}.
Phenomenology studies on heavy neutrino searches at $e p$ colliders can be found in 
Refs.~\cite{BUCHMULLER1991465,Buchmuller:1991tu,Buchmuller:1992wm,Ingelman:1993ve,deAlmeida:2002pr,Liang:2010gm,Blaksley:2011ey,Duarte:2014zea,Mondal:2016kof,Antusch:2016ejd,Lindner:2016lxq,Li:2018wut,Das:2018usr,Antusch:2019eiz,Cottin:2021tfo,Gu:2022muc,Batell:2022ogj}.

Among them, in Ref.~\cite{Gu:2022muc}, some of our authors develop the search strategy for a heavy Majorana neutrino via the lepton number violation signal process $p e^{-} \to \mu^{+} jjj$ at future electron-proton colliders. Assuming mixing parameters $|V_{\ell N}|^2=|V_{\mu N}|^2=|V_{e N}|^2$ and $|V_{\tau N}|^2 = 0$, the dominant SM background processes are considered, and 
discovery sensitivities
on the mixing parameter $|V_{\ell N}|^2$ are predicted for the heavy neutrino mass in the range of 
10-3000 GeV.
The results show that the 
sensitivities
at electron-proton colliders are much stronger than the current experimental limits at the LHC for $m_N$ above 30 GeV.

In this consecutive study, we concentrate on the signal process $p e^{-} \to \tau^{+} jjj$ with final state taus at the LHeC and FCC-eh.
Similar to Ref.~\cite{Gu:2022muc}, 
the LHeC (FCC-eh) is supposed to run with an electron beam energy of 60 GeV, a proton beam energy of 7 (50) TeV and an integrated luminosity of 1 (3) $\iab$;
the electron beam is considered to be unpolarized;
and to simplify the analyses, we consider the
phenomenological simplified Type-I 
model and the scenario that only one generation of heavy neutrinos $N$ is within the collider access.
The $N$ is assumed to mix with active neutrinos of tau and electron flavours, i.e.  $|V_{\tau N}|^2, |V_{e N}|^2 \neq 0$ and $|V_{\mu N}|^2 = 0$.

Since taus are unstable, they decay either leptonically into muons and electrons, or hadronically into mesons, leading to final state leptons or tau-jets at colliders.
Because leptonic and hadronic final states have different kinematics and background, to obtain and compare the 
sensitivities
for both states, we perform their analyses individually and forecast 
sensitivities
for heavy neutrinos in the range 
10-3000 GeV
at LHeC and FCC-eh. 

The organization of our paper is as follows. 
In Sec.~\ref{sec:sig}, we present the cross section of the signal process.
In Sec.~\ref{sec:Hdecay} and Sec.~\ref{sec:Ldecay}, we show the SM background processes and the analyses for the hadronic and leptonic final states, respectively.
In Sec.~\ref{sec:results}, we present sensitivity curves in the mixing parameters vs. heavy neutrino mass plane at the LHeC and FCC-eh.
The 
sensitivities
obtained from this study are compared with those from current experiments in Sec.~\ref{sec:limits}.
We conclude in Sec.~\ref{sec:discussion}.
Details of this study are listed in Appendices~\ref{app:EWPD}$-$\ref{app:Eff}.

\section{The signal production process}
\label{sec:sig}

\begin{figure}[h]
	\centering
	\includegraphics[width=7cm,height=5cm]{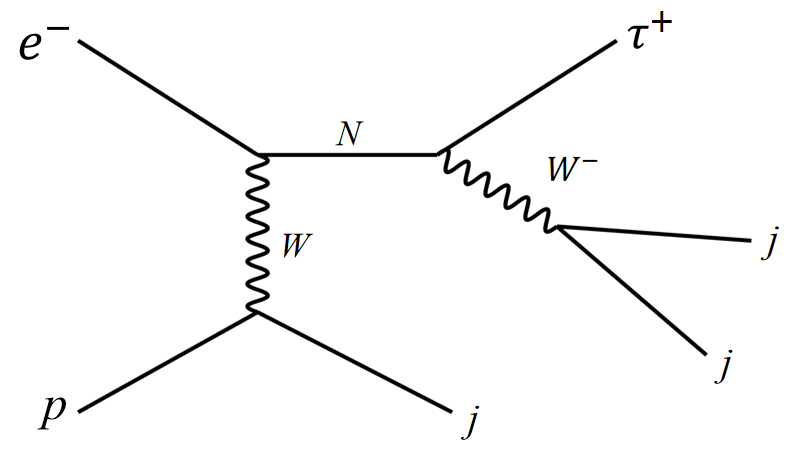}  
	\caption{The production process of the Lepton Number Violating (LNV) signal in the tau final state via the heavy Majorana neutrinos at $pe$ colliders. }
	\label{fig:Feynman}
\end{figure}

As shown in Fig.~\ref{fig:Feynman}, heavy Majorana neutrinos $N$ can be produced via the $t-$channel exchange of $W$ bosons at the $pe$ colliders, and decay finally into $\tau^+$ plus three jets.
The lepton number of this process changes from $+1$ to $-1$, so it violates the conservation of lepton number.

For the data simulation, we implement the Universal FeynRules Output model file~\cite{Alva:2014gxa,Degrande:2016aje} which extends the SM with additional heavy neutrinos interacting with active neutrinos into  MadGraph5~\cite{Alwall:2014hca} to generate the signal events.
When generating collision events, the default ``nn23lo1'' parton distribution function of the proton is used. 
To maximize the event acceptance, following loose 
requirements
are applied at the parton level  in MadGraph (implemented in the run\_card.dat file) for both the signal and background:
the minimal value of transverse momentum for the jet (lepton) is set to be 5 (2) GeV; 
the pseudorapidity ($\eta$) range for both the lepton and jet is set to be $|\eta|<10$;
the minimal value of the solid angular distance, $\Delta R = \sqrt{\Delta \eta^2 + \Delta\phi^2}$,  between final state objects is set to be 0.01.

Similar to our previous work~\cite{Azuelos:2019bwg,Gao:2019tio}, the Pythia6~\cite{Sjostrand:2006za} program is modified to perform the parton showering and the hadronization  for the $pe$ colliders, while the configuration card files~\cite{Delphes_cards} for the LHeC and FCC-eh detectors are implemented to the Delphes program~\cite{deFavereau:2013fsa} to complete the detector simulation.
For the jet reconstruction, jets are classified by using the FastJet package~\cite{Cacciari:2011ma} with anti-$k_t$ algorithm and cone radius $R$ = 0.4.
For the final state tau reconstruction, the default rates in the Delphes configuration card files are used, where the identification efficiency for tau-tagging is set to be 40\%, and the misidentification rate is set to be about 0.1\%.

\begin{figure}[ht]
\centering
\includegraphics[width=12cm,height=8cm]{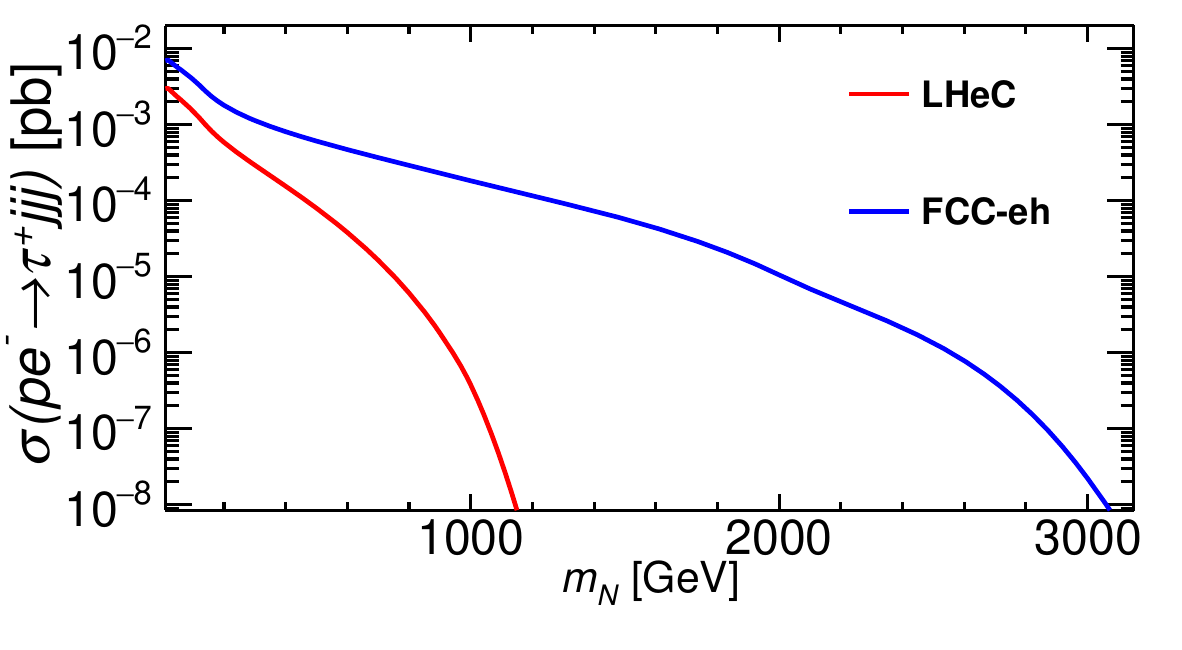}
\caption{
The production cross section of the LNV signal $p\, e^- \to \tau^+ jjj$ via the heavy Majorana neutrino $N$ when fixing 
$ |V_{\tau N}|^2\, |V_{eN}|^2 / \left( |V_{\tau N}|^2 + |V_{eN}|^2 \right) = 5 \times 10^{-5} $
and varying $m_N$ at the LHeC and FCC-eh.
}
\label{fig:crs}
\end{figure}

For the signal data simulation, 
we fix the benchmark mixing parameter $|V_{\ell N}|^2 = |V_{e N}|^2 = |V_{\tau N}|^2 = 10^{-4}$, and vary the heavy neutrino mass $m_N$.
At the LHeC (FCC-eh), when $m_N$ = 10 and 20 GeV, we generate 15.3 (7.8) and 3.3 (1.8) million signal events, respectively, while 0.3 (0.6) million signal events are generated when $m_N$ = 40, 60, 70, 100, 120, 200, 300, 400, 500, 600, 700, 800, 900, 1000 GeV.
Besides, at the LHeC, 0.3 million signal events are generated for one additional mass point with $m_N$ = 1150 GeV, while
at the FCC-eh, 0.6 million signal events are generated for two additional mass points with $m_N$ = 2000 and 3150 GeV.

To maintain consistency through this study, the production cross sections calculated by MadGraph5 are used to estimate the number of events for both the signal and the background processes.
In Fig.~\ref{fig:crs}, we plot the  cross sections of the LNV signal $p e^- \to \tau^+ jjj$ via the Majorana heavy neutrino N as a function of heavy neutrino mass $m_N$ at the LHeC and FCC-eh.
The signal productions are proportional to the parameter $|V_{\tau N}|^2\, |V_{eN}|^2 / \left( |V_{\tau N}|^2 + |V_{eN}|^2 \right)$ which is fixed to be  $5 \times 10^{-5}$ in Fig.~\ref{fig:crs}.
For large $m_N$ the cross sections for LHeC decrease much faster than those for FCC-eh. 
This behaviour can be understood from the parton distribution function of the proton, and is explained in details in Ref.~\cite{Gu:2022muc}.

\section{Hadronic $\tau_h$ final state} 
\label{sec:Hdecay}

\subsection{SM background processes} 
\label{subsec:HdecayBkg}

When tau decays hadronically, the final state contains one tau-jet $\tau_h$ with positive charge, at least three regular jets and missing energy.
There exist four types of background processes: $p e^- \to \tau^+ \tau^- e^- j j j$, $ \tau^+ \tau^- \nu_e j j j$, $ \tau^+ \nu_\tau  e^- j j j$, and $ \tau^+ \nu_\tau \nu_e j j j$, which we label as ``B1-B4'' in this article.
They can have the similar final state as the signal when the final state $\tau^-$ and $e^-$ are undetected.
Besides, because the jet production is still substantial at $pe$ colliders, the misidentified taus from jets needs to be considered.
The multi-jet process ``$p\, e^- \to \nu_e   j  j  j  j$ with a $j \to \tau_h^+$''  is the main background of such kind, and is labeled as ``B5a''. 
The Delphes configuration card files~\cite{Delphes_cards} are used for the LHeC and FCC-eh detector simulations in this study, and the misidentification rate for the tau-tagging is set to be about 0.1\% in these files. Therefore, we use this value as benchmark to test the effect of misidentified tau-jets.

Table~\ref{tab:crsHad} shows the production cross sections of background processes at the LHeC and FCC-eh for the hadronic $\tau_h$ final state. 
Among them, $\nu_e j j  j  j$ has the largest cross sections without any 
requirements,
followed by $\tau^+ \tau^- e^- j j j$ and $\tau^+ \nu_\tau  e^- j j j$.
For the remaining two background processes, due to their small cross sections, they do not play a dominant role.

\begin{table}[h]
	\centering
	\begin{tabular}{cccc}
		\hline 
		\hline 
	     &$\sigma$ [pb] & LHeC  & FCC-eh  \\ 	
		\hline 
		B1&$ \tau^+ \tau^- e^- j j j$& 0.22 & 0.86 \\ 
		B2&$ \tau^+ \tau^- \nu_e j j j$&0.052  &  0.29 \\  
		B3&$ \tau^+ \nu_\tau  e^- j j j$&0.28  & 1.6 \\  
		B4&$ \tau^+ \nu_\tau \nu_e j j j$&  $8.2\mltp10^{-6}$ & $9.3\mltp10^{-5}$  \\
		B5a&$  \nu_e j j jj$&  309 & 1159  \\ 
		\hline 
		\hline
	\end{tabular} 
	\caption{The production cross sections of 
		background processes at the LHeC and FCC-eh for the hadronic $\tau_h$ final state. 
	}
	\label{tab:crsHad}
\end{table}

The number of simulated events for each background process is determined according to its importance in reducing the statistical uncertainty of the final 
sensitivities.
At the LHeC (FCC-eh), we generate 2.1 (1.2) million $\tau^+ \tau^- e^- j j j$, 4.9 (3.0) million $\tau^+ \tau^- \nu_e j j j$, 21 (19) million $\tau^+ \nu_\tau  e^- j j j$, 0.5 (0.5) million  $\tau^+ \nu_\tau  \nu_e j j j$, and 32 (25) million $\nu_e j j j j$ events, respectively.
 
\subsection{Data analysis} 
\label{subsec:HdecayAna} 

We apply following 
preselection
to select the signal and reject the background events at the first stage.
(i) Exactly one $\tau$-jet with positive charge, i.e. $N(\tau_h^+) = 1$ and transverse momentum $p_T(\tau_h) >$ 5 GeV; events with final state electron(s) or muon(s) are vetoed.
(ii) All regular jets are sorted in descending order according to their transverse momenta and we require at least three jets, i.e. $N(j)\geq 3$; 
for the $p_T$ thresholds of jets, when heavy neutrino mass is below 80 GeV, the $p_T$ of the first three leading jets are required to be greater than 10 GeV, 
while when the mass is above 80 GeV, we require the first two leading jets have $p_T$ greater than 20 GeV and the third one has $p_T$ greater than 10 GeV.
(iii) To reject the background from misidentified taus, the solid angular distance $\Delta R$ between the regular jet and tau-jet is required to be larger than 0.8, i.e., $\Delta R (\tau_h, j) > 0.8$.

\textbf{\begin{figure}[h]
\centering
\includegraphics[width=7cm,height=5cm]{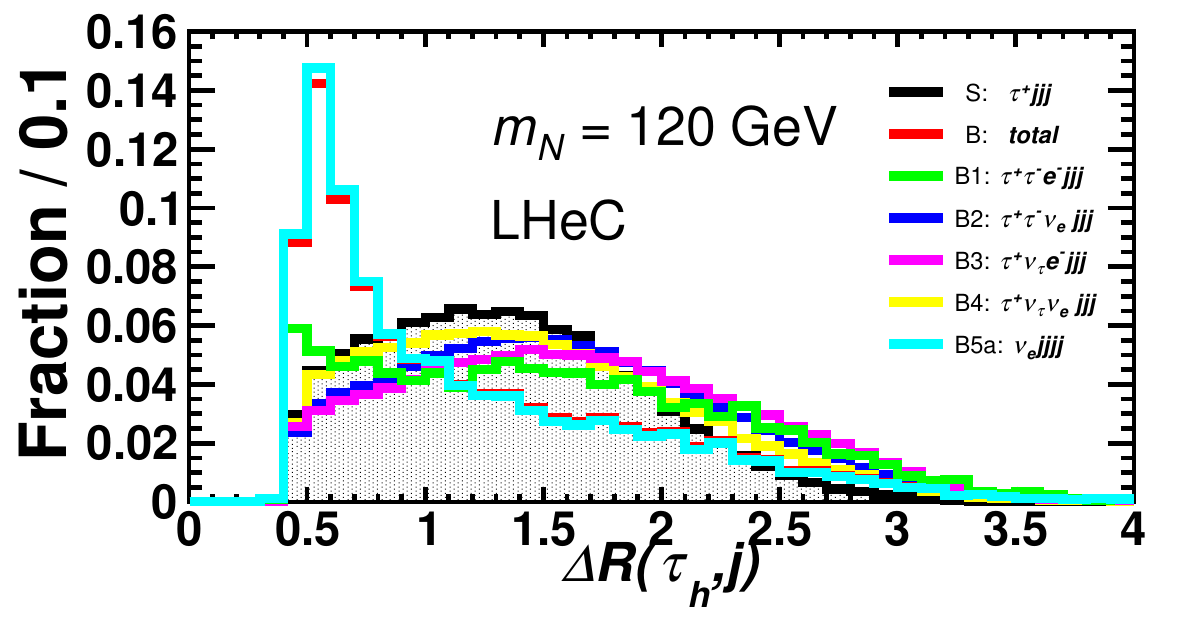}\,\,\,\,\,\,\,\,
\includegraphics[width=7cm,height=5cm]{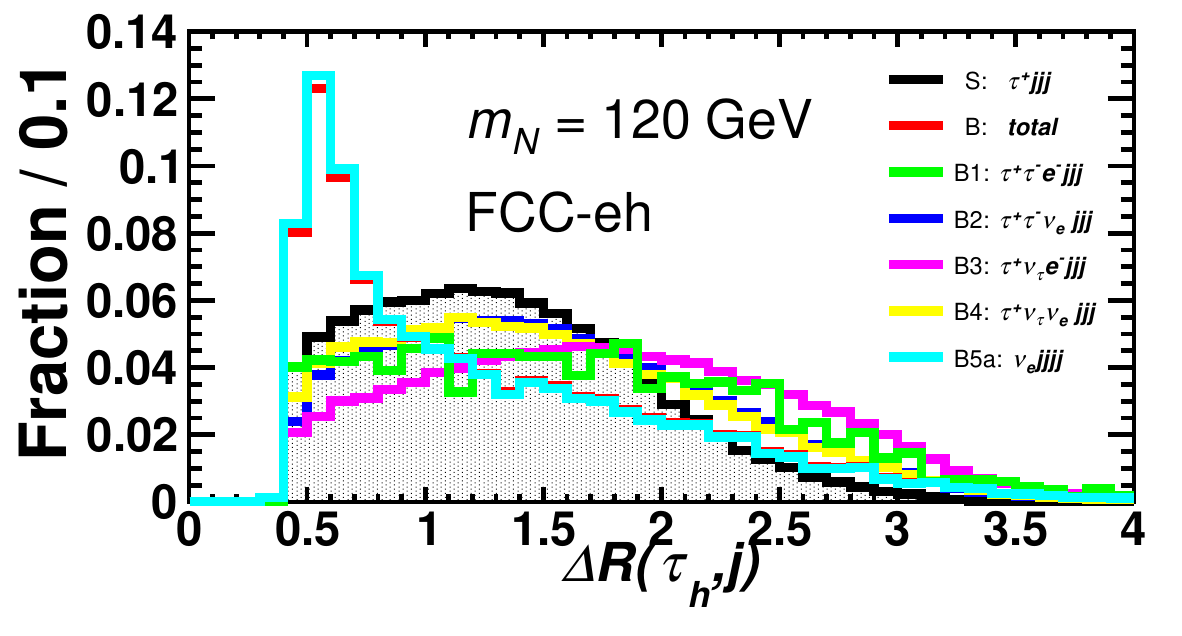}
\caption{
The $\Delta R(\tau_h, j)$ distributions for the signal with benchmark $m_N$ = 120 GeV and $ |V_{\tau N}|^2\, |V_{eN}|^2 / \left( |V_{\tau N}|^2 + |V_{eN}|^2 \right) = 5 \times 10^{-5}$, and for background processes after applying 
preselection
(i)-(ii) for the hadronic $\tau_h$ final state at the LHeC and FCC-eh. 
}
\label{fig:DeltaR}
\end{figure}}

To exploit the $\Delta R(\tau_h, j)$ observable, we calculate $\Delta R(\tau_h, j)$ values for all possible combinations of tau- and regular jets in the same event, and select the minimal value. 
Fig.~\ref{fig:DeltaR} shows the distributions of such minimal $\Delta R(\tau_h, j)$ 
for the signal with benchmark $m_N$ = 120 GeV 
and $ |V_{\tau N}|^2\, |V_{eN}|^2 / \left( |V_{\tau N}|^2 + |V_{eN}|^2 \right) = 5 \times 10^{-5}$, and for background
processes after applying 
preselection 
(i)-(ii).
One observes that the $\nu_e j j j j$ process still dominates after applying 
preselection
(i)-(ii), and can be rejected effectively by the $\Delta R(\tau_h, j)$ requirement.

In Table~\ref{tab:preselHad}, we show the number of events for the signal with benchmark $m_N=120$ GeV and background processes after applying the
preselection 
(i)-(iii) sequentially described above.  
One observes that, initially at the LHeC (FCC-eh), without any selection, B5a is a factor of 562 (417) larger than the sum of B1-B4, while the ratio is reduced to 21 (23) after preselection (i)-(ii) and becomes 14 (16) after preselection (iii), which means the preselection 
rejects 
B5a effectively.

\begin{table}[h]
	\centering
		\begin{tabular}{cccccccc}
			\hline
			\hline
			\multicolumn{2}{c}{}
			& signal & B1 & B2 & B3 & B4 & B5a  \\
			\hline
			\multirow{4}{*}{LHeC} 
			& initial &$1.2\mltp10^{3}$ & $2.2\mltp10^{5}$ & $5.2\mltp10^{4}$ & $2.8\mltp10^{5}$ & 8.2 & $3.1\mltp10^{8}$  \\ 
			& (i) & $2.8\mltp10^{2}$  & $1.6\mltp10^{3}$ & $4.7\mltp10^{3}$  & $5.5\mltp10^{3}$  & 1.9  & $4.7\mltp10^{5}$  \\ 
			& (ii) &  $2.2\mltp10^{2}$ &  $4.3\mltp10^{2}$ & $1.6\mltp10^{3}$  & $1.8\mltp10^{3}$ & 1.3 & $8.0\mltp10^{4}$   \\ 
		   & (iii) &  $1.8\mltp10^{2}$ &  $3.4\mltp10^{2}$ & $1.4\mltp10^{3}$  & $1.6\mltp10^{3}$      & 1.1   & $4.6\mltp10^{4}$  \\ 
			\hline
			\multirow{4}{*}{FCC-eh}
			&	initial & $1.0\mltp10^{4}$ &  $2.6\mltp10^{6}$ &  $ 8.7\mltp10^{5} $ & $4.9\mltp10^{6} $  & $2.8\mltp10^{2}$  & $3.5\mltp10^{9}$  \\ 
			& (i) & $2.3\mltp10^{3}$  &$1.4\mltp10^{4}$  &  $8.4\mltp10^{4}$ & $8.1\mltp10^{4}$  & 63  & $6.2\mltp10^{6}$    \\ 
			& (ii) & $1.9\mltp10^{3}$  & $4.5\mltp10^{3}$  &  $3.9\mltp10^{4}$ &  $3.5\mltp10^{4}$ & 52  & $1.8\mltp10^{6}$   \\ 
		   & (iii) &  $1.5\mltp10^{3}$ &  $3.8\mltp10^{3}$ & $3.3\mltp10^{4}$  & $3.1\mltp10^{4}$ &43 & $1.1\mltp10^{6}$  \\ 			
			\hline
			\hline
	\end{tabular}
\caption{
The number of events for the signal with benchmark $m_N$ = 120 GeV 
and $ |V_{\tau N}|^2\, |V_{eN}|^2 / \left( |V_{\tau N}|^2 + |V_{eN}|^2 \right) = 5 \times 10^{-5}$, and for background
processes after applying 
preselection 	
		(i)-(ii)
		sequentially for the hadronic $\tau_h$ final state. 
		The numbers correspond to the LHeC and
		FCC-eh with 1 and 3 $\iab$ integrated luminosity,
		respectively. 
	}
	\label{tab:preselHad}
\end{table}

To further reject the background, we input following nineteen observables into 
the Toolkit for Multivariate Analysis (TMVA) 
package~\cite{Hocker:2007ht} to perform the multivariate analysis (MVA).

\begin{enumerate}[label*=\Alph*.]
\item Four momenta of the final state 
$\tau_h$: 
$p_{x}(\tau_h)$, $p_{y}(\tau_h)$, $p_{z}(\tau_h)$, $E(\tau_h)$.
\item Four momenta of the first three final state regular jets:
$p_{x}(j_{1})$, $p_{y}(j_{1})$, $p_{z}(j_{1})$, $E(j_{1})$;
$p_{x}(j_{2})$, $p_{y}(j_{2})$, $p_{z}(j_{2})$, $E(j_{2})$;
$p_{x}(j_{3})$, $p_{y}(j_{3})$, $p_{z}(j_{3})$, $E(j_{3})$.
\item Number of regular jets:
$N(j)$.
\item Magnitude and azimuthal angle of the missing transverse momentum:
$\met$, $\phi(\met)$.
\item The minimal value of solid angular distances between the final state  tau- and regular jets:
$\Delta R (\tau_h, j)$. 
\end{enumerate}

\begin{figure}[h]
\centering
\includegraphics[width=7cm,height=5cm]{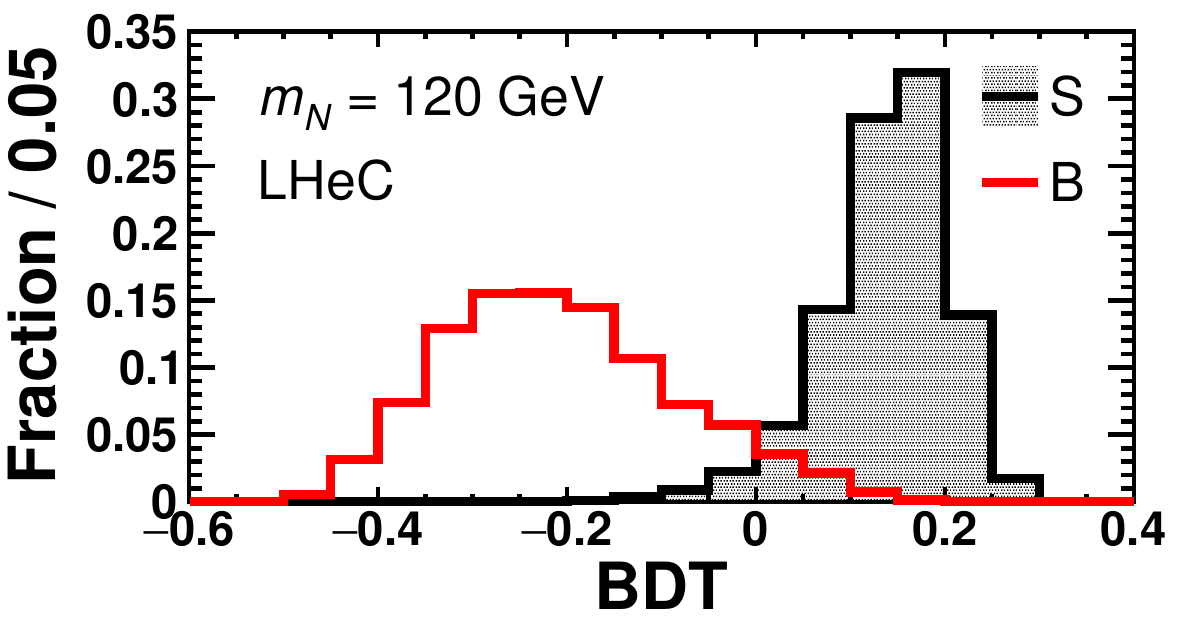}\,\,\,\,\,\,\,\,
\includegraphics[width=7cm,height=5cm]{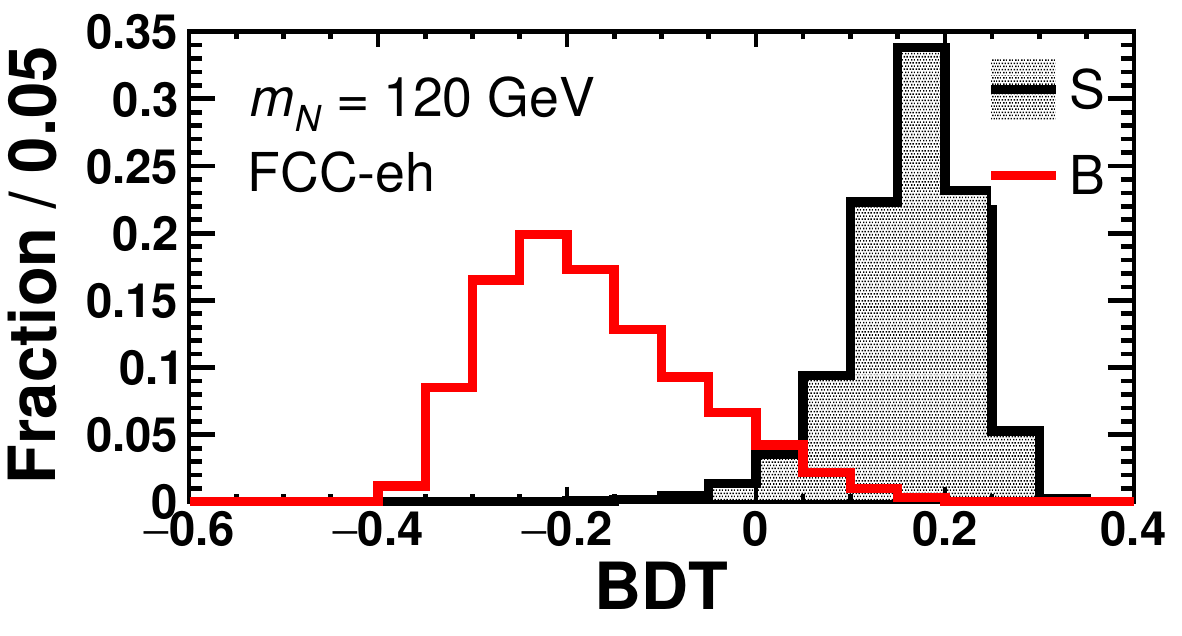}
\caption{
Distributions of BDT responses for the signal with 
benchmark $m_{N}$ = 120 GeV and $ |V_{\tau N}|^2\, |V_{eN}|^2 / \left( |V_{\tau N}|^2 + |V_{eN}|^2 \right) = 5 \times 10^{-5}$ (black, filled) and for
total SM background (red) at the LHeC (left) and FCC-eh (right) for the hadronic $\tau_h$ final state.}
\label{fig:BDTbenchHad}
\end{figure}

The Boosted Decision Tree (BDT) algorithm is adopted to perform the MVA and maximally reject the background.
The default setting in the TMVA package is used, where 
the number of trees in the forest ``NTrees'' is set to be 850, 
the minimum percentage of training events required in a leaf node ``MinNodeSize'' is set to be 2.5\%, 
the  maximal depth of the decision tree allowed ``MaxDepth'' is set to be 3, 
the learning rate for AdaBoost algorithm ``AdaBoostBeta'' is set to be 0.5, 
the relative size of bagged event sample to original size of the data sample ``BaggedSampleFraction'' is set to be 0.5,
and the number of grid points in variable range used in finding optimal cut in node splitting ``nCuts'' is set to be 20.
In Fig.~\ref{fig:BDTbenchHad}, we show BDT distributions for the total background and the benchmark signal with $m_{N}$ = 120 GeV at the LHeC and FCC-eh.
Since the kinematics of signal varies with $m_N$, distributions of BDT response also change with $m_N$. 
BDT distributions corresponding to other representative heavy neutrino masses are shown in 
Appendix~\ref{subapp:BDTtauH} for the LHeC and FCC-eh.

\section{Leptonic $\tau_\mu$ final state}
\label{sec:Ldecay}

\subsection{SM background processes}
\label{subsec:LdecayBkg}

When tau decays leptonically, we select the final state with tau decaying into muons
which we label as ``$\tau_\mu$'' in this article.
Thus, signature of final state has one positively charged muon, at least three regular jets plus missing energy.  
Besides the B1-B4 processes including taus, SM processes including muons can also contribute to the background in this scenario.
Therefore, four additional background processes: $\mu^+ \mu^- e^- j j j$, $\mu^+ \mu^- \nu_e j j j$, $\mu^+ \nu_\mu  e^- j j j$, and $\mu^+ \nu_\mu \nu_e j j j$ are included, and labelled as ``B5-B8'' in this article.
Table~\ref{tab:crsLep} shows the production cross sections of background processes at the LHeC and FCC-eh for the leptonic $\tau_\mu$ final state. 

\begin{table}[h]
	\centering
	\begin{tabular}{cccc}
		\hline 
		\hline 
		&$\sigma$ [pb] & LHeC  & FCC-eh  \\ 	
		\hline 
		B1	& $ \tau^+ \tau^- e^- j j j$& 0.22 & 0.86 \\  
		B2	& $ \tau^+ \tau^- \nu_e j j j$&0.052  &  0.29 \\ 
		B3	& $ \tau^+ \nu_\tau  e^- j j j$&0.28  & 1.6 \\  
		B4	& $ \tau^+ \nu_\tau  \nu_e j j j$& $8.2 \mltp 10^{-6}$  & $9.3 \mltp 10^{-5}$  \\ 
		B5 & $\mu^+ \mu^- e^- j j j$& 0.58 & 2.1 \\ 
		B6 & $\mu^+ \mu^- \nu_e j j j$& $8.6\mltp 10^{-2}$  &  0.39 \\ 
		B7 & $\mu^+ \nu_\mu  e^- j j j$&0.28  & 1.6 \\ 
		    B8 & $\mu^+ \nu_\mu \nu_e j j j$& $8.1 \mltp 10^{-6}$  & $9.3 \mltp 10^{-5}$  \\ 	
		\hline 
		\hline
	\end{tabular} 
	\caption{
    The production cross sections of background processes at the LHeC and FCC-eh for the leptonic $\tau_\mu$ final state. 	
	}
	\label{tab:crsLep}
\end{table}

For the background 
at the LHeC (FCC-eh),  
we generate 2.1 (1.2) million $\tau^+ \tau^- e^- j j j$,  4.9 (3.0) million $\tau^+ \tau^- \nu_e j j j$, 21 (19) million $\tau^+ \nu_\tau  e^- j j j$,  0.5 (0.5) million  $\tau^+ \nu_\tau  \nu_e j j j$, 2.1 (2.0) million $\mu^+ \mu^- e^- j j j$, 10.5 (6.0) million $\mu^+ \mu^- \nu_e j j j$, 27.4 (24.6) million $\mu^+ \nu_\mu e^- j j j$, 6.0 (6.4) million  $\mu^+ \nu_\mu  \nu_e j j j$, respectively.
The number of simulated events for each background process is determined according to its importance in reducing the statistical uncertainty of the final 
sensitivities.
 
\subsection{Data analysis}
\label{subsec:LdecayAna}

We apply following 
preselection 
to select the signal and reject the background events at the first stage.
(i) Exactly one muon with positive charge, i.e. 
$N(\tau_\mu^+) = 1$ 
and transverse momentum 
$p_T(\tau_\mu) >$ 5 GeV; 
events with final state electron(s) or tau-jet(s) are vetoed.
(ii) All regular jets are sorted in descending order according to their transverse momenta and we require at least three jets, i.e. $N(j)\geq 3$; 
for the $p_T$ thresholds of jets, when heavy neutrino mass is below 80 GeV, the $p_T$ of the first three leading jets are required to be greater than 10 GeV, 
while when the mass is above 80 GeV, we require the first two leading jets have $p_T$ greater than 20 GeV and the third one has $p_T$ greater than 10 GeV.
In Table~\ref{tab:preselLep}, we show the number of events for the signal with benchmark $m_N=120$ GeV and  dominant background processes after applying the 
preselection 
(i)-(ii) sequentially described above.  

\begin{table}[h]
	\centering
		\begin{tabular}{ccccccccc}
			\hline
			\hline
			\multicolumn{2}{c}{}
			& signal & B1 & B2 & B3 & B4 & B5 & B7  \\
			\hline
			\multirow{4}{*}{LHeC} 
			& initial &$1.2\mltp10^{3}$ & $2.2\mltp10^{5}$ & $5.2\mltp10^{4}$   & $2.8\mltp10^{5}$ & 8.2 & $5.8\mltp10^{5}$ & $2.8\mltp10^{5}$  \\ 
			& (i)      & $1.2\mltp10^{2}$ & $4.1\mltp10^{2}$ &  $2.1\mltp10^{3}$ & $2.6\mltp10^{3}$ &  $7.6\mltp10^{-1}$ & $2.6\mltp10^{3}$  & $1.9\mltp10^{4}$  \\ 
			& (ii)     & $9.9\mltp10^{1}$ & $1.4\mltp10^{2}$ & $7.9\mltp10^{2}$  & $8.7\mltp10^{2}$ &  $5.4\mltp10^{-1}$ & $8.0\mltp10^{2}$ & $6.6\mltp10^{3}$  \\ 
			\hline
			\multirow{4}{*}{FCC-eh}
			&	initial & $1.0\mltp10^{4}$ & $2.6\mltp10^{6}$ &  $8.7\mltp10^{5}$&$4.9\mltp10^{6}$ &  $2.8\mltp10^{2}$ & $6.2\mltp10^{6}$  & $4.9\mltp10^{6}$  \\ 
			& (i)       & $1.1\mltp10^{3}$ & $4.5\mltp10^{3}$ &  $4.0\mltp10^{4}$ & $3.8\mltp10^{4}$ &  $15$ & $1.3\mltp10^{4}$  & $2.8\mltp10^{5}$    \\ 
			& (ii)      & $8.7\mltp10^{2}$ & $1.5\mltp10^{3}$ &  $1.9\mltp10^{4}$ & $1.7\mltp10^{4}$ &  $12$ &  $4.1\mltp10^{3}$ &   $1.2\mltp10^{5}$ \\ 
			\hline
			\hline
			
	\end{tabular}
	\caption{
	The number of events for the signal with
		benchmark $m_N$ = 120 GeV
and $ |V_{\tau N}|^2\, |V_{eN}|^2 / \left( |V_{\tau N}|^2 + |V_{eN}|^2 \right) = 5 \times 10^{-5}$, and for six dominant background
processes after applying 
preselection 
(i)-(ii)
		sequentially for the leptonic $\tau_\mu$ final state. 
		The numbers correspond to the LHeC and
		FCC-eh with 1 and 3 $\iab$ integrated luminosity,
		respectively. 
	}
	\label{tab:preselLep}
\end{table}

To further reject the background, we input following nineteen observables to perform the BDT-MVA.
\begin{enumerate}[label*=\Alph*.]
\item Four momenta of the final state 
$\tau_\mu$:
$p_{x}(\tau_\mu)$, $p_{y}(\tau_\mu)$, $p_{z}(\tau_\mu)$, $E(\tau_\mu)$. 
\item Four momenta of the first three final state regular jets:
$p_{x}(j_{1})$, $p_{y}(j_{1})$, $p_{z}(j_{1})$, $E(j_{1})$;
$p_{x}(j_{2})$, $p_{y}(j_{2})$, $p_{z}(j_{2})$, $E(j_{2})$;
$p_{x}(j_{3})$, $p_{y}(j_{3})$, $p_{z}(j_{3})$, $E(j_{3})$.
\item Number of regular jets:
$N(j)$.
\item Magnitude and azimuthal angle of the missing transverse momentum:
$\met$, $\phi(\met)$.
\end{enumerate}

\begin{figure}[h]
	\centering
		\includegraphics[width=7cm,height=5cm]{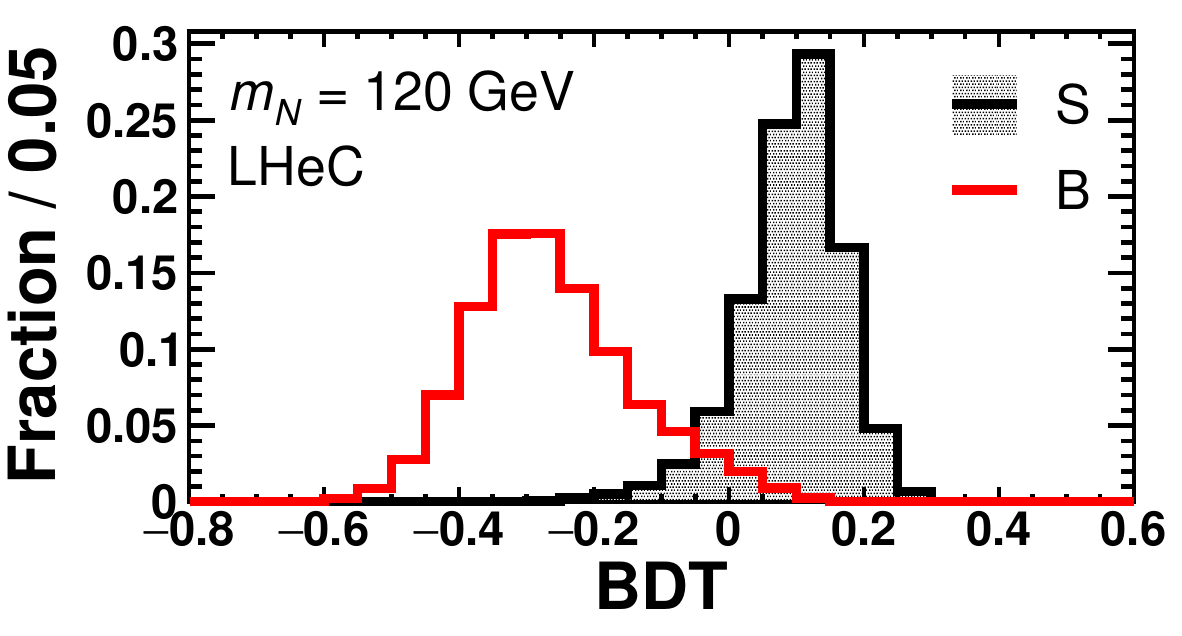}\,\,\,\,\,\,\,\,\,\,
		\includegraphics[width=7cm,height=5cm]{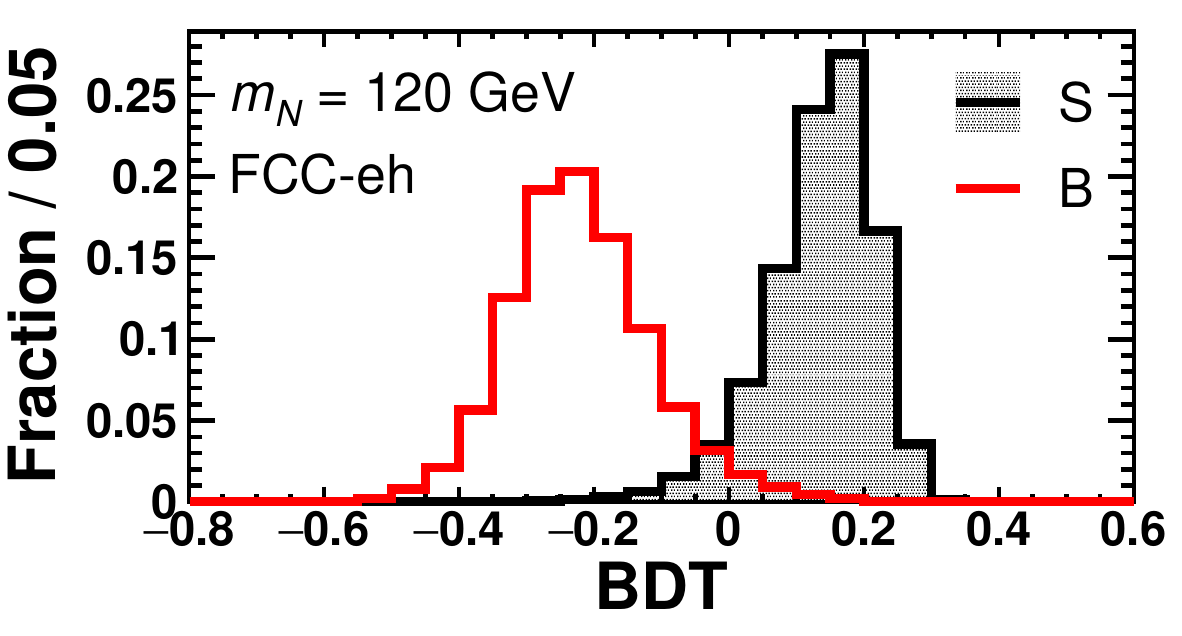}
	\caption{Distributions of BDT responses for the signal with 
benchmark $m_{N}$ = 120 GeV and $ |V_{\tau N}|^2\, |V_{eN}|^2 / \left( |V_{\tau N}|^2 + |V_{eN}|^2 \right) = 5 \times 10^{-5}$ (black, filled) and for
total SM background (red) at the LHeC (left) and FCC-eh (right) for the leptonic $\tau_\mu$ final state.}
	\label{fig:BDTbenchLep}
\end{figure}

In Fig.~\ref{fig:BDTbenchLep}, we show BDT distributions for the total background and the benchmark signal with $m_{N}$ = 120 GeV at the LHeC and FCC-eh.
Since the kinematics of signal varies with $m_N$, distributions of BDT response also change with $m_N$. 
BDT distributions corresponding to other representative heavy neutrino masses are shown in 
Appendix~\ref{subapp:BDTtauL} for the LHeC and FCC-eh.

\section{Results}
\label{sec:results}

\begin{figure}[h]
\centering
\includegraphics[width=7cm,height=5cm]{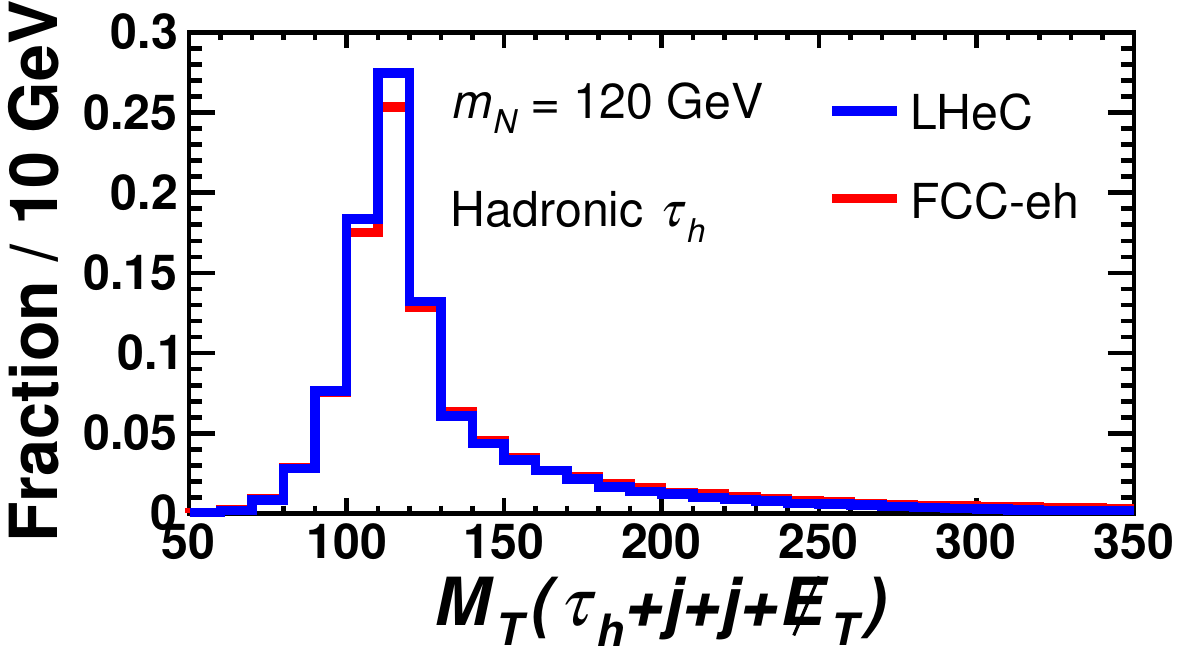}\,\,\,\,\,\,\,\,\,\,
\includegraphics[width=7cm,height=5cm]{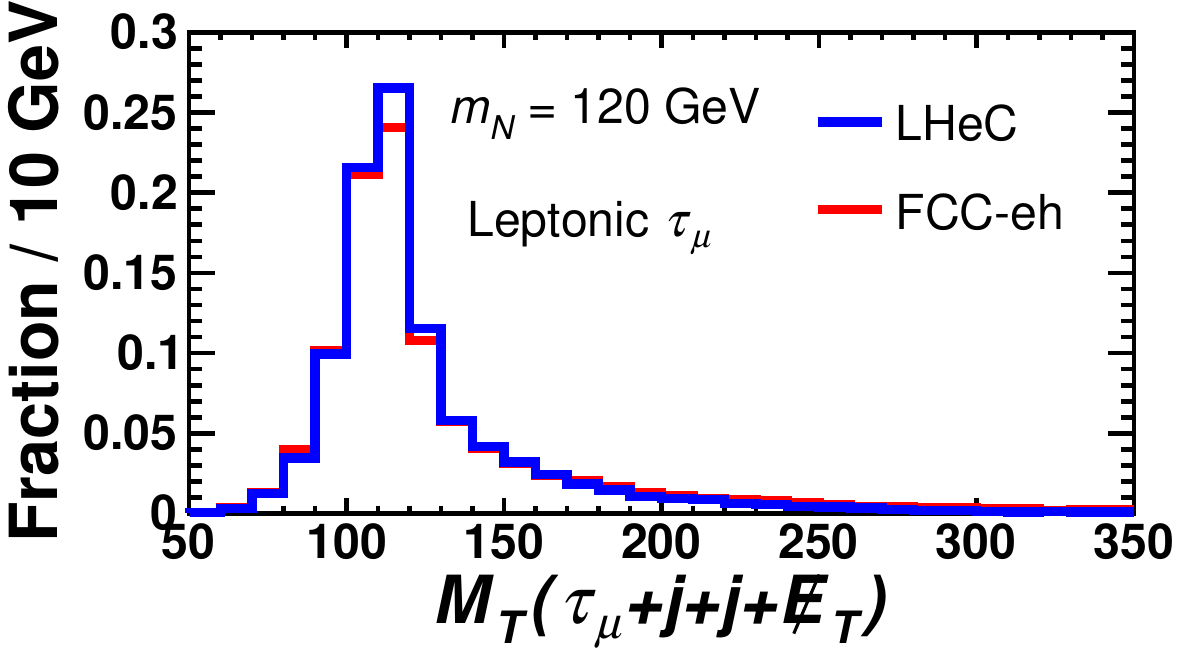}
\caption{
Distributions of transverse mass $M_T$ of the $(\tau+j+j+\met)$ system for the signal with 
benchmark $m_{N}$ = 120 GeV and $ |V_{\tau N}|^2\, |V_{eN}|^2 / \left( |V_{\tau N}|^2 + |V_{eN}|^2 \right) = 5 \times 10^{-5}$
for the hadronic 
$\tau_h$ (left) and leptonic $\tau_\mu$ (right) 
final states, respectively.
}
\label{fig:mN120}
\end{figure}

In this section, based on our former analyses, we show the 
discovery sensitivities
on the parameter
$|V_{\tau N}|^2\, |V_{eN}|^2 / \left( |V_{\tau N}|^2 + |V_{eN}|^2 \right)$ for the heavy neutrino mass $m_N$ in the range of 
10 to 3000 GeV.
We also develop the strategy to 
reconstruct 
the heavy neutrino mass.
Considering the heavy neutrino decays into one tau plus two jets for our signal, when $m_N$ is above $W$-boson mass, we check di-jet combinations for all jets with $p_T > 10$ GeV,
and recognize the combination with its invariant mass closest to 
$W$-boson
mass as the di-jet $(j+j)$ from the $N$ decay.
This di-jet is then combined with final state $\tau_h$ (
$\tau_\mu$ 
) and missing energy to reconstructed transverse mass $M_T$ of the whole 
$(\tau_h +j+j+\met)$ ($(\tau_\mu+j+j+\met)$) 
system.
Here, transverse mass $M_T \equiv \sqrt{ (E_T^{\rm vis.}+\met)^2 - ( \vec{p}_T^{\rm \,\,vis.} + \pmiss )^2 }$\,, where $E_T^{\rm vis.}$ ($\vec{p}_T^{\rm \,\,vis.}$) is the transverse energy (momentum) of the visible object or system, $\met$ ($\pmiss$) is the missing transverse energy (momentum).
$E_T^{\rm vis.} = \sqrt{ (\vec{p}_T^{\rm \,\,vis.})^2 + (m^{\rm vis.})^2 }$\,, where $m^{\rm vis.}$ is the invariant mass of the visible object or system.
$\met = |\pmiss|$, assuming the invariant mass of the invisible object is zero.
Fig~\ref{fig:mN120} shows distributions of 
$M_T(\tau_h+j+j+\met)$ and  $M_T(\tau_\mu+j+j+\met)$ 
for the signal with benchmark $m_{N}$ = 120 GeV after preselection at the LHeC  and FCC-eh  for the hadronic 
$\tau_h$ (left) and leptonic $\tau_\mu$ (right) 
final states, respectively.
One observes that the transverse mass has sharp peak around $m_N$, which means that it can be used to reconstruct the heavy neutrino mass.

We note that above input observables listed in Sec.~\ref{subsec:HdecayAna} and Sec.~\ref{subsec:LdecayAna} including the four-momenta and angles are very basic and usually called low-level variables for the MVA.
One can also construct some complicated observables and input such high-level variables to perform the MVA analysis.
The BDT-MVA combines the information from all input observables with 
correlations. 
Because the low-level observables have already included the information of all final state objects, inputting the high-level observables will give similar BDT distributions and will not improve the final 
sensitivities
a lot, c.f. Ref.~\cite{Gu:2022muc}.
However, distributions of high-level observables are helpful for researchers to understand the kinematics of both signal and background.
They are usually more distinct between the signal and background, and can be used to perform the cut-based analysis.
Therefore, in Appendix~\ref{app:HLobs},
we show distributions of some high-level observables after preselection for the signal and dominant background processes assuming $m_N = 120$ GeV for the hadronic $\tau_h$ and leptonic $\tau_\mu$ final states at the LHeC and FCC-eh, respectively.

After the preselection, the BDT 
selection 
is optimized according to the signal statistical significance calculated by Eq.~(\ref{eqn:statSgf})~\cite{cowan2012discovery, ATLAS:2020yaz, Bhattiprolu:2020mwi}
for each mass case.
\begin{equation}
\sigma_{\rm stat} = \sqrt{2 \left[ \left( N_s+N_b \right) \, {\rm ln}\left(1+\frac{N_s}{N_b} \right) - N_s \right] }\,\, ,
\label{eqn:statSgf}
\end{equation}
where $N_s$ ($N_b$) is the number of signal (total background) events after all 
selections. 

In Table~\ref{tab:allEffHad}, we show selection efficiencies of preselection and BDT 
requirements. 
for both the signal with representative $m_N$ assumptions and background processes at the LHeC and FCC-eh for the hadronic $\tau_h$ final state.
Selection efficiencies for the leptonic $\tau_\mu$ final state are shown in Table~\ref{tab:allEffLep}.
The total selection efficiency is the product of preselection and BDT 
selection 
efficiencies. 
The number of signal or background events after all 
selections 
can be calculated by multiplying the production cross section, collider luminosity and total selection efficiency.

\begin{figure}[h]   
\centering
\includegraphics[width=12cm,height=8cm]{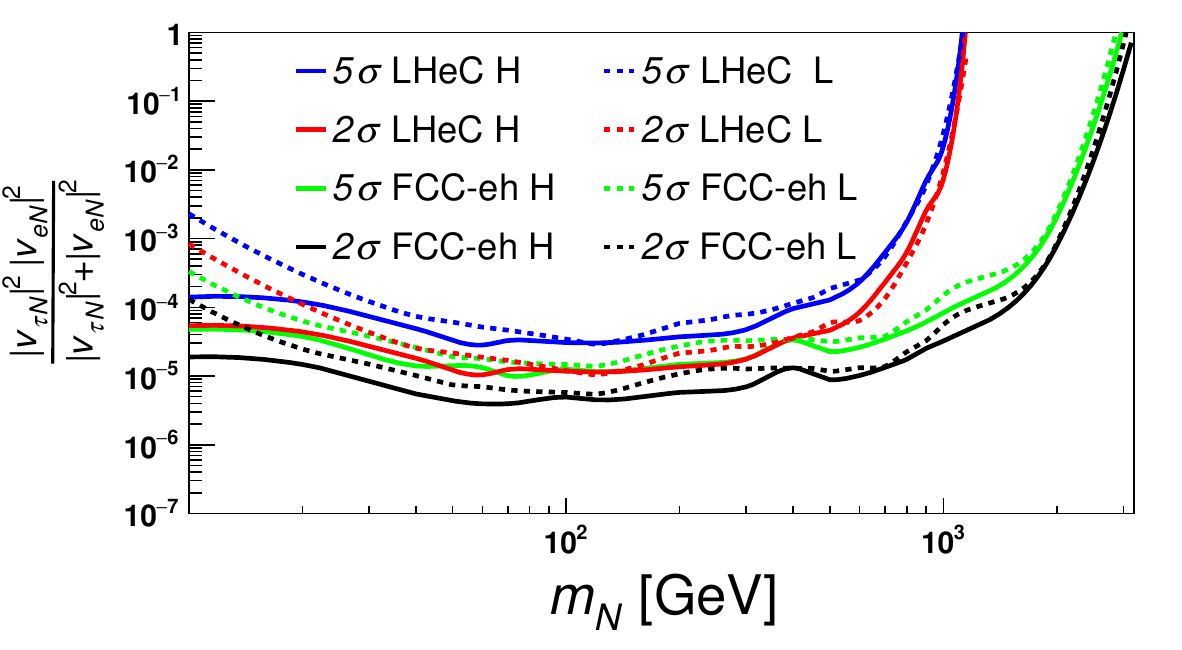}   
\caption{
2- and 5-$\sigma$ 
discovery sensitivities
on the parameter $ |V_{\tau N}|^2\, |V_{eN}|^2 / \left( |V_{\tau N}|^2 + |V_{eN}|^2 \right) $ as varying the heavy neutrino mass in the range of 
10 to 3000 GeV
for the hadronic $\tau_h$ (H) and leptonic $\tau_\mu$ (L) final states at the LHeC and FCC-eh.
}
\label{fig:sensitivity}
\end{figure}

Fig.~\ref{fig:sensitivity} presents 
discovery sensitivities
at the 2- and 5-$\sigma$ significances on the parameter
$|V_{\tau N}|^2 |V_{eN}|^2 / ( |V_{\tau N}|^2 + |V_{eN}|^2 )$ for the heavy neutrino mass in the range of 
10 to 3000 GeV.
Both 
sensitivities
for the hadronic 
$\tau_h$ and leptonic $\tau_\mu$ 
final states at the LHeC and FCC-eh are shown.
At the LHeC, as $m_N$ changes from 10 GeV to 120 GeV, for the leptonic $\tau_\mu$ final state, the 2-$\sigma$ 
discovery sensitivities
on $ |V_{\tau N}|^2\, |V_{eN}|^2 / \left( |V_{\tau N}|^2 + |V_{eN}|^2 \right) $ decrease from $8.4 \times 10^{-4}$  to $1.0 \times 10^{-5}$ , then begin to increase  
afterwards.
The 2-$\sigma$ 
discovery sensitivity
at the FCC-eh has similar behavior as that at the LHeC, but its varying range is much smaller, i.e., between $1.3 \times 10^{-4}$  and $5.4\times 10^{-6}$.
At both colliders, the 5-$\sigma$ 
discovery sensitivities
are slightly weaker than those for 2-$\sigma$.

For the hadronic $\tau_h$ final state,
at the LHeC, as $m_N$ changes from 10 GeV to 100 GeV, the 2-$\sigma$ 
discovery sensitivities
on $ |V_{\tau N}|^2\, |V_{eN}|^2 / \left( |V_{\tau N}|^2 + |V_{eN}|^2 \right) $ decrease from $5.4\times 10^{-5}$ to $1.2 \times 10^{-5}$, then begin to increase  
afterwards.
The 2-$\sigma$ 
discovery sensitivity
at the FCC-eh also has similar behavior as that at the LHeC, but its varying range is smaller, i.e. between $1.9 \times 10^{-5}$  and $5.0 \times 10^{-6}$.
Compared with the leptonic $\tau_\mu$ final state, the hadronic $\tau_h$ final state has bigger signal rate due to larger branching ratio for hadronic tau decay mode, but it suffers from the misidentified tau background which restricts its discovery sensitivities.
Therefore, sensitivities for both final states are found to be similar for most of the parameter space investigated in this study.

\section{Sensitivities compared with current experimental limits}
\label{sec:limits}

\begin{figure}[h]   
\centering
\includegraphics[width=12cm,height=8cm]{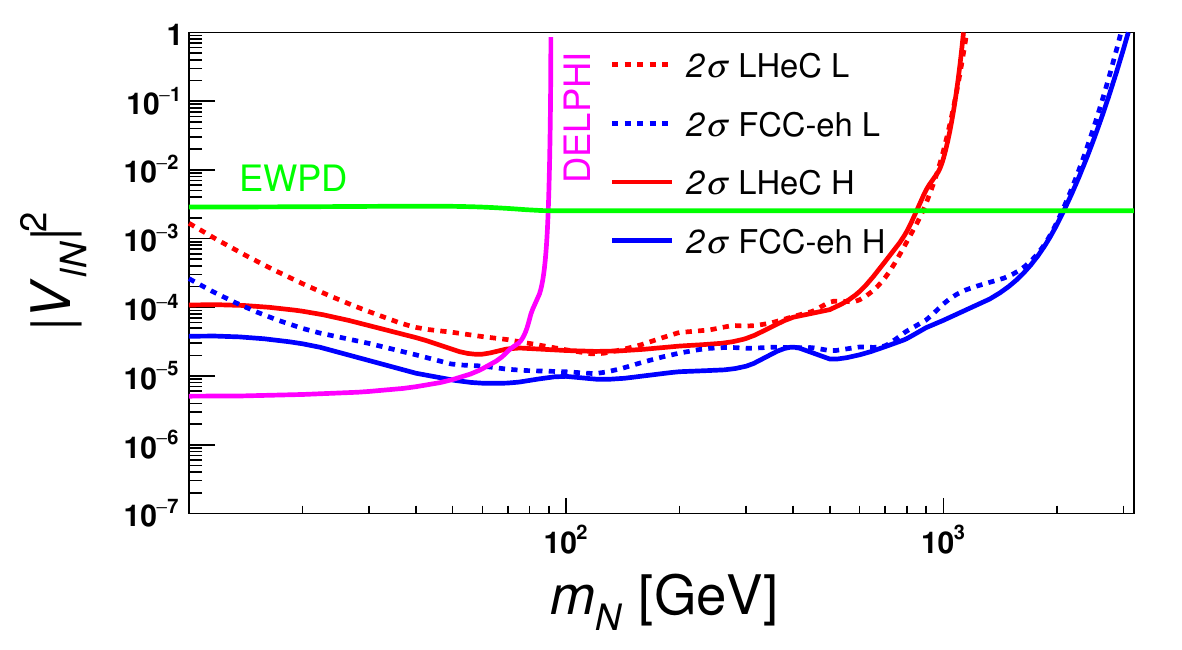}   
\caption{
Assuming the mixing parameter $|V_{\ell N}|^2 = |V_{\tau N}|^2 = |V_{eN}|^2$, 
the 
discovery sensitivities
on $|V_{\ell N }|^2$ at 2-$\sigma$ significance as the heavy neutrino mass changes from 
10 to 3000 GeV
for Hadronic $\tau_h$ (H) and Leptonic $\tau_\mu$ (L) final states at the LHeC and FCC-eh.
The constraints from EWPD~\cite{delAguila:2008pw,Akhmedov:2013hec,Basso:2013jka,deBlas:2013gla,Antusch:2015mia,Cheung:2020buy} (green solid line) and DELPHI experiments~\cite{DELPHI:1996qcc} (purple solid line) are derived and displayed in the same plot for the comparison, see more details in 
Appendix~\ref{app:EWPD}.
}
\label{fig:sensitivity1}
\end{figure}

In Fig. \ref{fig:sensitivity1}, assuming mixing parameters $|V_{\tau N}|^2 = |V_{eN}|^2 = |V_{\ell N}|^2$, the 2-$\sigma$ 
discovery sensitivities
on $|V_{\ell N}|^2$ are shown for both the hadronic $\tau_h$ and leptonic $\tau_\mu$ final states at the LHeC and FCC-eh.
With the same assumption,
we also derive the limits on $|V_{\ell N}|^2$ from Electroweak Precision Data (EWPD)~\cite{delAguila:2008pw,Akhmedov:2013hec,Basso:2013jka,deBlas:2013gla,Antusch:2015mia,Cheung:2020buy} and DELPHI experiment~\cite{DELPHI:1996qcc}, and show them in the same plot for the comparison. 
The details of derivation process are shown in Appendix~\ref{app:EWPD}.  
We observe that sensitivity bounds from the LHeC and FCC-eh are stronger than those from EWPD when $m_N \lesssim  900$ GeV,
and also stronger than those from DELPHI when $m_N \gtrsim 70$ GeV.

In this study, we require mixing parameters $|V_{\tau N}|^2, |V_{e N}|^2 \neq 0$ and $|V_{\mu N}|^2 = 0$, and set 
discovery sensitivities
on the parameter $|V_{eN}|^2|V_{\tau N}|^2/\left(|V_{eN}|^2+|V_{\tau N}|^2\right)$, which is a function of both  $|V_{\tau N}|^2$ and $|V_{e N}|^2$.
Because of the challenges in detecting the final state taus, current experimental limits on mixing parameter $|V_{\tau N}|^2$ are much weaker than those on $|V_{e N}|^2$ and $|V_{\mu N}|^2$.
Therefore, it is interesting to investigate the individual 
sensitivities
on $|V_{\tau N}|^2$.

\begin{figure}[h]   
\centering
\includegraphics[width=7cm,height=5cm]{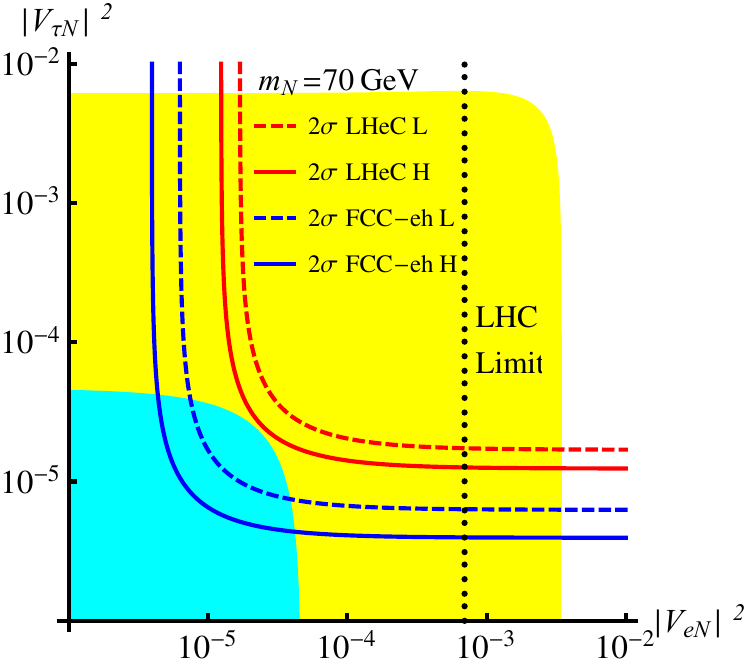} \,\,\,\,\,\,\,\,\,\,
\includegraphics[width=7cm,height=5cm]{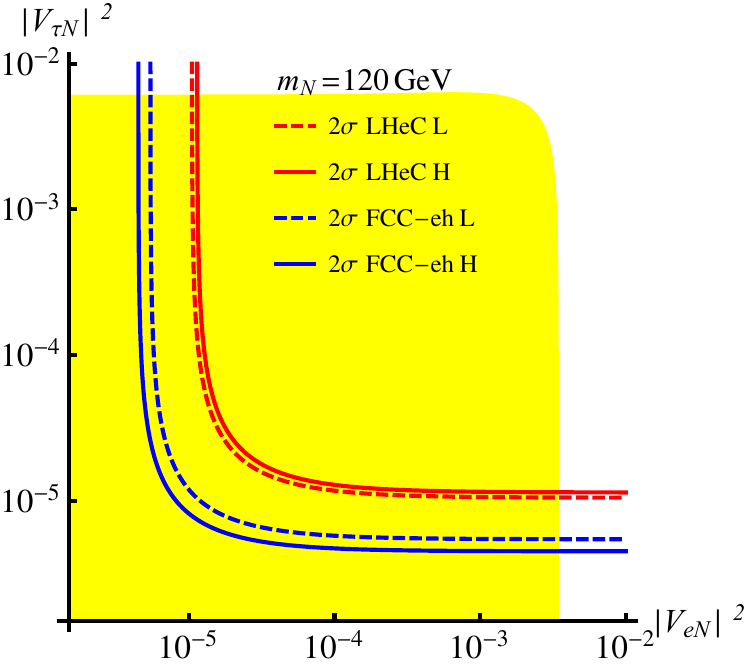}  
\includegraphics[width=7cm,height=5cm]{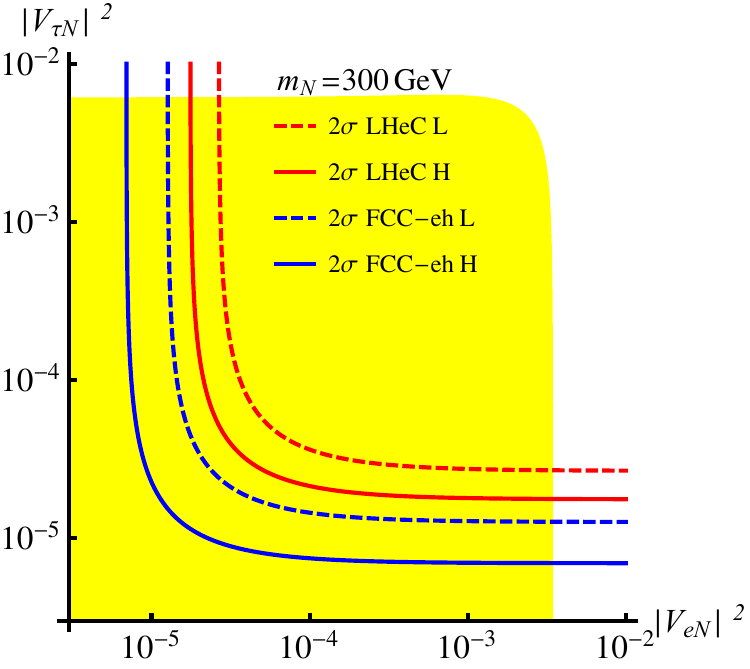}\,\,\,\,\,\,\,\,\,\,
\includegraphics[width=7cm,height=5cm]{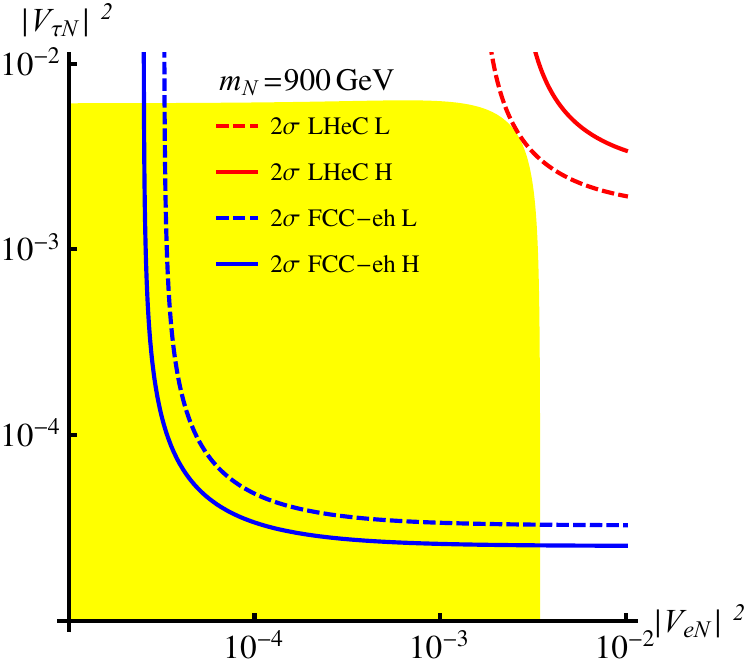}  
\caption{
Assuming $m_N =$ 70, 120, 300, 900 GeV, the 2-$\sigma$ 
sensitivity
curves for the Hadronic $\tau_h$ (H, solid) and Leptonic $\tau_\mu$ (L, dashed) final states at the LHeC (red) and FCC-eh (blue) in the $|V_{\tau N}|^2$ vs. $|V_{e N}|^2$ plane, where the top right parameter space of the curves are excluded.
From Appendix~\ref{app:EWPD}, yellow regions in all plots are allowed by the EWPD~\cite{delAguila:2008pw,Akhmedov:2013hec,Basso:2013jka,deBlas:2013gla,Antusch:2015mia,Cheung:2020buy}, while the cyan region in the top left plot is allowed by the DELPHI measurement~\cite{DELPHI:1996qcc}.
Constraints on $|V_{eN}|^2$ from the LHC experiments~\cite{CMS:2018iaf,CMS:2018jxx,ATLAS:2019kpx} are shown as black dotted line, which appears only in the top left plot, because these constraints are too weak for large $m_N$.  
}
\label{fig:VtauN2_mN1}
\end{figure}

In Fig.~\ref{fig:VtauN2_mN1}, assuming representative heavy neutrino masses, we show the expected 2-$\sigma$ 
discovery sensitivities
at future LHeC and FCC-eh colliders in the $|V_{\tau N}|^2$ vs. $|V_{e N}|^2$ plane, together with existed limits as a comparison. 
For any given mass, the expected 2-$\sigma$ 
discovery sensitivity
for $|V_{eN}|^2|V_{\tau N}|^2/\left(|V_{eN}|^2+|V_{\tau N}|^2\right)$ is a fixed number (as shown in Fig. \ref{fig:sensitivity}), which behaves as a hyperbola curve in the $|V_{\tau N}|^2$ vs. $|V_{e N}|^2$ plane.
The top right parameter space of the curves is excluded.
From Appendix~\ref{app:EWPD}, yellow regions in all plots are allowed by the EWPD~\cite{delAguila:2008pw,Akhmedov:2013hec,Basso:2013jka,deBlas:2013gla,Antusch:2015mia,Cheung:2020buy}, while the cyan region in the top left plot is allowed by the DELPHI measurement~\cite{DELPHI:1996qcc}.
Constraints on $|V_{eN}|^2$ from the LHC experiments~\cite{CMS:2018iaf,CMS:2018jxx,ATLAS:2019kpx} are shown as black dotted line, which appears only in the top left plot, because this constraint is too weak for large $m_N$. 

In the top left plot, we choose $m_N=70~\textrm{GeV}$ as a typical benchmark point when $m_N<m_Z$. 
In this case, results show that the future FCC-eh~(both leptonic $\tau_\mu$ and hadronic $\tau_h$ final states) and LHeC experiments (only hadronic $\tau_h$ final state) are both possible to probe additional parameter space compared with the DELPHI experiment. 
In other three plots, we choose $m_N=120$, $300$, and $900~\textrm{GeV}$ as benchmark points, and all of them locate in the range where $m_N>m_Z$. In these cases,  the combination of measurements from EWPD and searches at $pe$ collider constrain the parameter space synergistically, while the limits on $|V_{eN}|^2$ from the LHC experiments are much weaker and not shown. 
Our results show that except at very large $m_N$ region, for example, $m_N\gtrsim900~\textrm{GeV}$, the expected 
sensitivity
from LHeC experiment will be similar to that from EWPD, due to the quickly decreasing signal production cross section.
For other mass ranges, future LHeC and FCC-eh experiments are expected to 
significantly enhance the discovery potential for a large portion of the $|V_{\tau N}|^2$ vs. $|V_{e N}|^2$ plane,
compared with the EWPD and LHC searches.

\section{Conclusion and discussion}
\label{sec:discussion}

In this paper, we utilize the lepton number violation signal process of $p\, e^- \to \tau^+ jjj$ to search for heavy Majorana neutrinos at the future electron-proton colliders.
We consider the LHeC (FCC-eh) running with an electron beam energy of 60 GeV, a proton beam energy of 7 (50) TeV and an integrated luminosity of 1 (3) $\iab$.
The electron beam is considered to be unpolarized.
To simplify the analyses, we consider the
phenomenological simplified Type-I 
and assume that only one generation of heavy neutrinos $N$ is within the collider access and mixes with active neutrinos of tau and electron flavours, i.e. $|V_{\tau N}|^2, |V_{e N}|^2 \neq 0$ and $|V_{\mu N}|^2 = 0$.
We perform analyses for both hadronic $\tau_h$ and leptonic $\tau_\mu$ final states, and forecast 
discovery sensitivities
for heavy neutrinos mass $m_N$ in the range 
10-3000 GeV 
at LHeC and FCC-eh. 

The cross sections of the LNV signal $p e^- \to \tau^+ jjj$ are presented in Fig.~\ref{fig:crs} as a function of $m_N$ at the LHeC and FCC-eh.
We apply detector configurations and simulate signal and related SM background events for both hadronic $\tau_h$ and leptonic $\tau_\mu$ final states.
When tau decays hadronically (leptonically), we select the final state containing the tau-jet (muon).
The 
preselections 
are applied to select the signal and reject the background events at the first stage.
Then various observables are input, and the BDT algorithm is adopted to perform the multivariate analysis and maximally reject the background.
The BDT distributions with different $m_N$ assumptions are presented and some high-level observables corresponding benchmark $m_N$ case are shown in appendices.

The strategy to 
reconstruct 
the heavy neutrino mass is developed and distributions of reconstructed mass are presented.
After the preselection, the BDT 
selection 
is optimized to maximize the signal statistical significance.
In Appendix~\ref{app:Eff}, we show selection efficiencies of preselection and BDT 
requirements. 
for both the signal with representative $m_N$ assumptions and background processes at the LHeC and FCC-eh.
The 
discovery sensitivities
on parameter $|V_{\tau N}|^2 |V_{eN}|^2 / ( |V_{\tau N}|^2 + |V_{eN}|^2 )$ for the heavy neutrino mass in the range of 
10 to 3000 GeV 
are presented in Fig.~\ref{fig:sensitivity}.
At the 2-$\sigma$ significance, the best sensitivity
is $\sim 1.2 \times 10^{-5}\,\,(5.0 \times 10^{-6})$ at the LHeC (FCC-eh) when $m_N \sim 100$ GeV for the hadronic $\tau_h$ final state.
The discovery sensitivities for the leptonic $\tau_\mu$ final state are found to be similar for most of the parameter space investigated than those for the hadronic $\tau_h$ final state. 

We derive the limits on the mixing parameters from EWPD and DELPHI experiment in Appendix~\ref{app:EWPD}. 
Assuming $|V_{\tau N}|^2 = |V_{eN}|^2 = |V_{\ell N}|^2$, 
sensitivities
on $|V_{\ell N}|^2$ from $pe$ collider searches are compared with those from the EWPD and DELPHI experiment in Fig. \ref{fig:sensitivity1}.
We find that sensitivity bounds from the LHeC and FCC-eh are stronger than those from EWPD for almost all the mass range, and also stronger than those from DELPHI when $m_N \gtrsim 70$ GeV.

To investigate the individual 
sensitivities
on $|V_{\tau N}|^2$, in Fig.~\ref{fig:VtauN2_mN1}, assuming representative $m_N$, we show the expected 2-$\sigma$ 
sensitivity
curves from $pe$ collider searches in the $|V_{\tau N}|^2$ vs. $|V_{e N}|^2$ plane, together with existed limits from EWPD, DELPHI, and LHC experiments for comparison.
Our results show that compared with current experimental limits, 
future $pe$ experiments can probe large additional regions in the parameter space formed by $|V_{\tau N}|^2$ and $|V_{e N}|^2$, and 
thus significantly enhance the discovery potential for a large portion of the $|V_{\tau N}|^2$ vs. $|V_{e N}|^2$ plane.

Compared our sensitivity results with those in Ref~\cite{Gu:2022muc}, at $pe$ colliders, sensitivities for the $\tau^+$ final state are weaker than the $\mu^+$ final state. This is mainly because that (i) due to the challenges in detecting the final state taus, the signal rate is much lower for the $\tau^+$ final state; (ii) the $\tau^+$ final state also suffers from larger background.
Besides, as pointed in Ref~\cite{Gu:2022muc}, 
because the total background cross section for the $e^+$ final state is about three times larger than that for the $\mu^+$ final state, sensitivities for the $e^+$ final state is expected to be weaker than those for the $\mu^+$ final state.
However, because the $p\, e^- \to e^+ jjj$ signal process can depend on the mixing parameter $|V_{e N}|^2$ only, it is a unique channel to probe $|V_{e N}|^2$ independent of other mixing parameters.
In this sense, the detailed analyses of the $e^+$ final state are still meaningful, and we leave it for future studies.

A jet can be misidentified as a hadronic $\tau_h$ in the detector, which can contribute to the background for the hadronic $\tau_h$ final state.
Because of the huge production cross section, even a small misidentification rate of jet to hadronic tau can still lead to large background.
To estimate its effects, we include the multi-jet  process ``$p\, e^- \to \nu_e   j  j  j  j$'' in the background.
Our analysis indicates that the $\Delta R(\tau, j)$ and BDT 
requirements 
can reject such background effectively.
However, because this background is greatly affected by the detector's performance and the detectors of future $pe$ colliders are still under development, it should be carefully considered for future  studies.

\appendix

\section{Constraints from $Z\rightarrow N\nu, N\bar{\nu}$ rare decay channels and EWPD}
\label{app:EWPD}

In this appendix, we show details about the constraints from $Z$-boson rare decay \cite{DELPHI:1996qcc} and EWPD \cite{delAguila:2008pw,Akhmedov:2013hec,Basso:2013jka,deBlas:2013gla,Antusch:2015mia,Cheung:2020buy}. 
In this paper, since we consider the signal process $p\, e^-
\rightarrow N(\rightarrow \tau^+W^-)\, j$ corresponding to both $W$-$N$-$e$ and $W$-$N$-$\tau$ vertices, thus we assume mixing parameters $|V_{eN}|^2$ and $|V_{\tau N}|^2$ are both nonzero. We also assume $|V_{\mu N}|^2=0$ here for simplicity.

First, we consider the $Z$-boson rare decay processes. If $m_N<m_Z$, the two-body rare decay channels $Z\rightarrow N\nu,N\bar{\nu}$ are open \footnote{If $m_N<m_Z/2$, the decay channel $Z\rightarrow NN$ is also open. However, its 
partial decay width behaves as $\Gamma_{Z\rightarrow NN}\propto|V_{\ell N}|^4$, which is always ignorable comparing
with the $Z\rightarrow N\nu,N\bar{\nu}$ decay channels.}
with the partial width 
\begin{equation}
\frac{\Gamma_{Z\rightarrow N\nu_{\ell}}}{\Gamma_{Z\rightarrow\nu_{\ell}\bar{\nu}_{\ell}}}=
\frac{\Gamma_{Z\rightarrow N\bar{\nu}_{\ell}}}{\Gamma_{Z\rightarrow\nu_{\ell}\bar{\nu}_{\ell}}}=
\left|V_{\ell N}\right|^2f\left(\frac{m_N}{m_Z}\right),
\end{equation}
with the function 
\begin{equation}
f(x)\equiv\left\{\begin{array}{cc}\left(1-x^2\right)^2 \left(1+x^2/2\right),&(x\leq1);\\0&(x>1).\end{array}\right.
\end{equation}
Finally, we obtain the branching ratio for exotic decay 
\begin{equation}
\label{eq:Br}
\textrm{Br}_{\textrm{exo}}\simeq0.13\left(|V_{eN}|^2+|V_{\tau N}|^2\right)f\left(\frac{m_N}{m_Z}\right).
\end{equation}

Experimentally, $\textrm{Br}_{\textrm{exo}}\lesssim1.3\times10^{-6}$ at $95\%$ C.L. \cite{DELPHI:1996qcc}, and thus Eq.~(\ref{eq:Br}) corresponds to the upper limit
\begin{equation}
|V_{eN}|^2+|V_{\tau N}|^2\lesssim\frac{10^{-5}}{f(m_N/m_Z)}
\end{equation}
for $m_N<m_Z$. Assuming $\left|V_{\tau N}\right|^2 = \left|V_{eN}\right|^2 = \left|V_{\ell N}\right|^2$, we show the $95\%$ C.L. upper limit on $\left|V_{\ell N}\right|^2$ in Fig. \ref{fig:sensitivity1} as a comparison with the expected future LHeC or FCC-eh 
sensitivities
in the $m_N<m_Z$ region.

When $m_N>m_Z$, the two body $Z\rightarrow N\nu,N\bar{\nu}$ decay channels are closed and thus we need to find
the constraints from EWPD. In the EWPD studies, the observables $\alpha$, $G_F$, and $m_Z$ which have the best accuracy are treated as inputs \cite{Zyla:2020zbs,Workman:2022ynf}
\begin{eqnarray}
m_Z=91.2~\textrm{GeV},&\quad\quad&G_F=1.166\times10^{-5}~\textrm{GeV}^{-2},\nonumber\\
\alpha^{-1}(0)=137.0,&\quad\quad&\alpha^{-1}(m_Z)=128.0.
\end{eqnarray}
Other observables are derived from $\alpha$, $G_F$, and $m_Z$. For example, in the SM, we have the tree level relation 
\footnote{Here $\sin\theta_W$, $\cos\theta_W$, and $\tan\theta_W$ are separately denoted as $s_W$, $c_W$, 
and $t_W$ for simplicity.}
\begin{equation}
\label{eq:sw2}
s^2_W c^2_W=\frac{\pi \alpha}{ \sqrt{2}G_F m^2_Z }\,\,.
\end{equation}
The weak coupling is denoted as $G_W\equiv\frac{g^2}{4\sqrt{2}m^2_W}$, and is equivalent to the
Fermi constant $G_F$ in the SM. However, experimentally the Fermi constant $G_F$ is extracted from the $\mu\rightarrow e\nu_e\bar{\nu}_{\mu}$ decay process, which should be modified from $G_W$ by a nonzero $|V_{eN}|^2$ as 
\begin{equation}
G_F=G_W\sqrt{1-|V_{eN}|^2}\simeq G_W\left(1-\frac{|V_{eN}|^2}{2}\right).
\end{equation}
Based on Eq.~(\ref{eq:sw2}), we have the modification of $s^2_W$ comparing with its SM value as 
\begin{equation}
\frac{s^2_W}{s^2_{W,\,\textrm{SM}}}-1=-\frac{c^2_{W,\,\textrm{SM}}}{2c_{2W,\,\textrm{SM}}}|V_{eN}|^2=-0.714\, |V_{eN}|^2.
\end{equation}
The partial decay widths of a $Z$-boson to fermion pairs should be
\begin{equation}
\Gamma_{Z\rightarrow f\bar{f}}=\frac{N_c\, G_W\, m^3_Z}{6\sqrt{2}\pi}\left(g^2_{V,f}+g^2_{A,f}\right),
\end{equation}
where $g_{V,f}=T_{3,f}-2Q_fs^2_W$ and $g_{A,f}=T_{3,f}$. Thus the partial decay widths of $Z$ to fermion pairs
are also modified due to the modifications in $G_W$ and $s^2_W$ as following.
\begin{eqnarray}
\frac{\Gamma_{Z\rightarrow\ell^+\ell^-}}{\Gamma_{Z\rightarrow\ell^+\ell^-,\, \textrm{SM}}}&=&
\frac{G_W}{G_F}\left(1+0.101\left|V_{eN}\right|^2\right);\\
\frac{\Gamma_{Z\rightarrow\nu_{e}\bar{\nu}_e,\nu_{\tau}\bar{\nu}_{\tau}}}{\Gamma_{Z\rightarrow\nu_{e}\bar{\nu}_e,\, \nu_{\tau}\bar{\nu}_{\tau},\, \textrm{SM}}}&=&\frac{G_W}{G_F}\left(1-\left|V_{eN,\, \tau N}\right|^2\right)^2;\\
\frac{\Gamma_{Z\rightarrow\nu_{\mu}\bar{\nu}_{\mu}}}{\Gamma_{Z\rightarrow\nu_{\mu}\bar{\nu}_{\mu},\, \textrm{SM}}}&=&
\frac{G_W}{G_F};\\
\frac{\Gamma_{Z\rightarrow u_i\bar{u}_i}}{\Gamma_{Z\rightarrow u_i\bar{u}_i,\, \textrm{SM}}}&=&
\frac{G_W}{G_F}\left(1+0.294\left|V_{eN}\right|^2\right);\\
\frac{\Gamma_{Z\rightarrow d_i\bar{d}_i}}{\Gamma_{Z\rightarrow u_i\bar{u}_i,\, \textrm{SM}}}&=&
\frac{G_W}{G_F}\left(1+0.206\left|V_{eN}\right|^2\right);\\
\frac{\Gamma_{Z\rightarrow\textrm{had}}}{\Gamma_{Z\rightarrow\textrm{had},\, \textrm{SM}}}&=&
\frac{G_W}{G_F}\left(1+0.235\left|V_{eN}\right|^2\right).
\end{eqnarray}
In above equations, $\ell$ denotes charged leptons, $\nu$ denotes neutrinos, $u_i$ denotes up-type quarks, $d_i$ denotes down-type quarks, and the index ``had'' denotes all hadrons. We then obtain the modification of the 
$Z$-boson
total decay width as
\begin{equation}
\frac{\Gamma_Z}{\Gamma_{Z,\, \textrm{SM}}}=\frac{G_W}{G_F}\Bigg[1+0.175\left|V_{eN}\right|^2 
-0.133\left(\left|V_{eN}\right|^2+\left|V_{\tau N}\right|^2\right)\left(1-f\left(\frac{m_N}{m_Z}\right)\right)\Bigg].
\label{eq:GaZ}
\end{equation}
If $m_N>m_Z$, Eq.~(\ref{eq:GaZ}) becomes
\begin{equation}
\frac{\Gamma_Z}{\Gamma_{Z,\, \textrm{SM}}}=1+0.541\left|V_{eN}\right|^2-0.133\left|V_{\tau N}\right|^2.
\end{equation}
Some useful observables in EWPD study are defined as \cite{Zyla:2020zbs,Workman:2022ynf}
\begin{eqnarray}
R_{\ell}\equiv\frac{\Gamma_{Z\rightarrow\textrm{had}}}{\Gamma_{Z\rightarrow\ell^+\ell^-}},\quad
R_q\equiv\frac{\Gamma_{Z\rightarrow q\bar{q}}}{\Gamma_{Z\rightarrow\textrm{had}}},\quad\textrm{and}\quad\nonumber\\
\sigma_H\equiv\sigma_{e^+e^-\rightarrow Z\rightarrow\textrm{had}}=
\frac{12\pi\, \Gamma_{Z\rightarrow e^+e^-}\, \Gamma_{Z\rightarrow\textrm{had}}}{m^2_Z\, \Gamma^2_Z}.
\end{eqnarray}
The modifications of these quantities can be derived as 
\begin{eqnarray}
\frac{R_{\ell}}{R_{\ell,\, \textrm{SM}}}-1&=&0.134\left|V_{eN}\right|^2;\\
\frac{R_c}{R_{c,\, \textrm{SM}}}-1&=&0.059\left|V_{eN}\right|^2;\\
\frac{R_c}{R_{c,\, \textrm{SM}}}-1&=&-0.029\left|V_{eN}\right|^2;\\
\frac{\sigma_H}{\sigma_{H,\, \textrm{SM}}}-1&=&0.267\left(\left|V_{eN}\right|^2+\left|V_{\tau N}\right|^2\right) \left(1-f\left(\frac{m_N}{m_Z}\right)\right) -0.013\left|V_{eN}\right|^2.
\label{eq:sigH}
\end{eqnarray}
If $m_N>m_Z$, Eq.~(\ref{eq:sigH}) becomes
\begin{equation}
\frac{\sigma_H}{\sigma_{H,\, \textrm{SM}}}-1=0.254\left|V_{eN}\right|^2+0.267\left|V_{\tau N}\right|^2.
\end{equation}

We perform the global-fit using the observables $R_{q=b,c}$, $R_{\ell=e,\mu,\tau}$, $\sigma_H$, $\Gamma_Z$ \cite{Zyla:2020zbs,Workman:2022ynf}, and also $s^2_W$ extracted from Tevatron \cite{CDF:2018cnj} and LHC \cite{ATLAS:2018gqq}. Assuming $\left|V_{\tau N}\right|^2 = \left|V_{eN}\right|^2 =\left|V_{\ell N}\right|^2$,
we show the $95\%$ C.L. upper limit on $\left|V_{\ell N}\right|^2$ in Fig. \ref{fig:sensitivity1} as a comparison with the expected future LHeC or FCC-eh 
sensitivities.
It depends weakly on $m_N$, and in the $m_N>m_Z$ region, the limit is about
$\left|V_{\ell N}\right|^2\lesssim2.5\times10^{-3}$. Our results are consistent with those in \cite{Antusch:2015mia,Cheung:2020buy} \footnote{Numerical study shows that the constraint from $\tau\rightarrow e\gamma$ rare decay is always weaker than that from EWPD and $Z\rightarrow N\nu,N\bar{\nu}$ rare decay in the whole $m_N$ region, and thus we do not show it in our results.}.

\newpage
\section{Distributions of BDT responses}
\label{app:BDT}

\subsection{Hadronic $\tau_h$ final state }
\label{subapp:BDTtauH}

\begin{figure}[h]
\centering
\includegraphics[width=7cm,height=5cm]{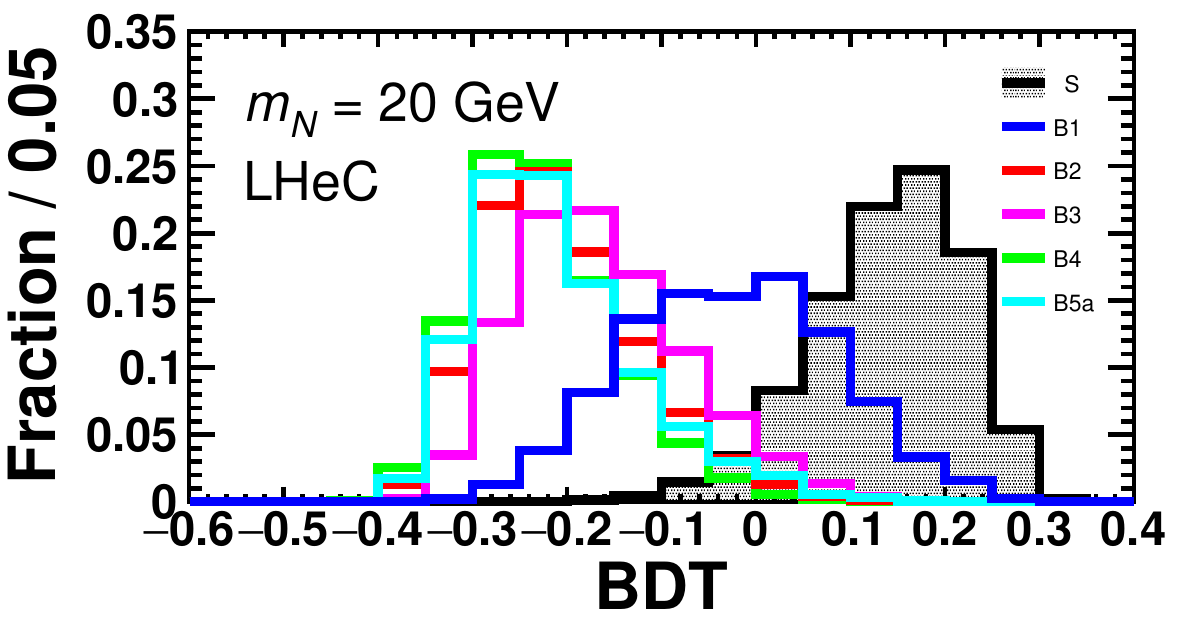}\,\,\,\,\,\,\,\,
\includegraphics[width=7cm,height=5cm]{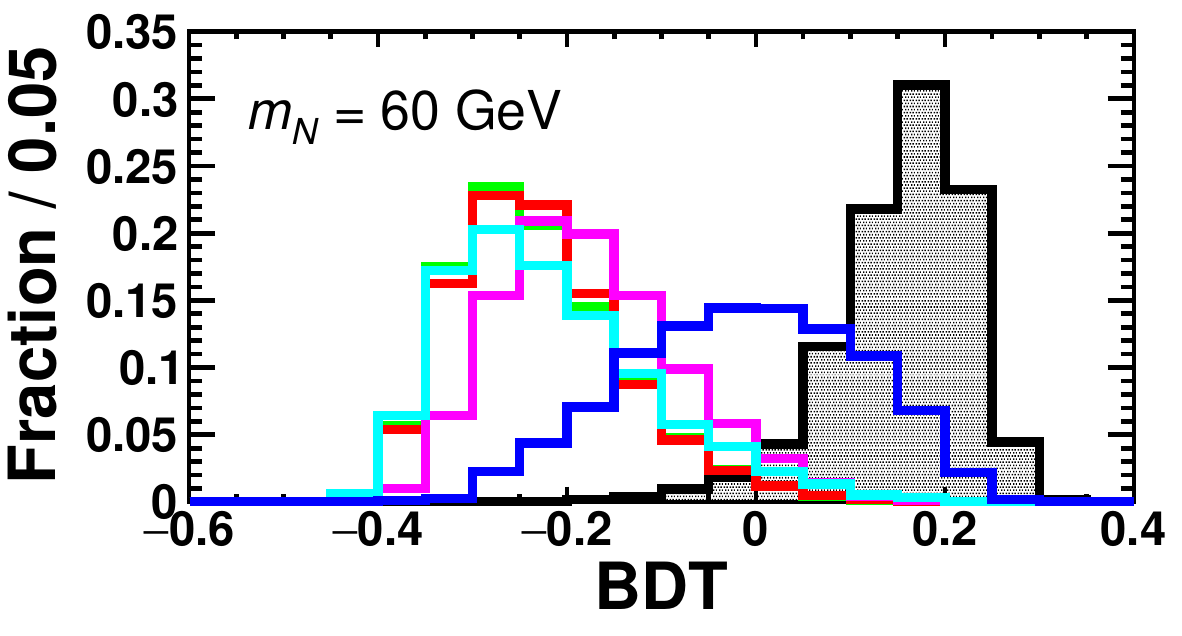}
\includegraphics[width=7cm,height=5cm]{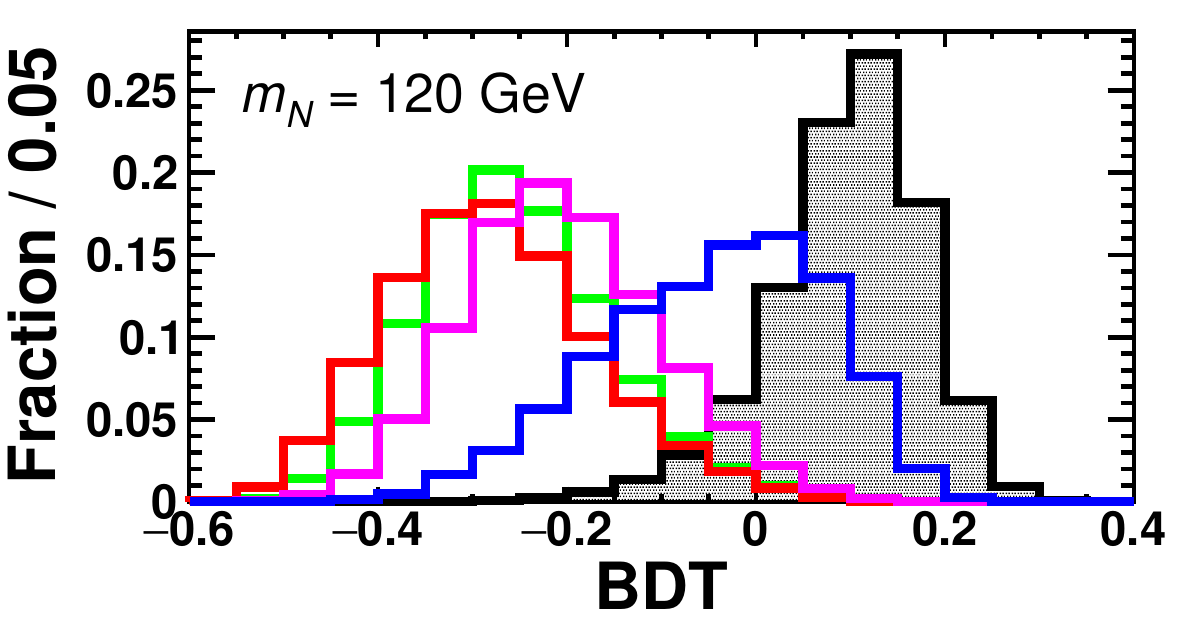}\,\,\,\,\,\,\,\,
\includegraphics[width=7cm,height=5cm]{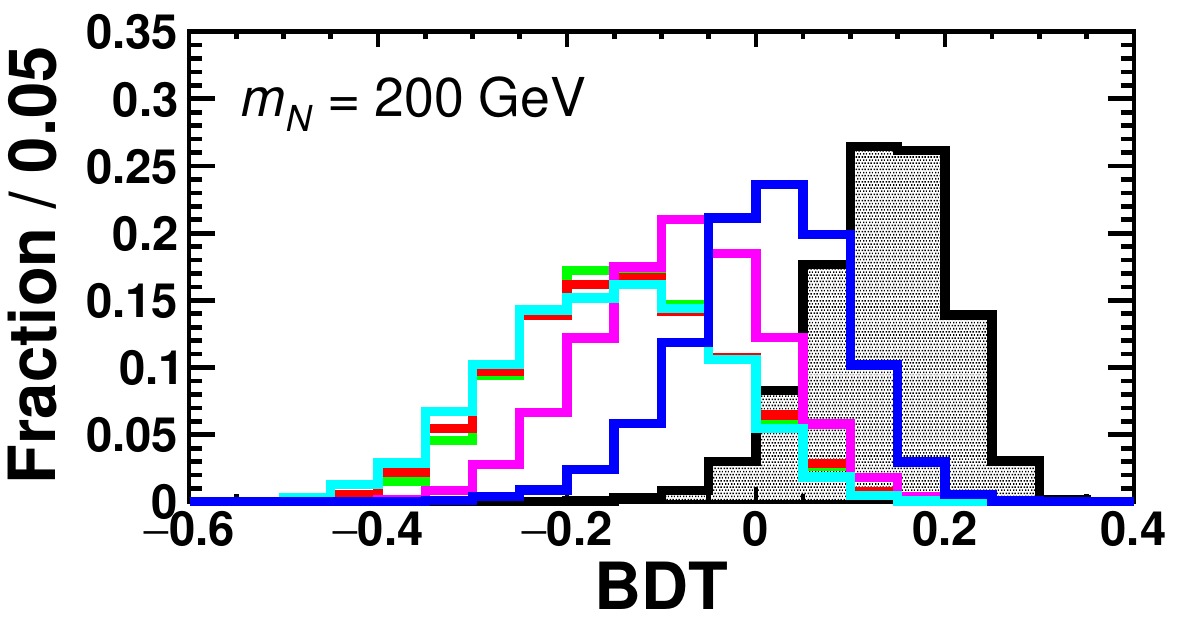}
\includegraphics[width=7cm,height=5cm]{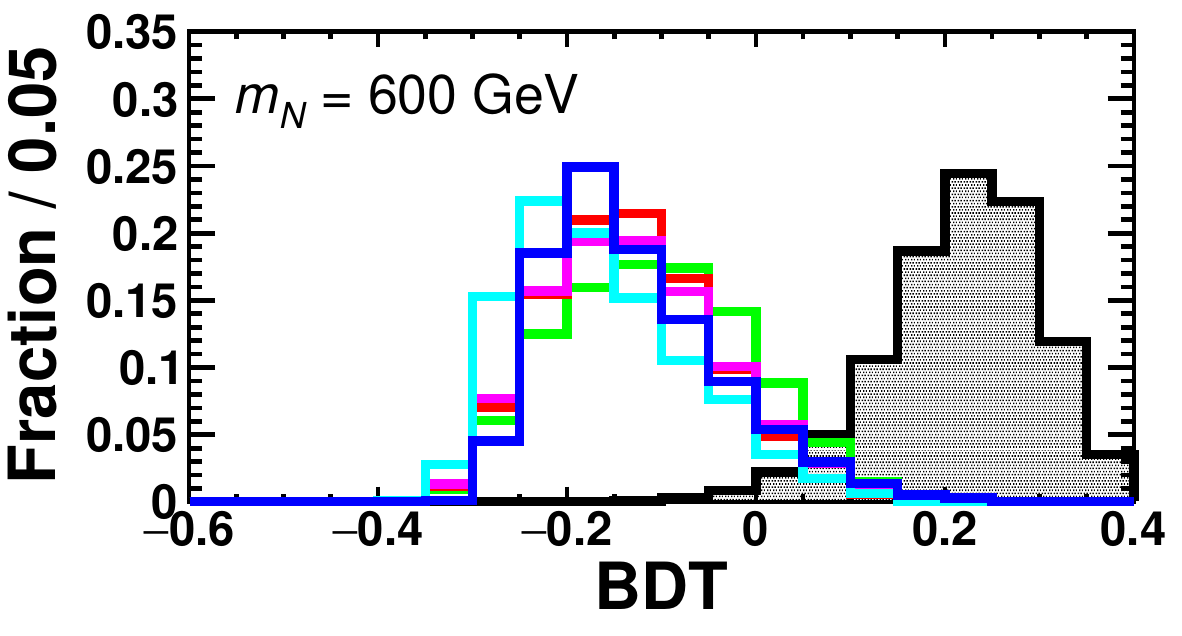}\,\,\,\,\,\,\,\,
\includegraphics[width=7cm,height=5cm]{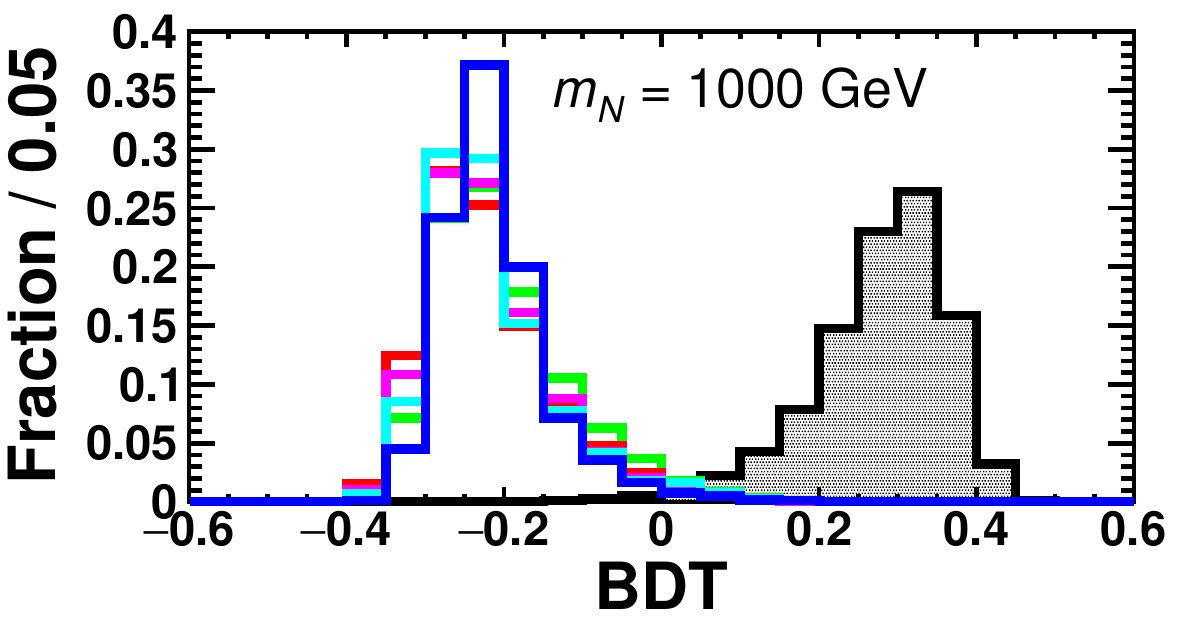}
\caption{
Distributions of BDT responses for the signal (black, filled) and background processes at the LHeC with different $m_N$ assumptions
and fixing $ |V_{\tau N}|^2\, |V_{eN}|^2 / \left( |V_{\tau N}|^2 + |V_{eN}|^2 \right) = 5 \times 10^{-5}$,
for the hadronic $\tau_h$ final state.
}
\label{fig:BDTLHeCHad}
\end{figure}

\begin{figure}[h] 
\centering
\includegraphics[width=7cm,height=5cm]{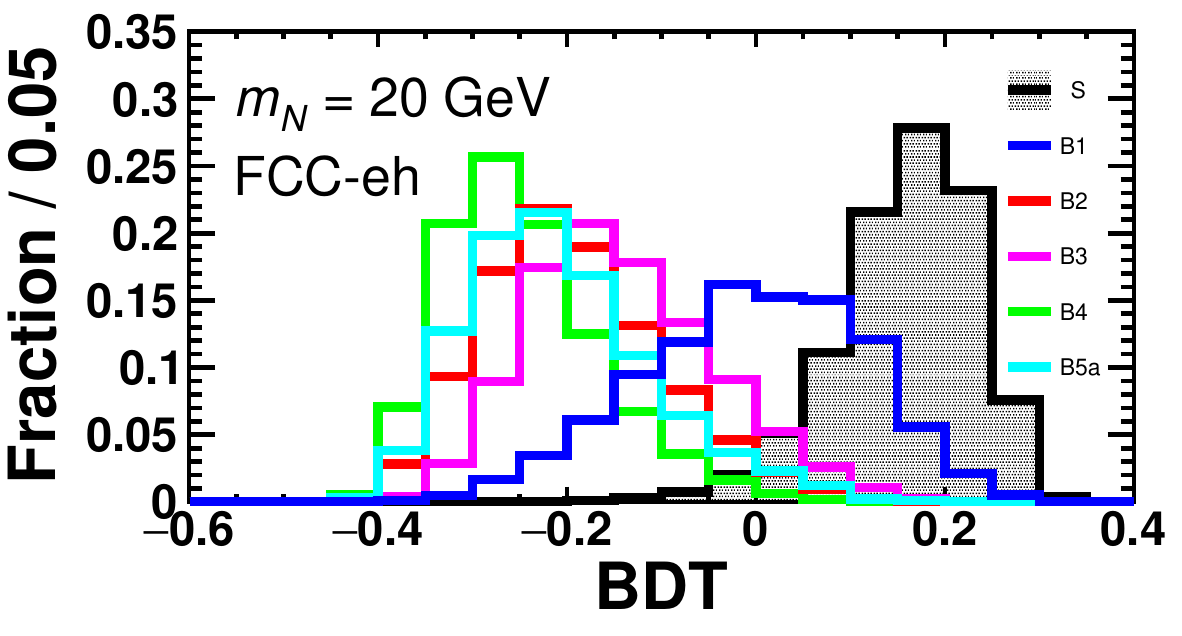}\,\,\,\,\,\,\,\,
\includegraphics[width=7cm,height=5cm]{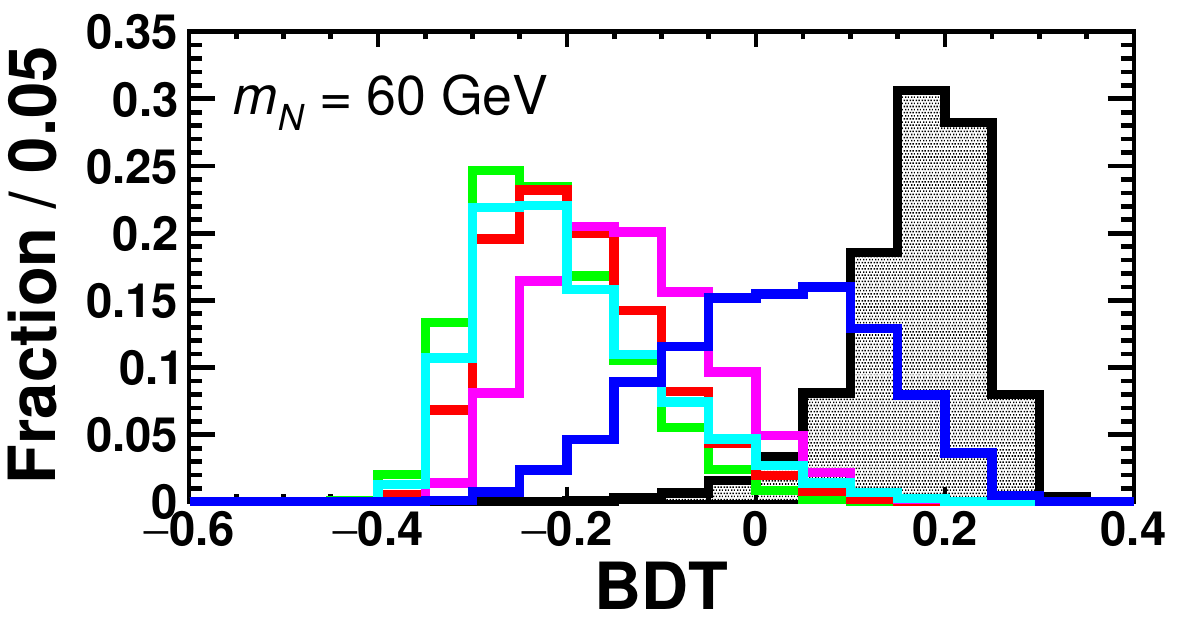}
\includegraphics[width=7cm,height=5cm]{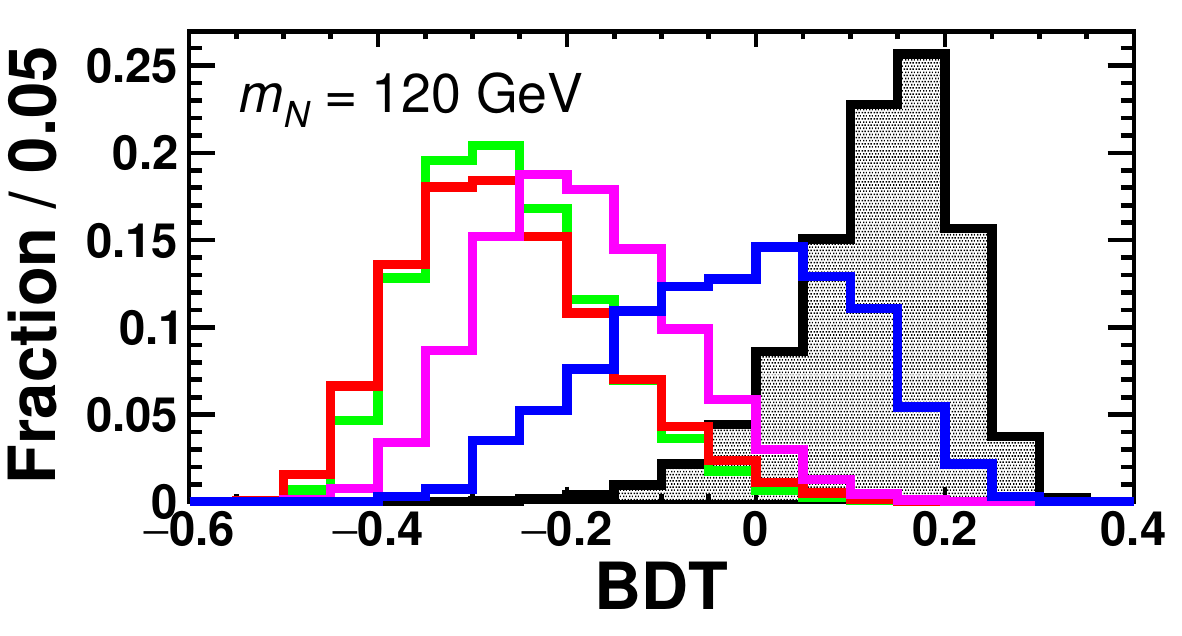}\,\,\,\,\,\,\,\,
\includegraphics[width=7cm,height=5cm]{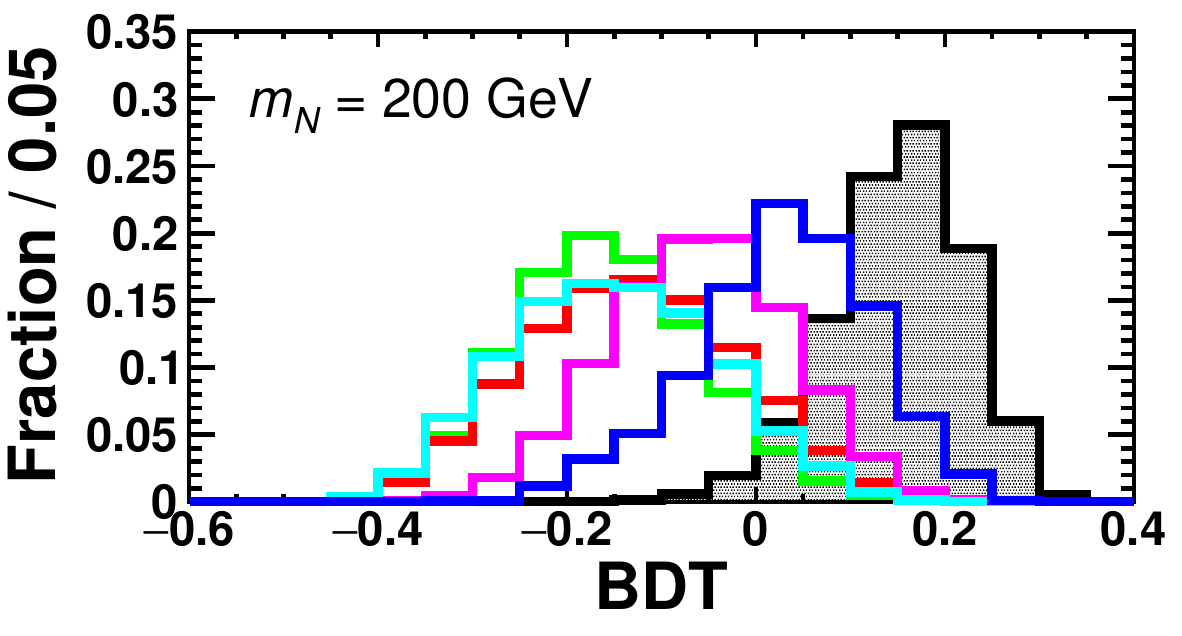}
\includegraphics[width=7cm,height=5cm]{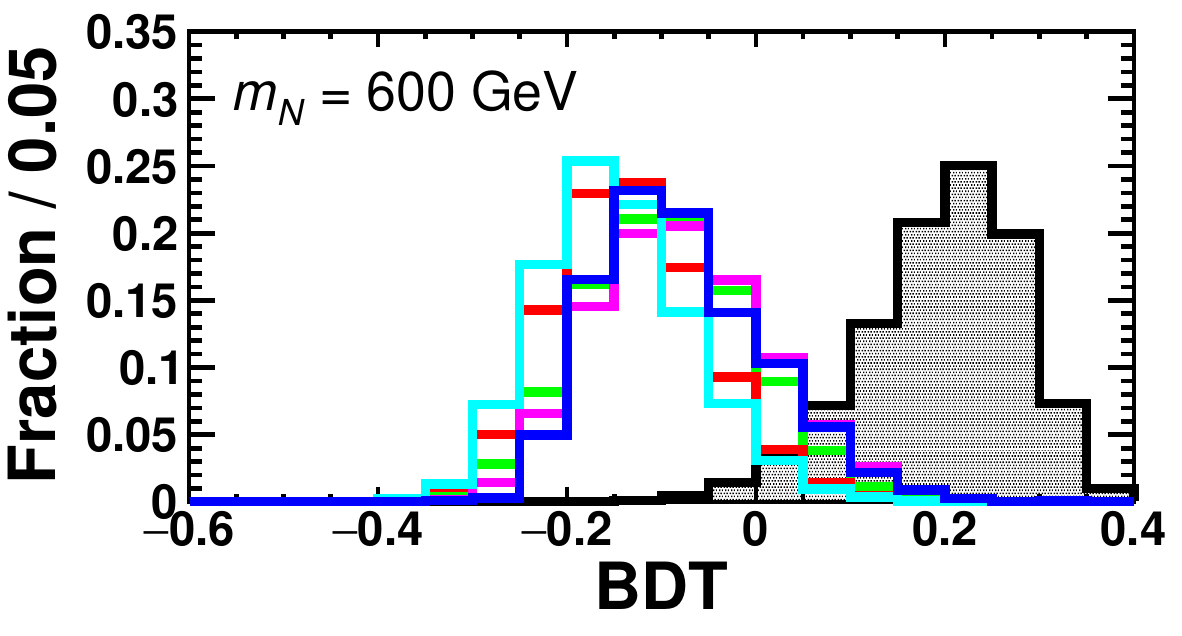}\,\,\,\,\,\,\,\,
\includegraphics[width=7cm,height=5cm]{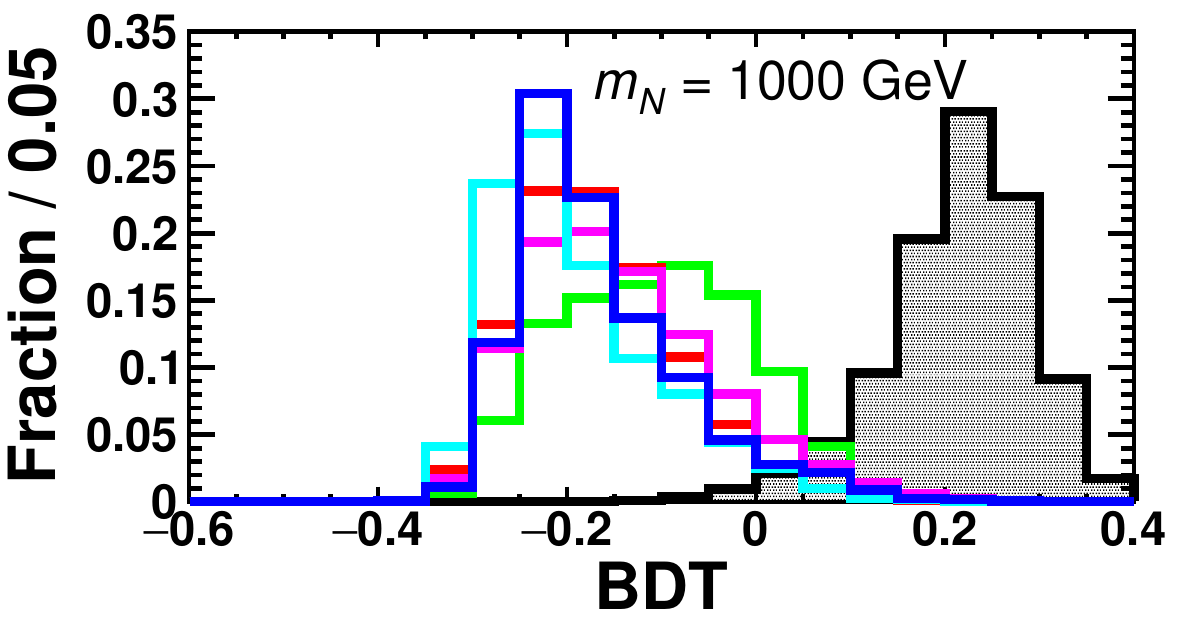}
\caption{
Similar as Fig.~\ref{fig:BDTLHeCHad}, but at the FCC-eh.
}
\label{fig:BDTFCC-ehHad}
\end{figure}

\newpage
\subsection{Leptonic $\tau_\mu$ final state }
\label{subapp:BDTtauL}

\begin{figure}[h] 
\centering
\includegraphics[width=7cm,height=5cm]{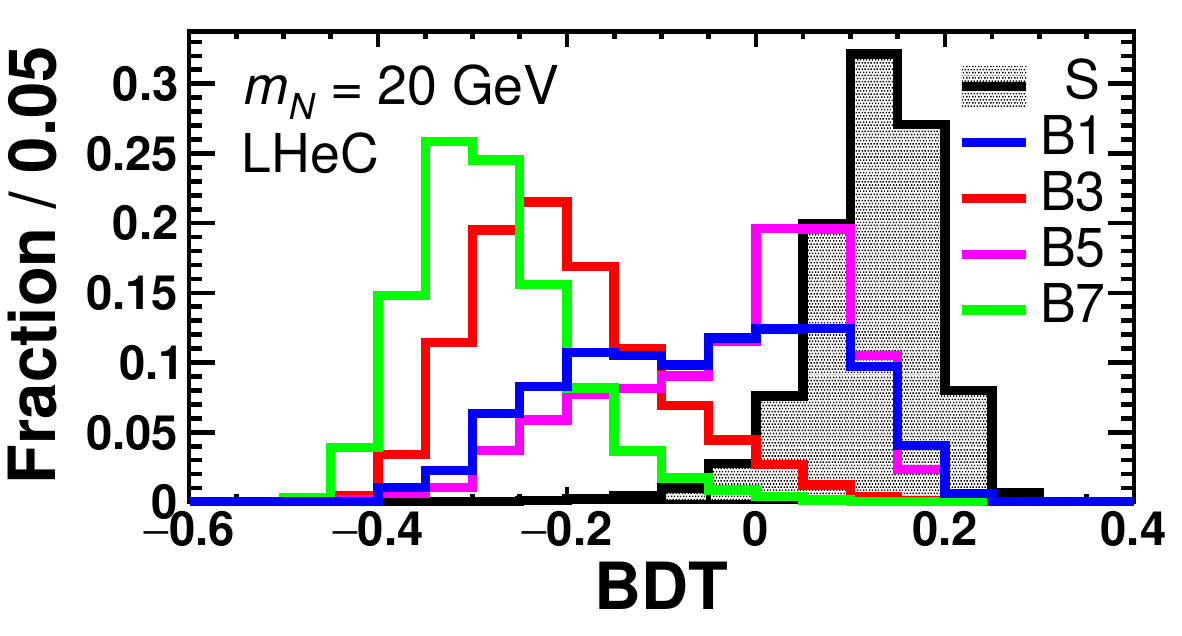}\,\,\,\,\,\,\,\,
\includegraphics[width=7cm,height=5cm]{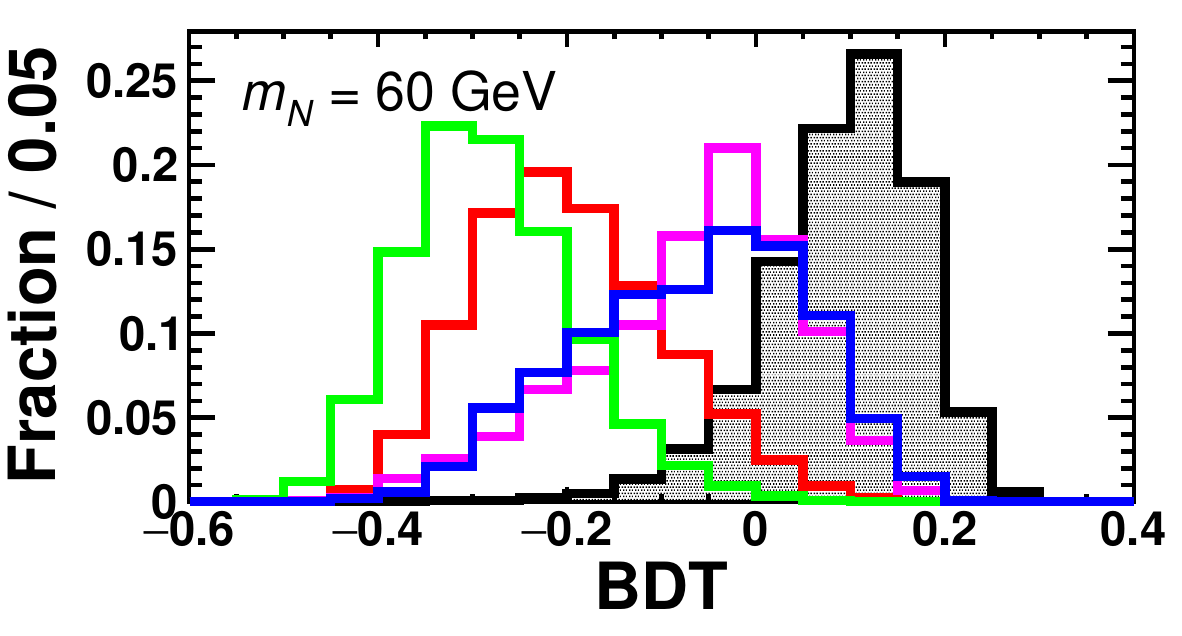}
\includegraphics[width=7cm,height=5cm]{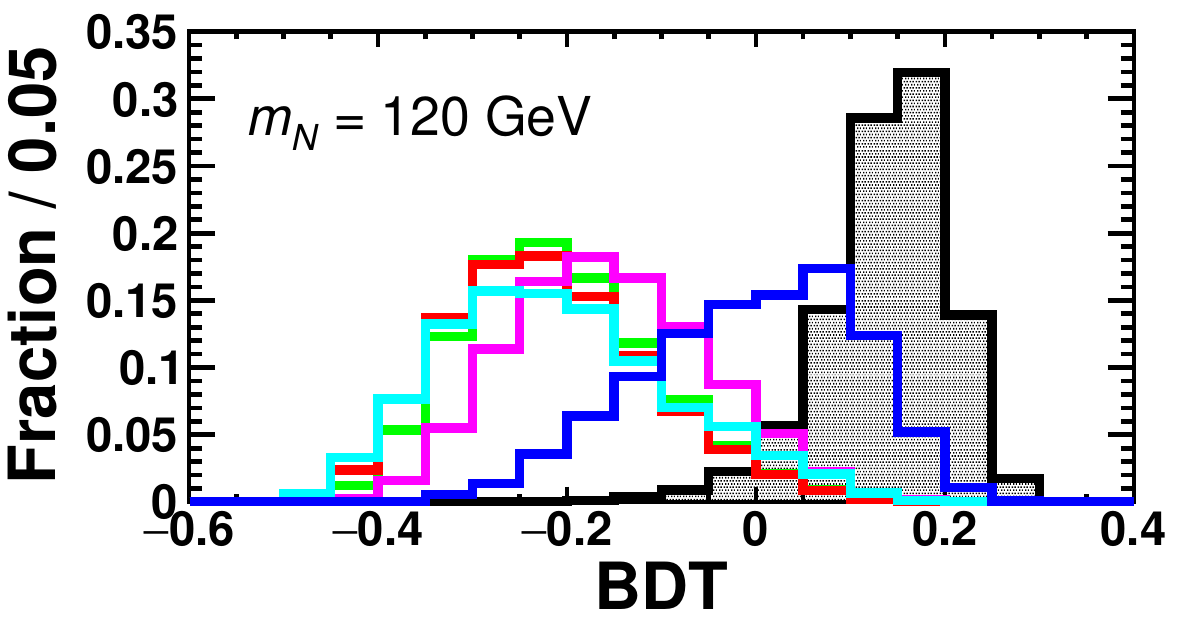}\,\,\,\,\,\,\,\,
\includegraphics[width=7cm,height=5cm]{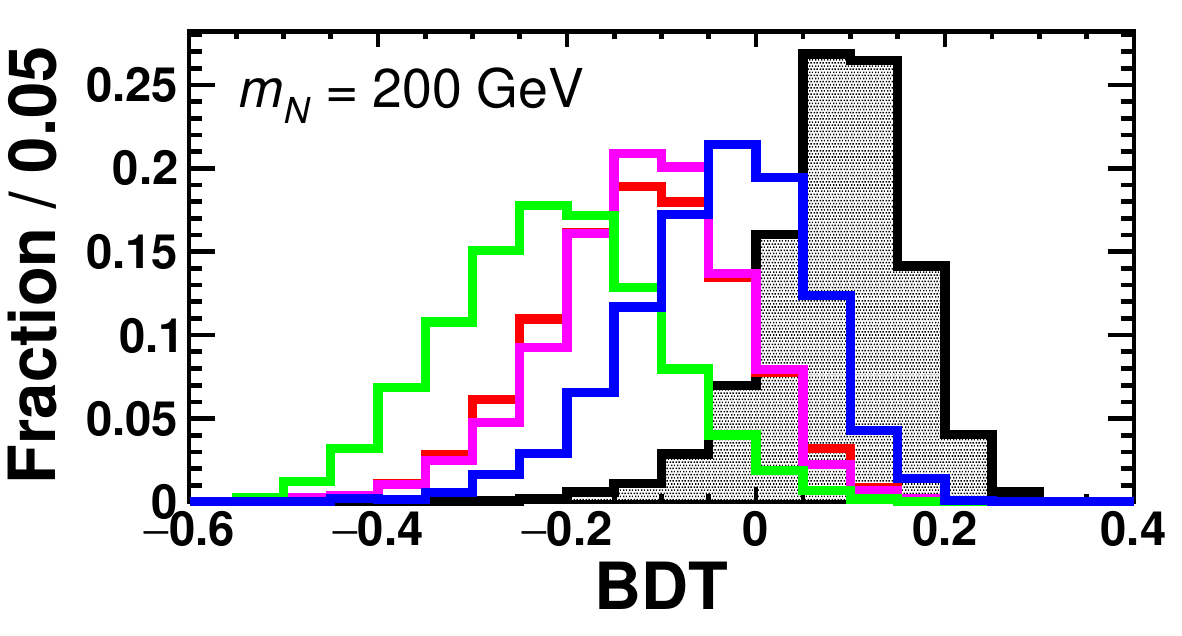}
\includegraphics[width=7cm,height=5cm]{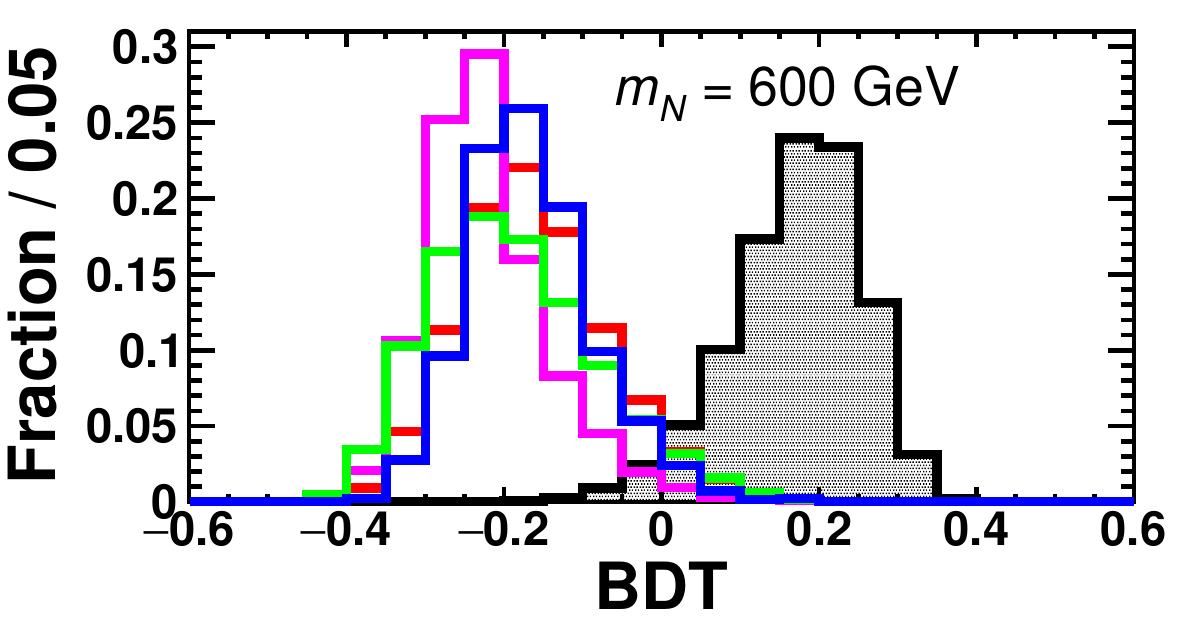}\,\,\,\,\,\,\,\,
\includegraphics[width=7cm,height=5cm]{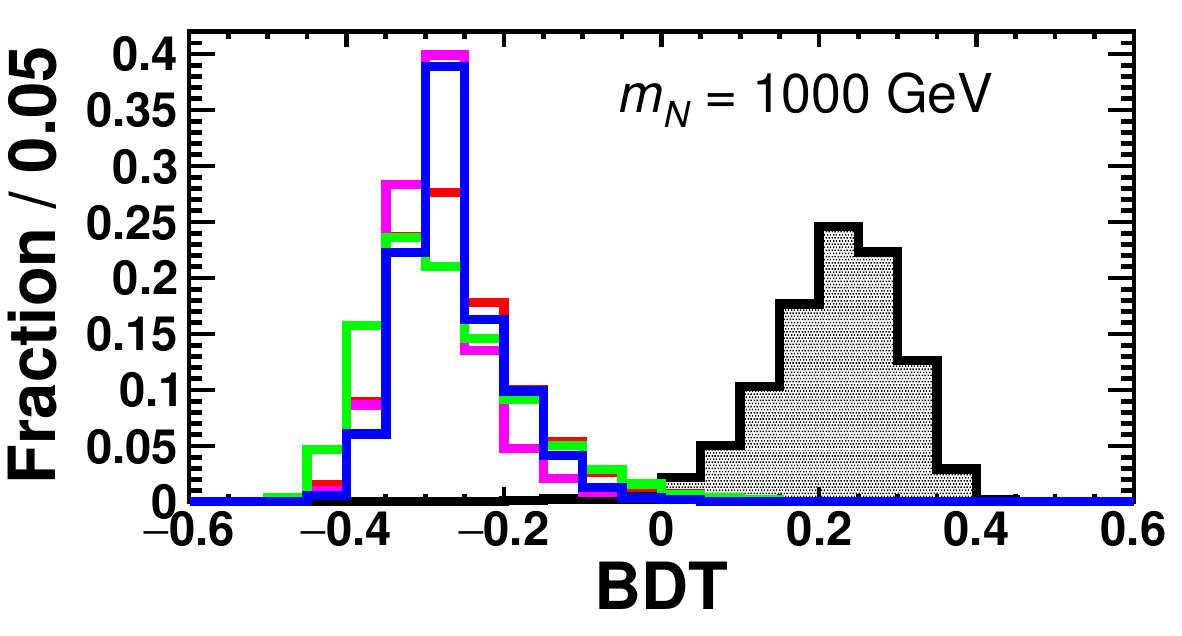}
\caption{
Distributions of BDT responses for the signal (black, filled) and dominant background processes at the LHeC with different $m_N$ assumptions
and fixing $ |V_{\tau N}|^2\, |V_{eN}|^2 / \left( |V_{\tau N}|^2 + |V_{eN}|^2 \right) = 5 \times 10^{-5}$,
for the leptonic $\tau_\mu$ final state.
}
\label{fig:BDTLHeCLep}
\end{figure}

\begin{figure}[h] 
\centering
\includegraphics[width=7cm,height=5cm]{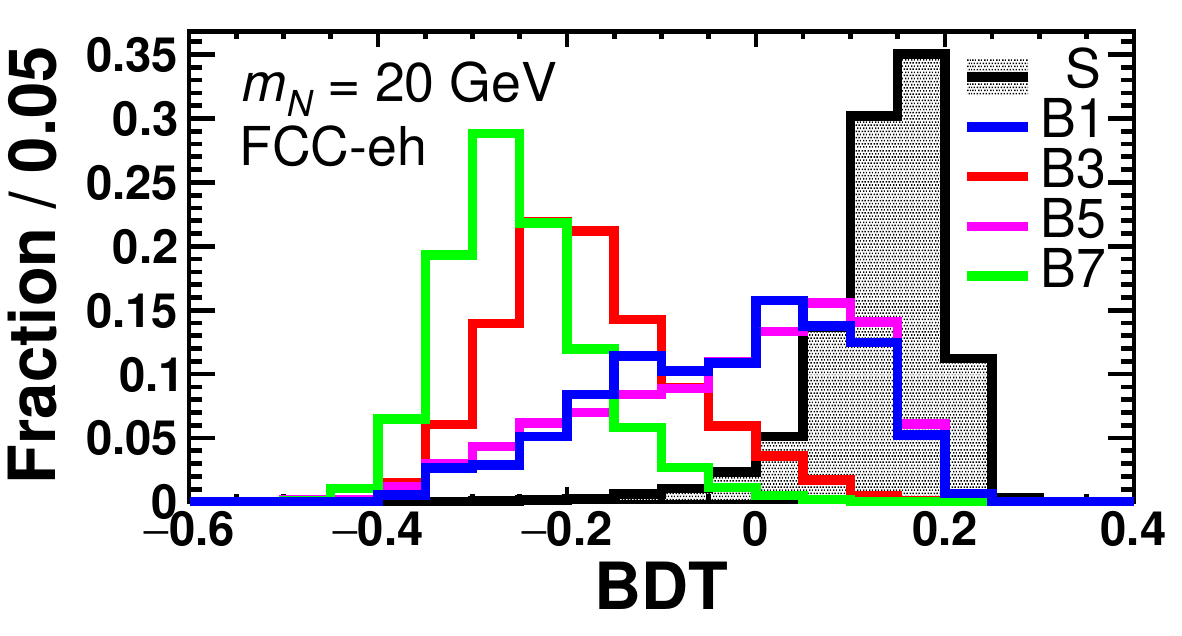}\,\,\,\,\,\,\,\,
\includegraphics[width=7cm,height=5cm]{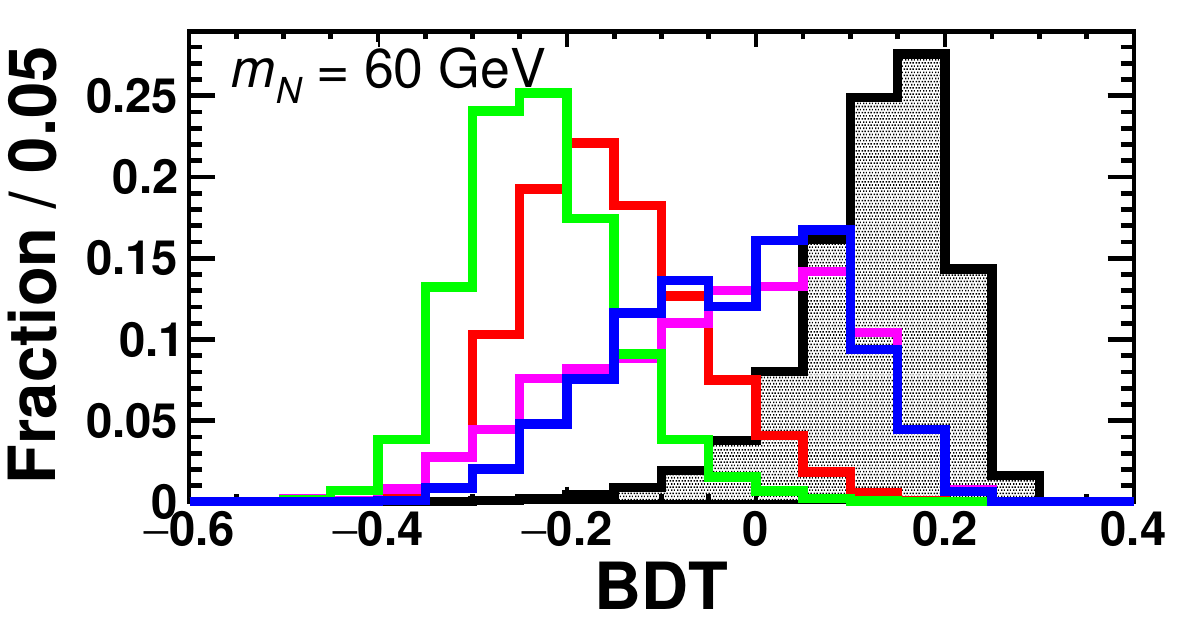}
\includegraphics[width=7cm,height=5cm]{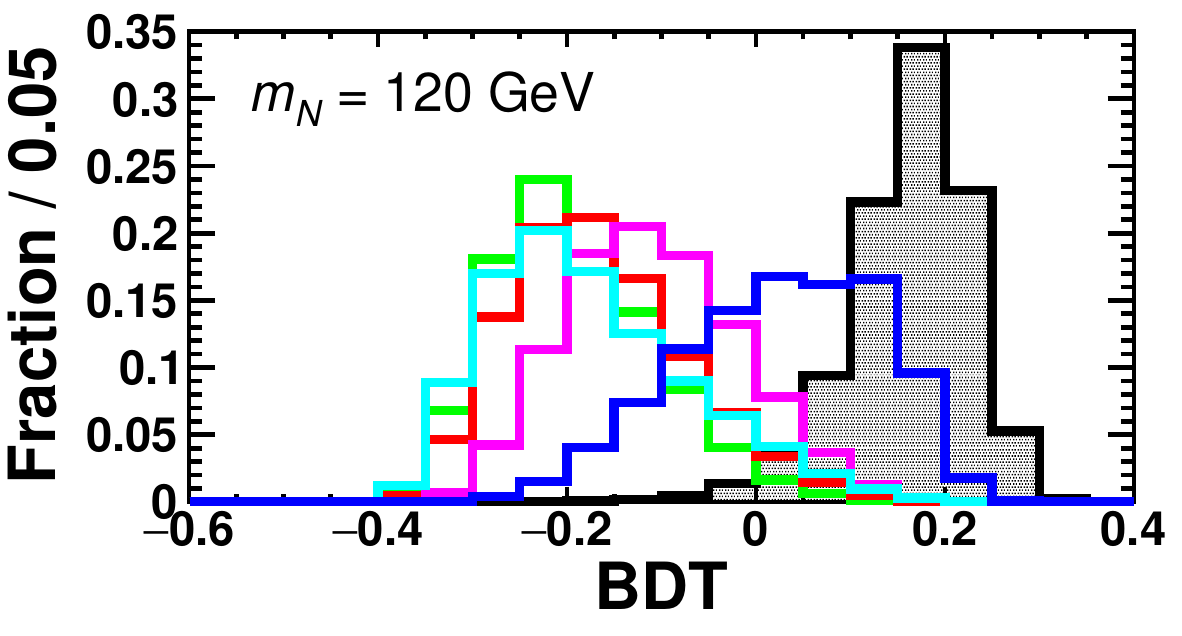}\,\,\,\,\,\,\,\,
\includegraphics[width=7cm,height=5cm]{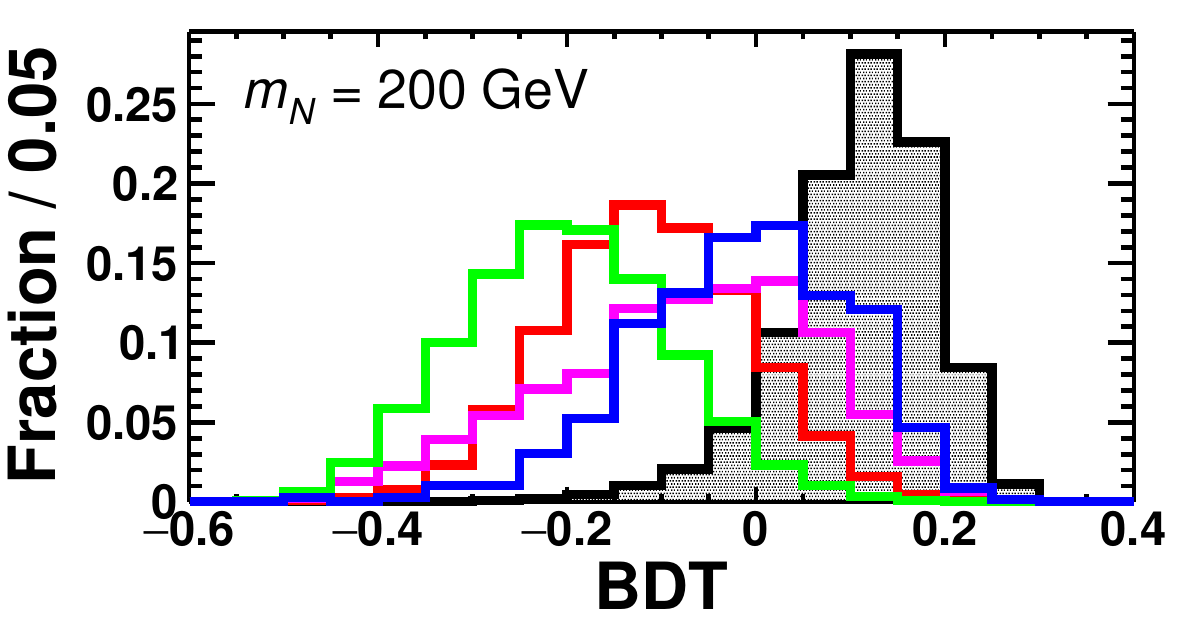}
\includegraphics[width=7cm,height=5cm]{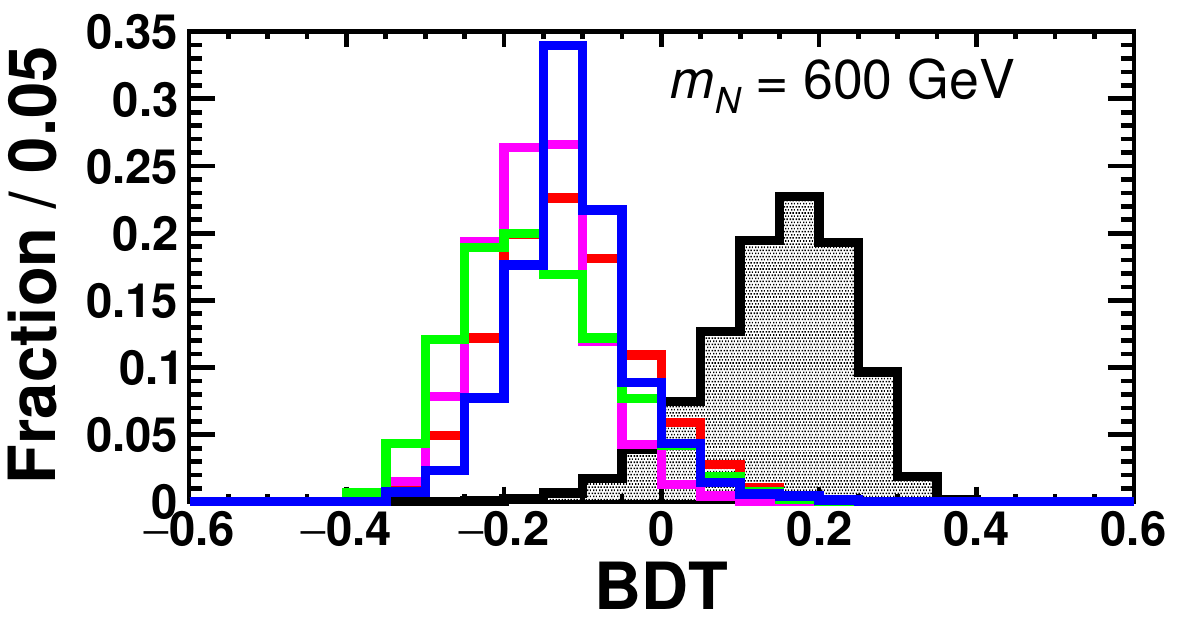}\,\,\,\,\,\,\,\,
\includegraphics[width=7cm,height=5cm]{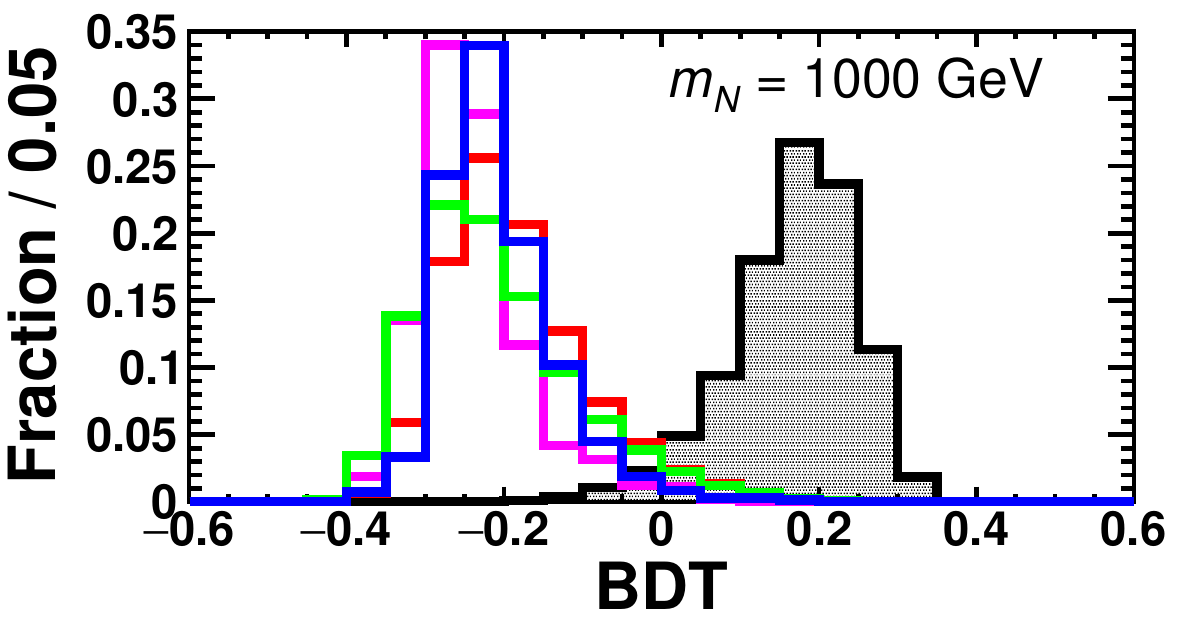}
\caption{
Similar as Fig.~\ref{fig:BDTLHeCLep}, but at the FCC-eh.
}
\label{fig:BDTFCC-ehLep}
\end{figure}

\section{Distributions of representative high-level observables}
\label{app:HLobs}	

Here, $\met$ is the magnitude of the missing transverse momentum.
$p_T$ and $\eta$ are the transverse momentum and pseudorapidity of the final state object, respectively.
$\Delta\phi(\tau_h / \tau_\mu, \met)$ 
is the azimuthal angle difference between the final state $\tau_h$ / 
$\tau_\mu$
and missing transverse momentum.
$M_T(\tau_h / \tau_\mu + \met)$ 
is the transverse mass of the system of final state $\tau_h$ / 
$\tau_\mu$
and missing transverse momentum.
The invariant mass of the di-jet $M(j+j)$ denotes the reconstructed $W$-boson mass, and the transverse mass 
$M_T(\tau_h / \tau_\mu +j+j+\met)$ 
measures the reconstructed $N$ mass, which are explained in Sec.~\ref{sec:results}.

\subsection{Hadronic $\tau_h$ final state }

\begin{figure}[]
\centering
\includegraphics[width=7cm,height=5cm]{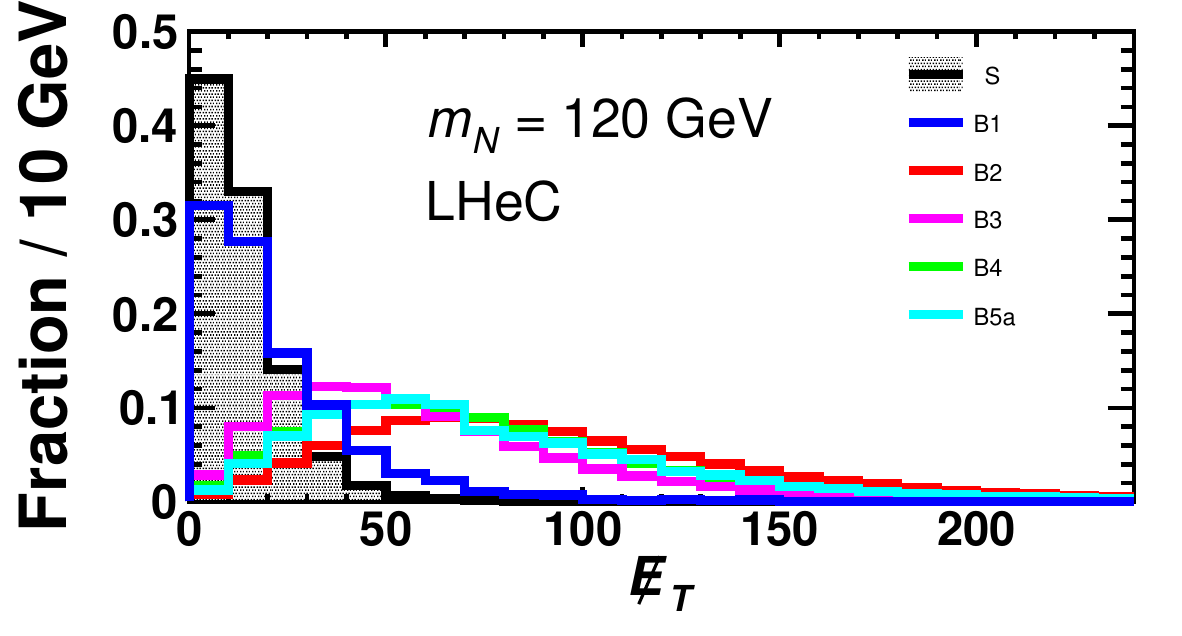}\,\,\,\,\,\,\,\,
\includegraphics[width=7cm,height=5cm]{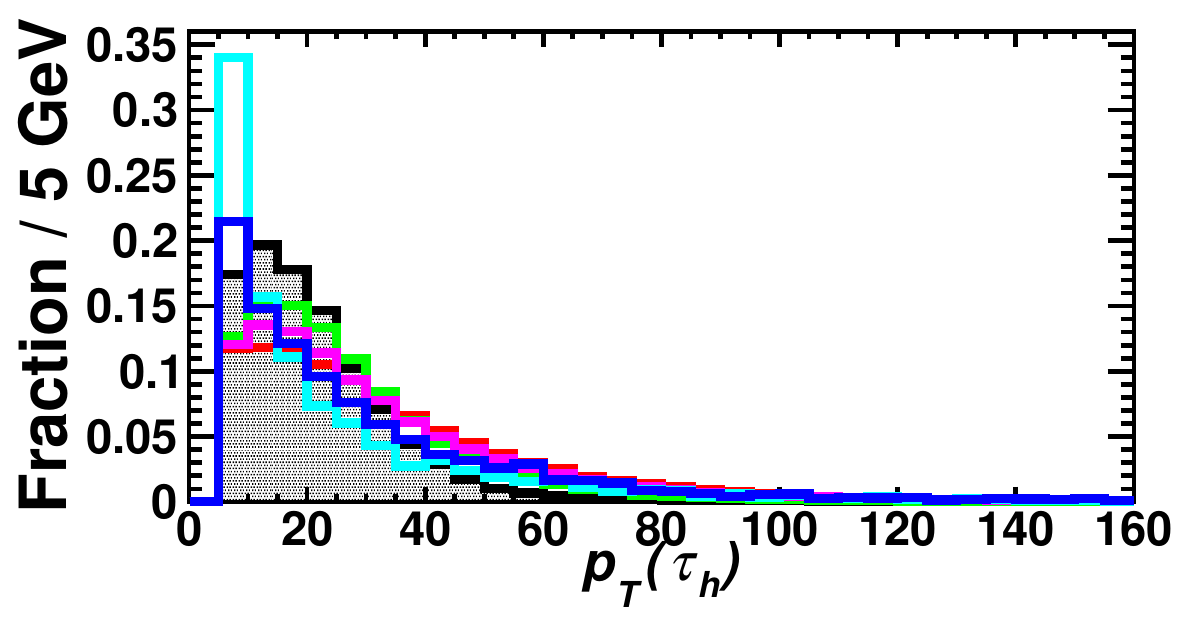}
\includegraphics[width=7cm,height=5cm]{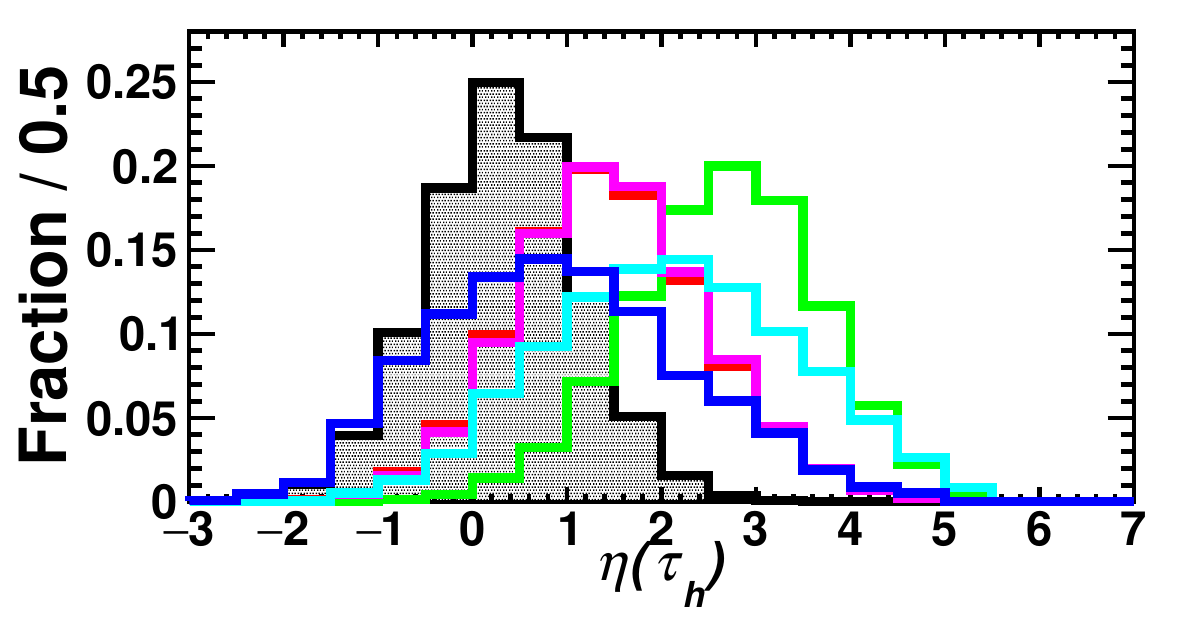}\,\,\,\,\,\,\,\,
\includegraphics[width=7cm,height=5cm]{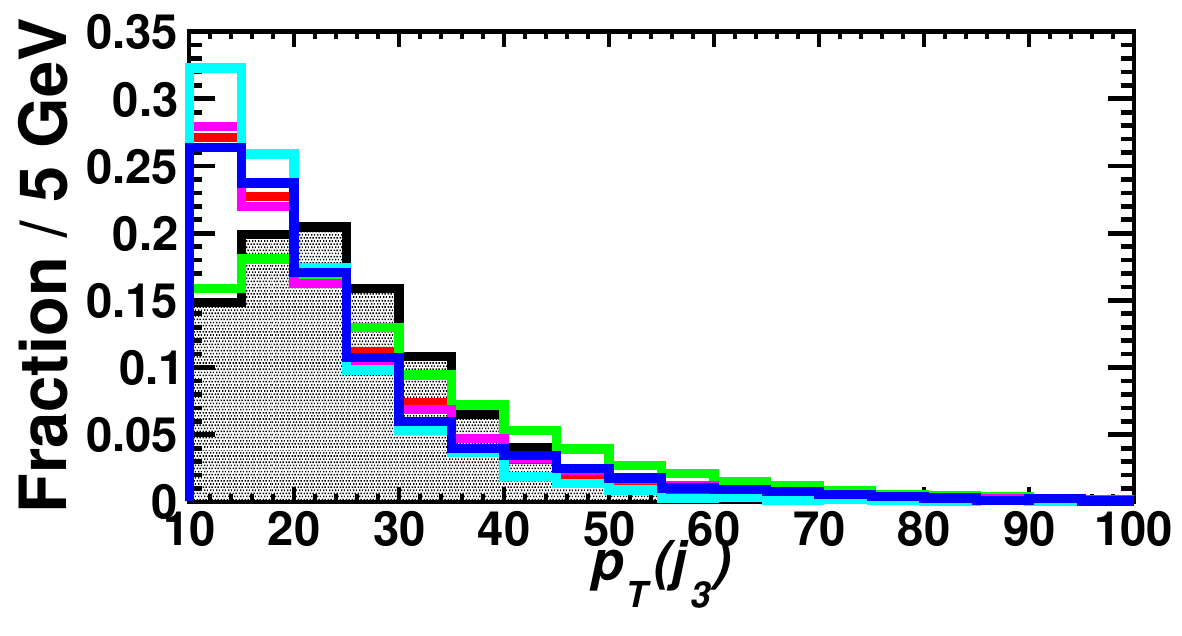} 
\includegraphics[width=7cm,height=5cm]{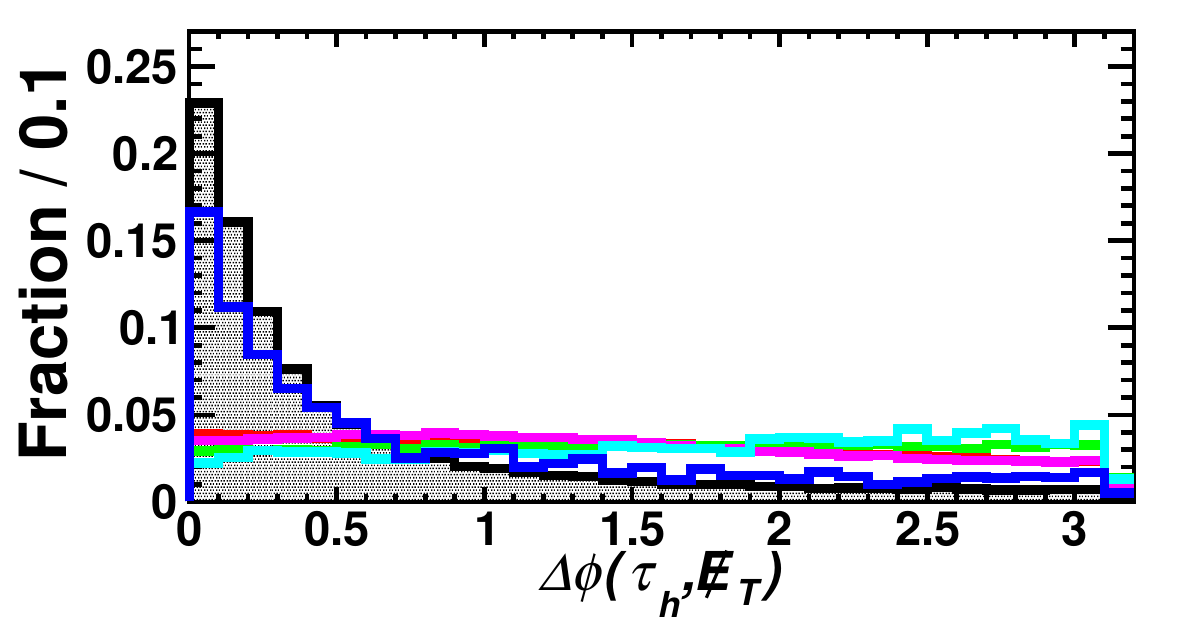}\,\,\,\,\,\,\,\,
\includegraphics[width=7cm,height=5cm]{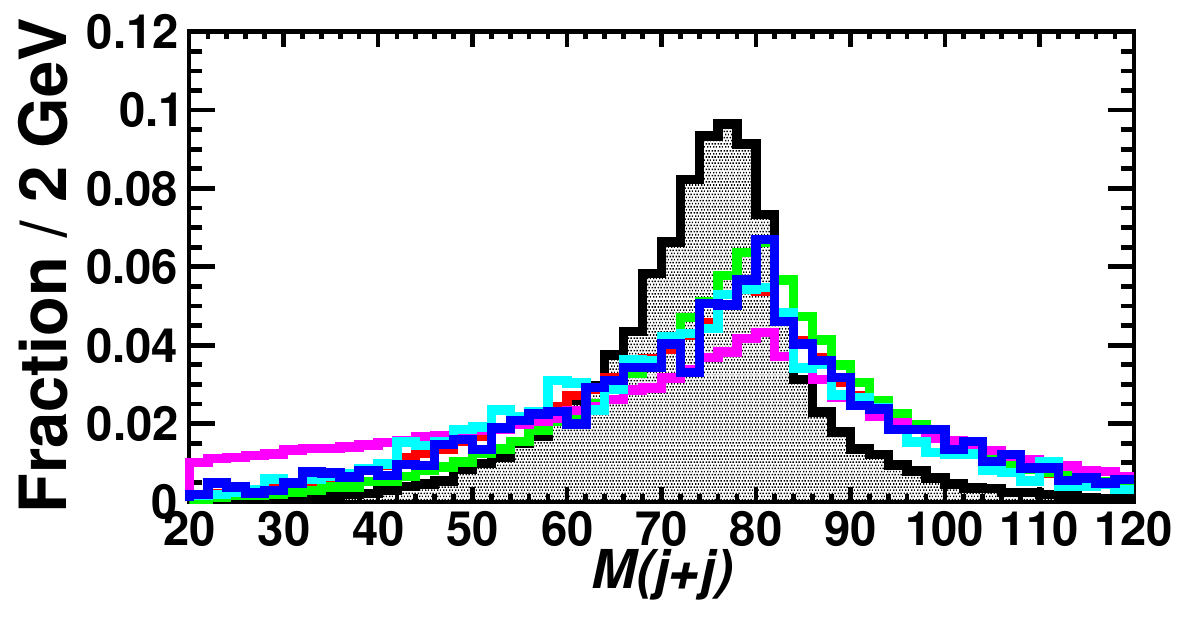}
\includegraphics[width=7cm,height=5cm]{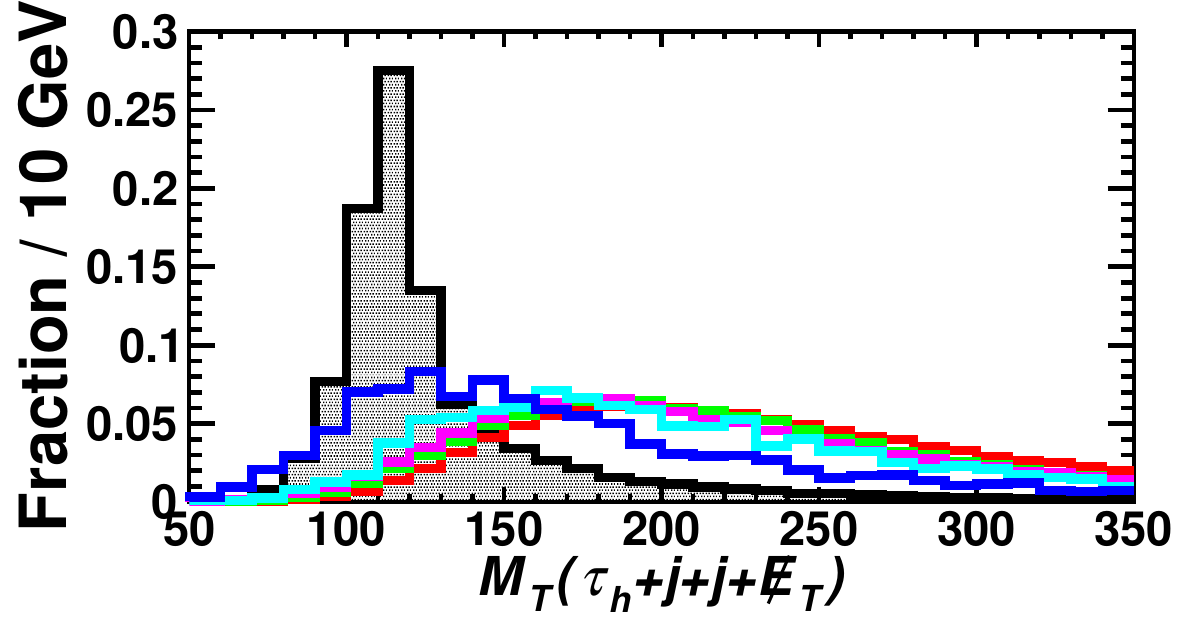}\,\,\,\,\,\,\,\,
\includegraphics[width=7cm,height=5cm]{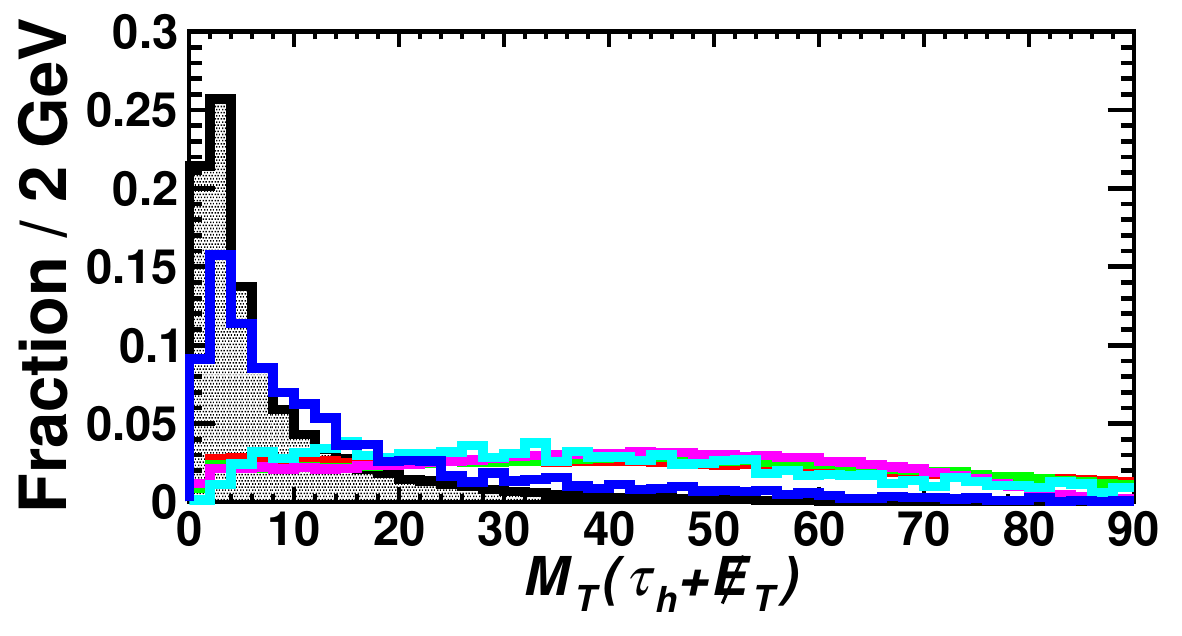}
\caption{
Distributions of some high-level observables after preselection for the signal (black, filled) and background processes at the LHeC assuming 
benchmark $m_{N}$ = 120 GeV and $ |V_{\tau N}|^2\, |V_{eN}|^2 / \left( |V_{\tau N}|^2 + |V_{eN}|^2 \right) = 5 \times 10^{-5}$,
for the hadronic $\tau_h$ final state.
}
\label{fig:ObsLHeCHad}
\end{figure}

\begin{figure}[]
\centering
\includegraphics[width=7cm,height=5cm]{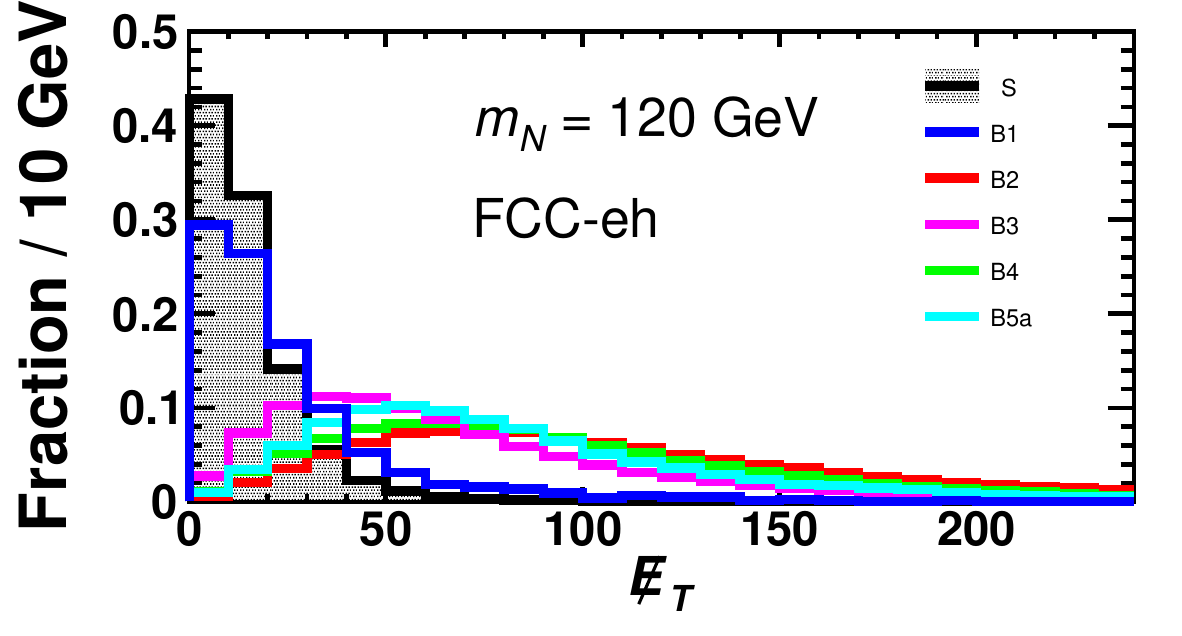}\,\,\,\,\,\,\,\,
\includegraphics[width=7cm,height=5cm]{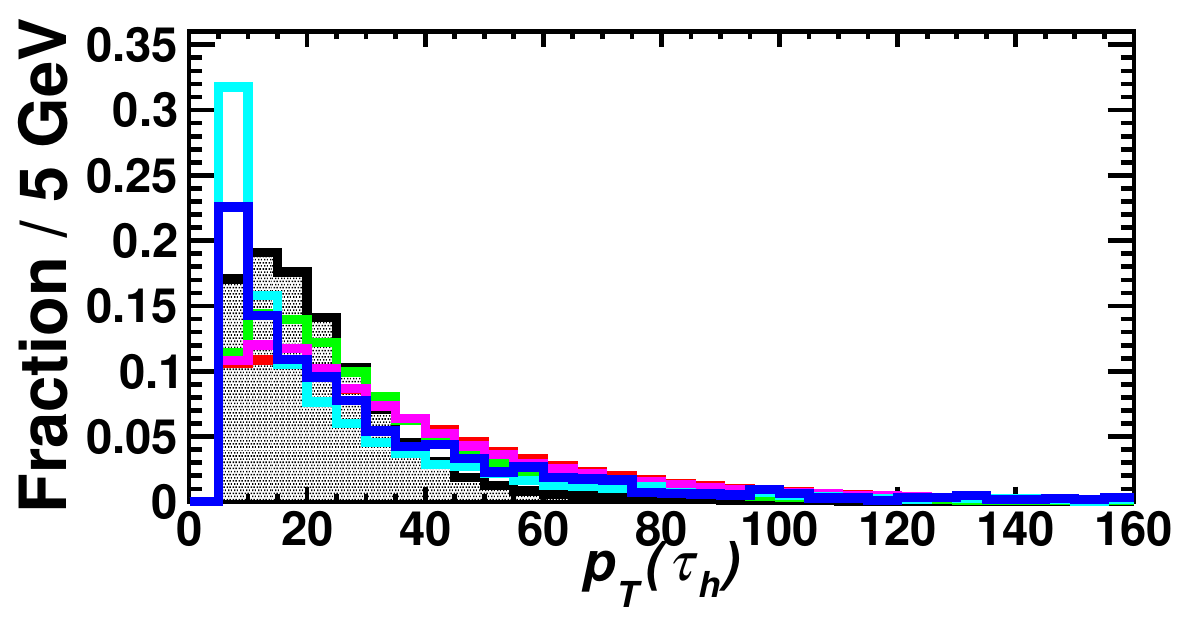}
\includegraphics[width=7cm,height=5cm]{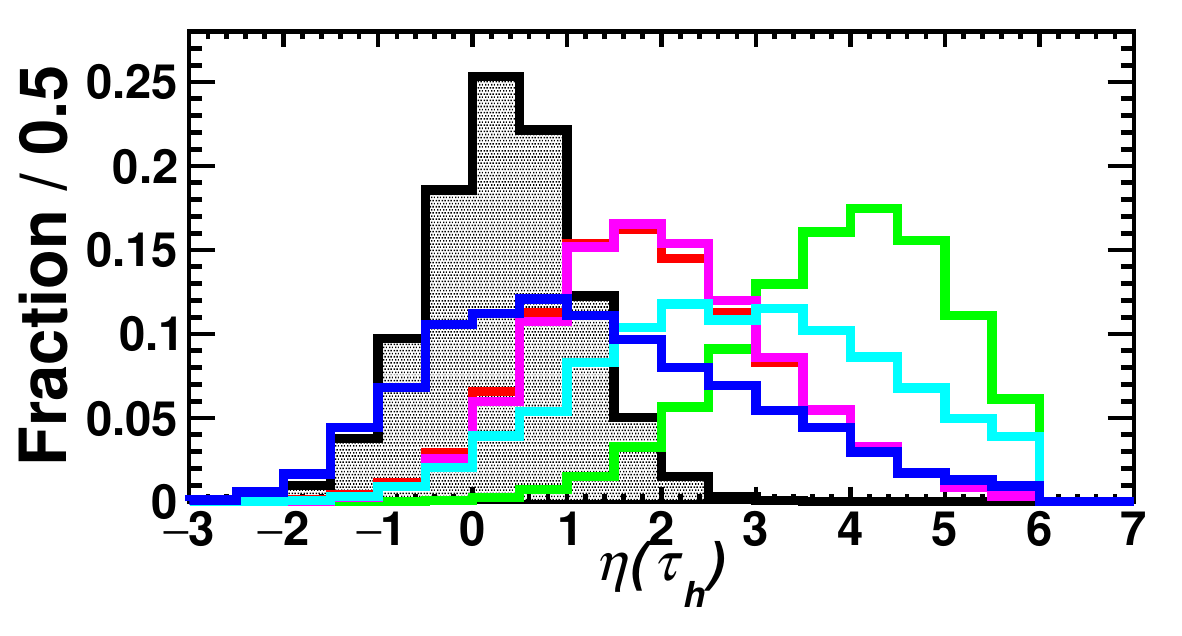}\,\,\,\,\,\,\,\,
\includegraphics[width=7cm,height=5cm]{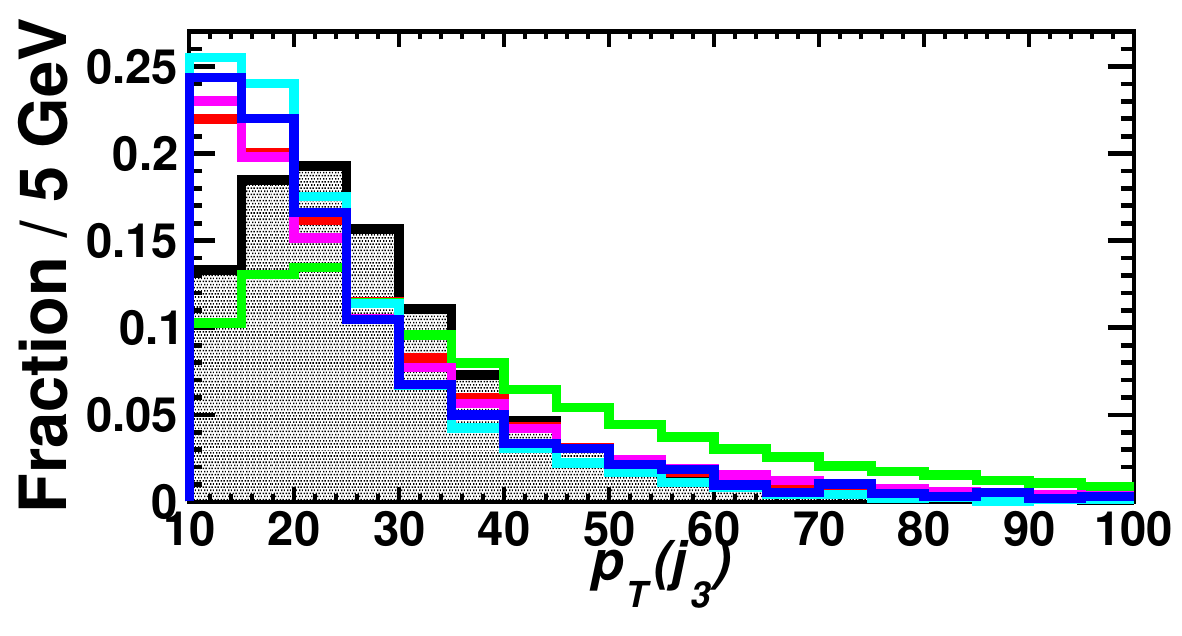} 
\includegraphics[width=7cm,height=5cm]{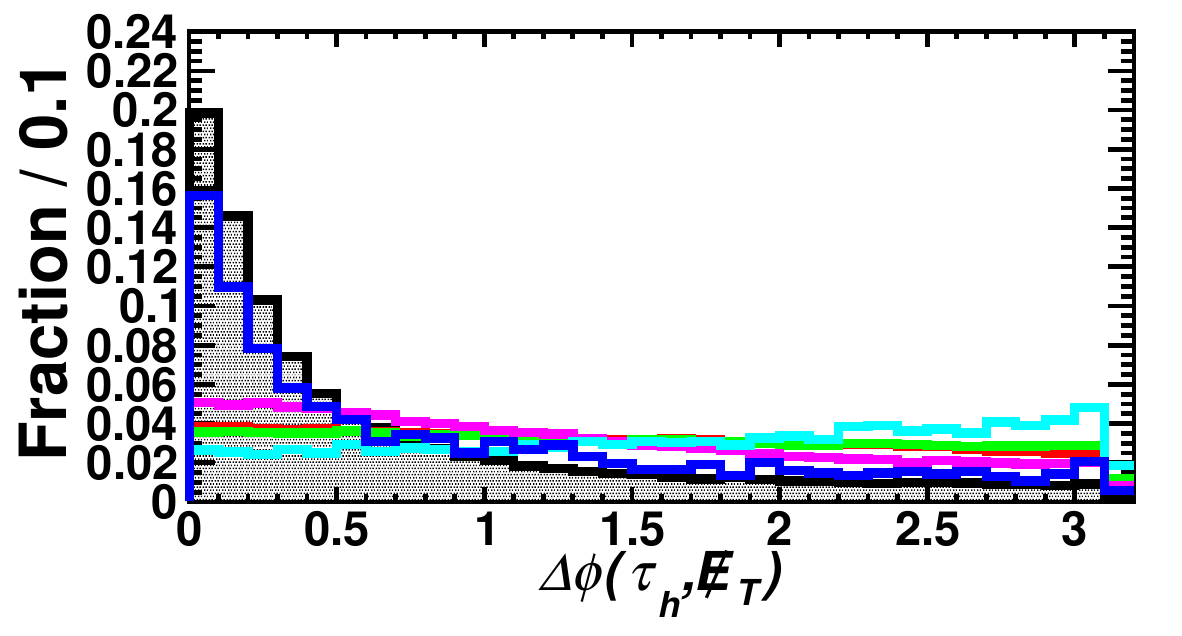}\,\,\,\,\,\,\,\,
\includegraphics[width=7cm,height=5cm]{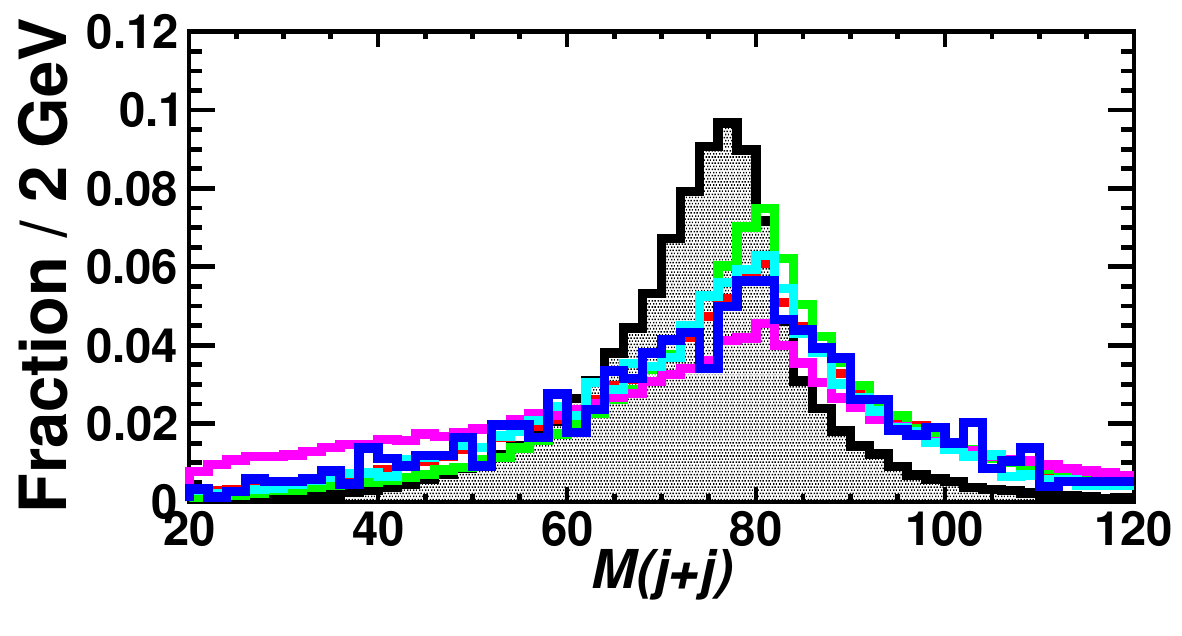} 
\includegraphics[width=7cm,height=5cm]{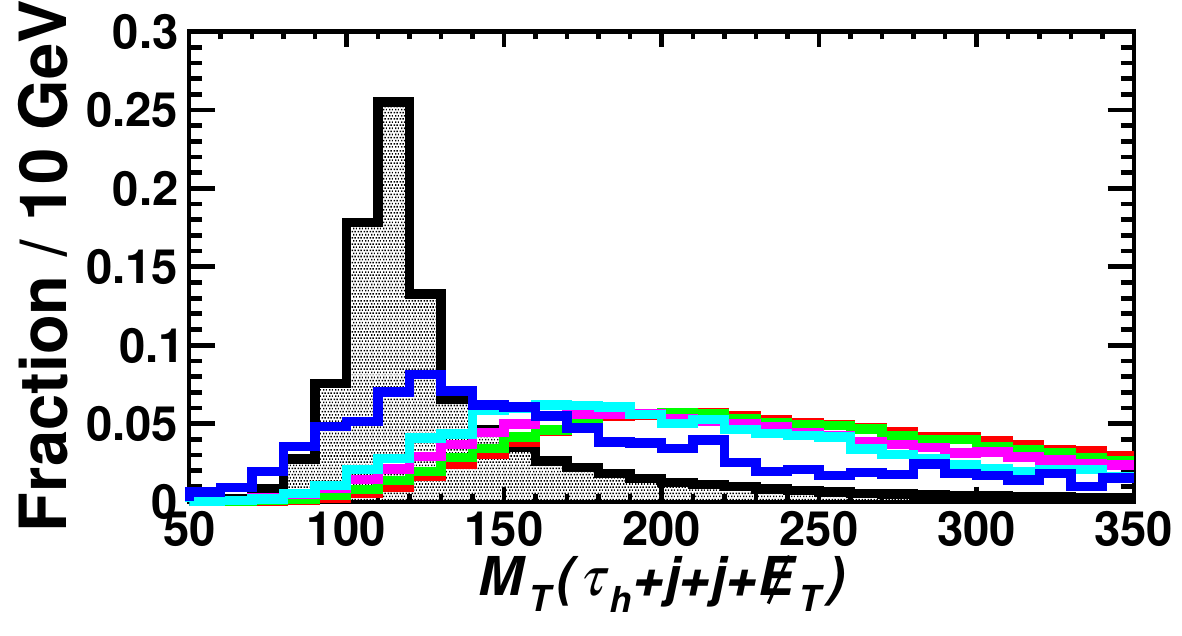}\,\,\,\,\,\,\,\,
\includegraphics[width=7cm,height=5cm]{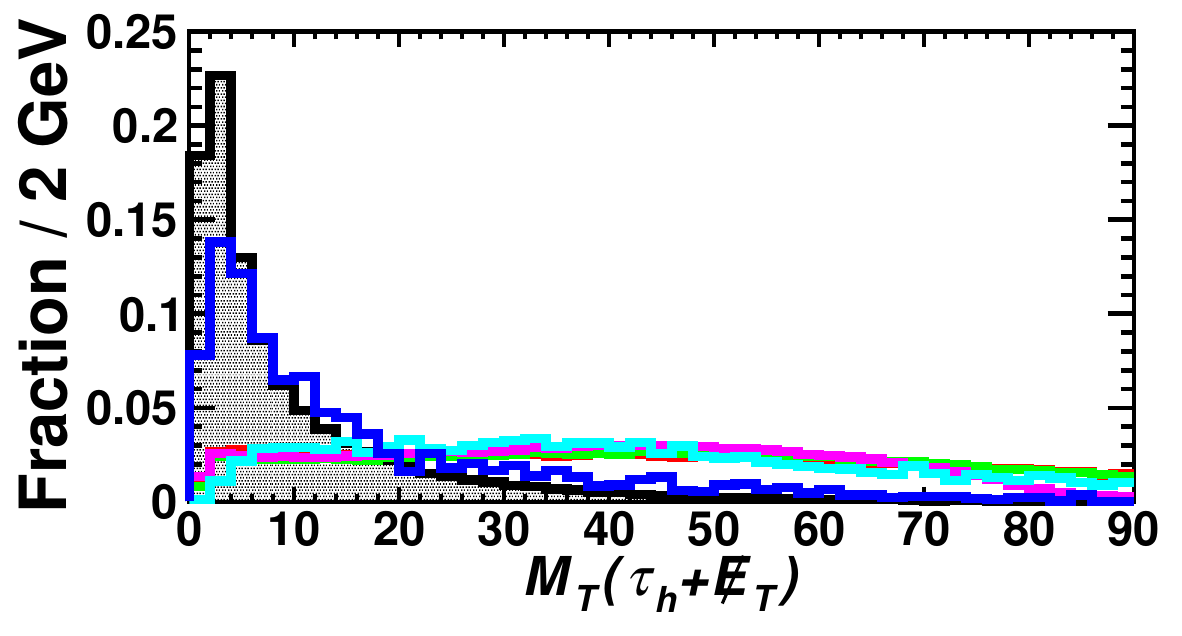} 
\caption{
Similar as Fig.~\ref{fig:ObsLHeCHad}, but at the FCC-eh.
}
\label{fig:ObsFCCehHad}
\end{figure}

\newpage
\subsection{Leptonic $\tau_\mu$ final state }
	
\begin{figure}[]
\centering
\includegraphics[width=7cm,height=5cm]{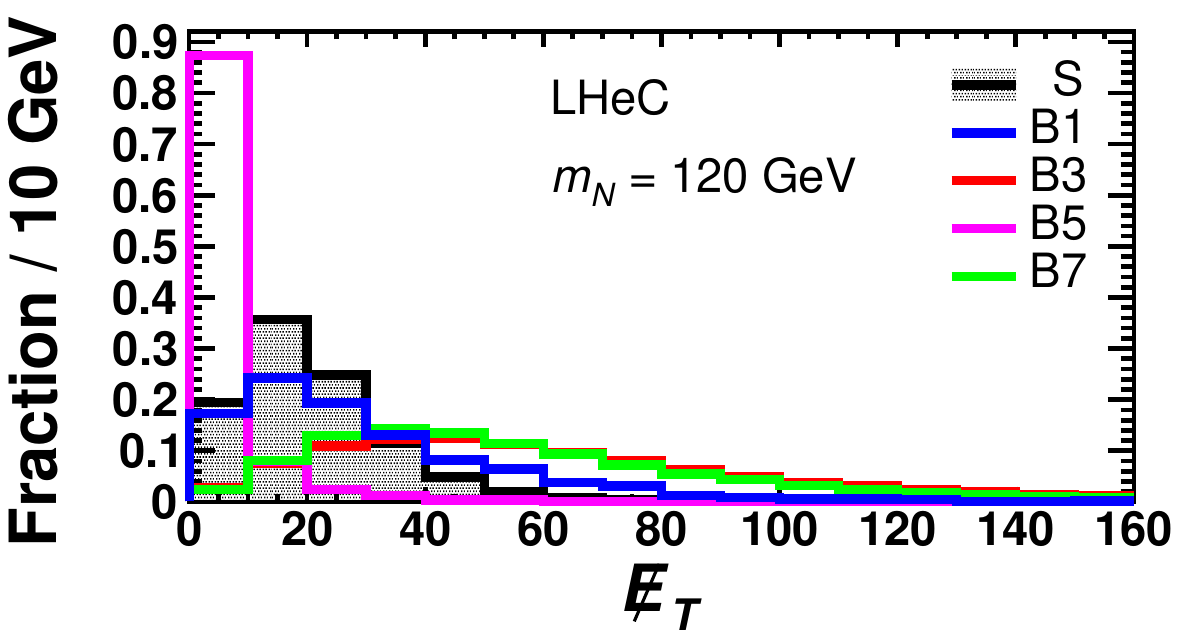}\,\,\,\,\,\,\,\,
\includegraphics[width=7cm,height=5cm]{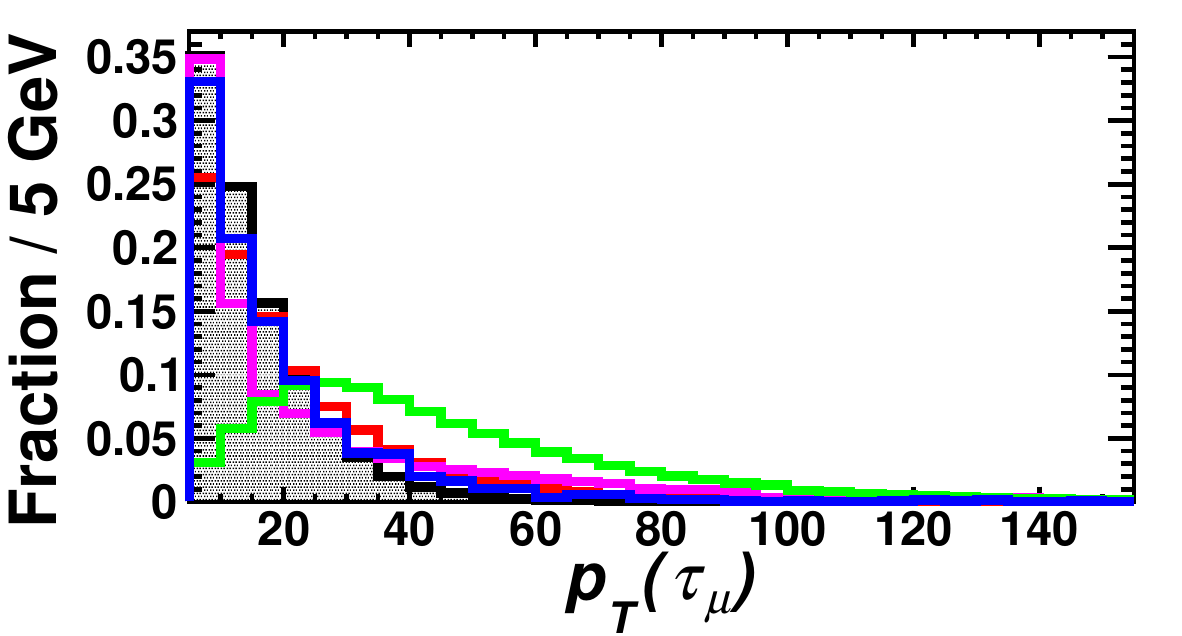}
\includegraphics[width=7cm,height=5cm]{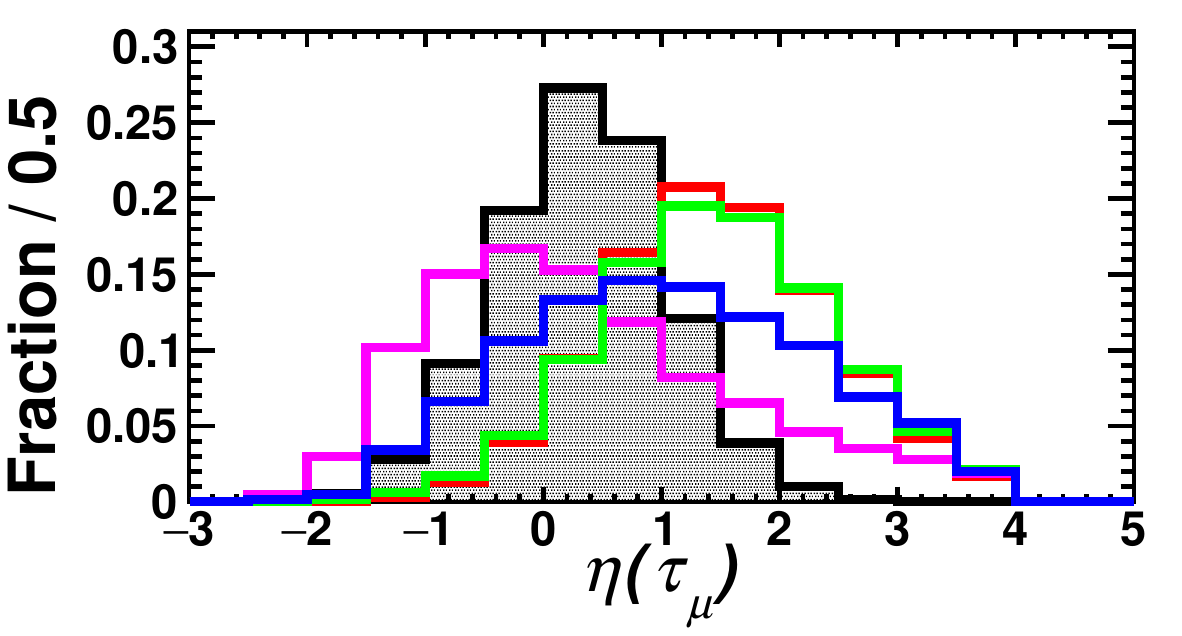}\,\,\,\,\,\,\,\,
\includegraphics[width=7cm,height=5cm]{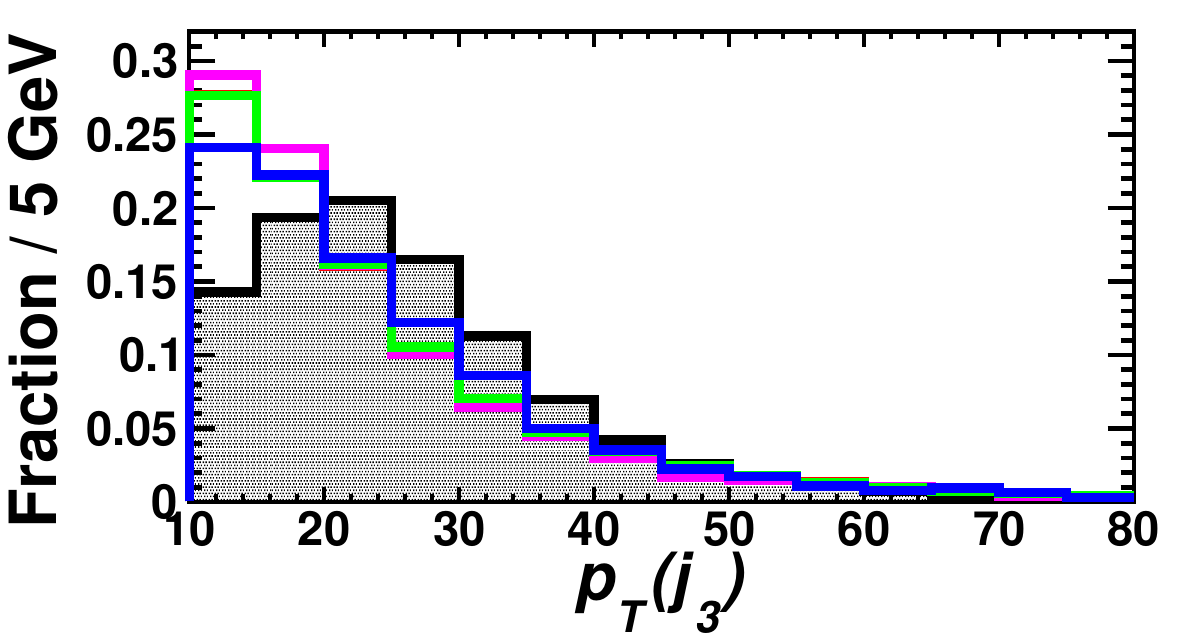}
\includegraphics[width=7cm,height=5cm]{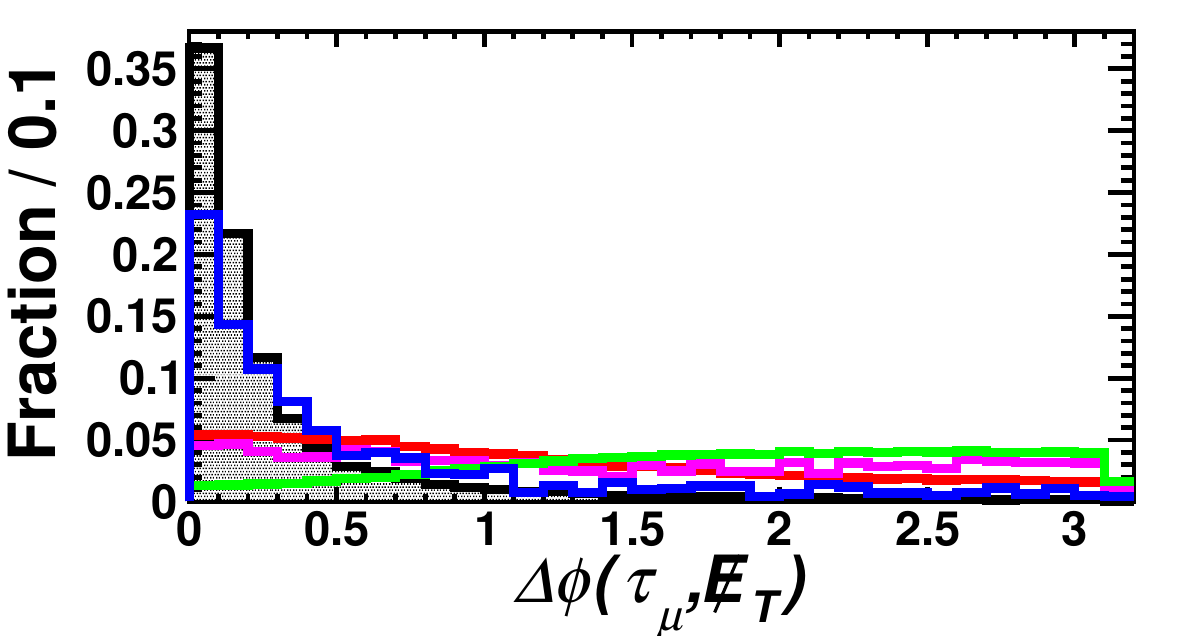}\,\,\,\,\,\,\,\,
\includegraphics[width=7cm,height=5cm]{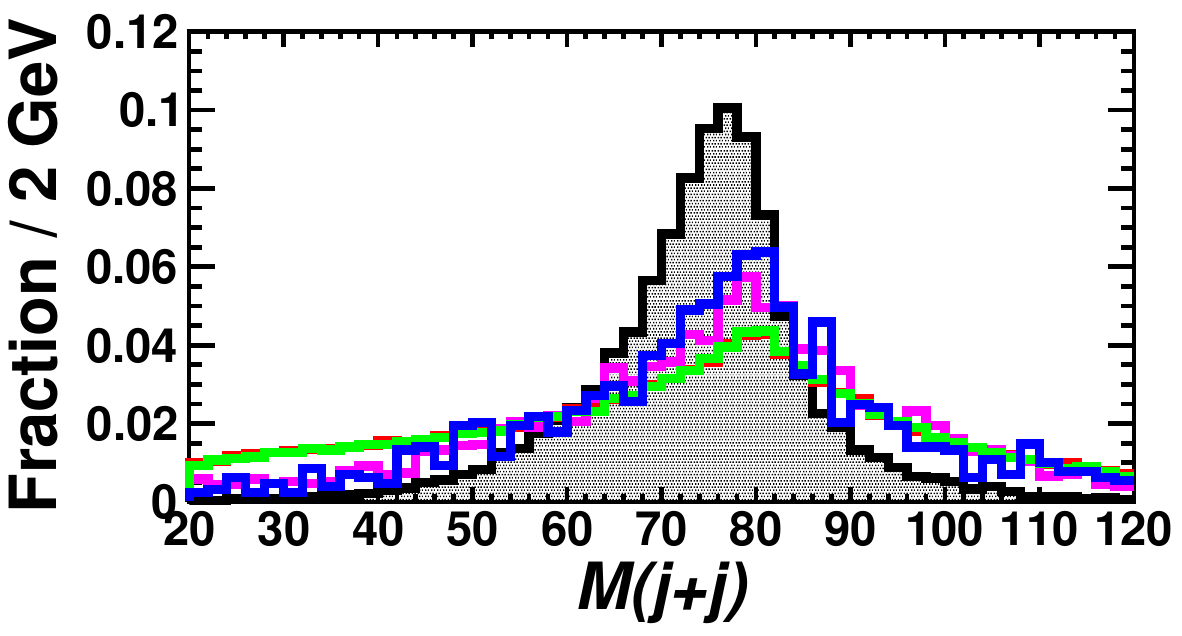}
\includegraphics[width=7cm,height=5cm]{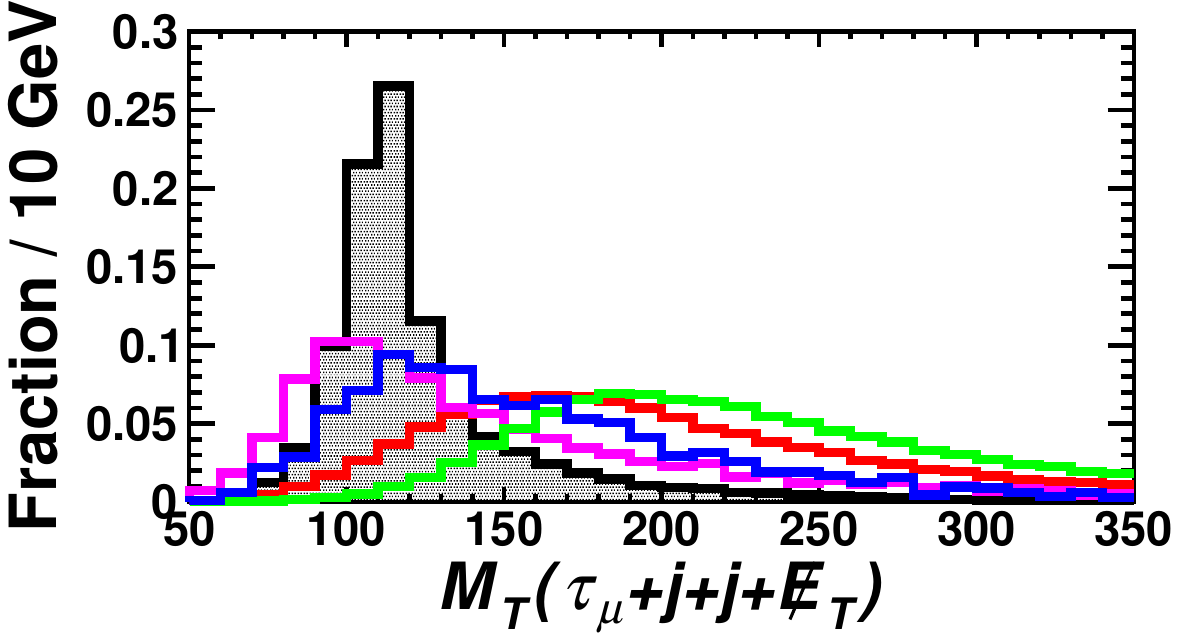}\,\,\,\,\,\,\,\,
\includegraphics[width=7cm,height=5cm]{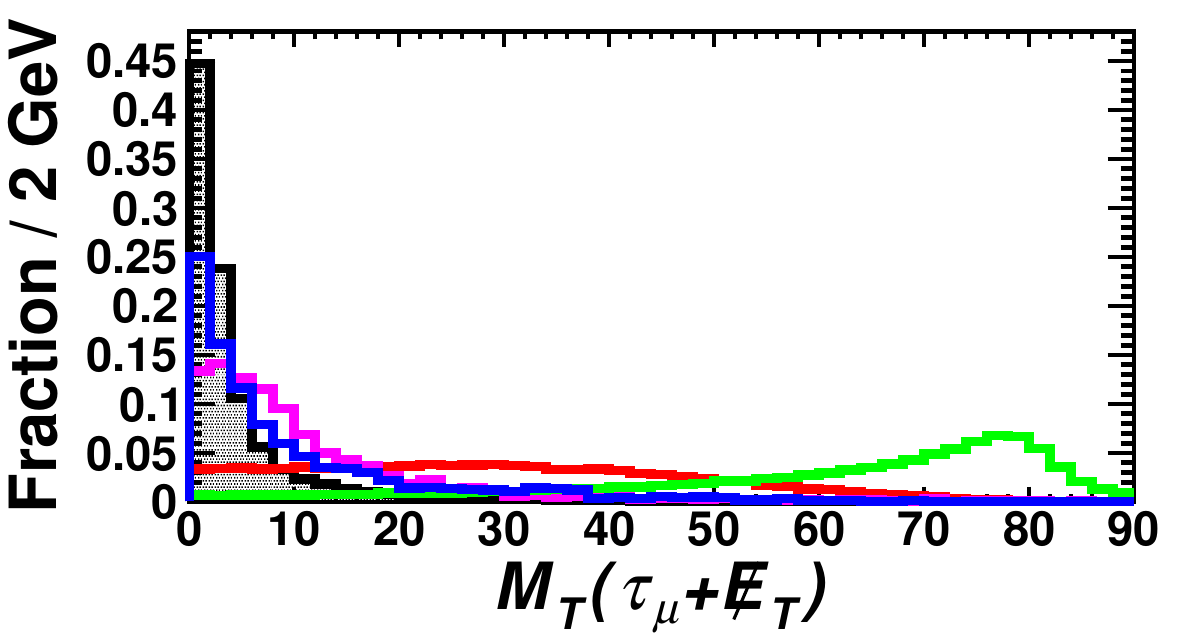}
\caption{
Distributions of some high-level observables after preselection for the signal (black, filled) and dominant background processes at the LHeC assuming 
benchmark $m_{N}$ = 120 GeV and $ |V_{\tau N}|^2\, |V_{eN}|^2 / \left( |V_{\tau N}|^2 + |V_{eN}|^2 \right) = 5 \times 10^{-5}$, 
for the leptonic $\tau_\mu$ final state.
}
\label{fig:ObsLHeCLep}
\end{figure}

\begin{figure}[]
\centering
\includegraphics[width=7cm,height=5cm]{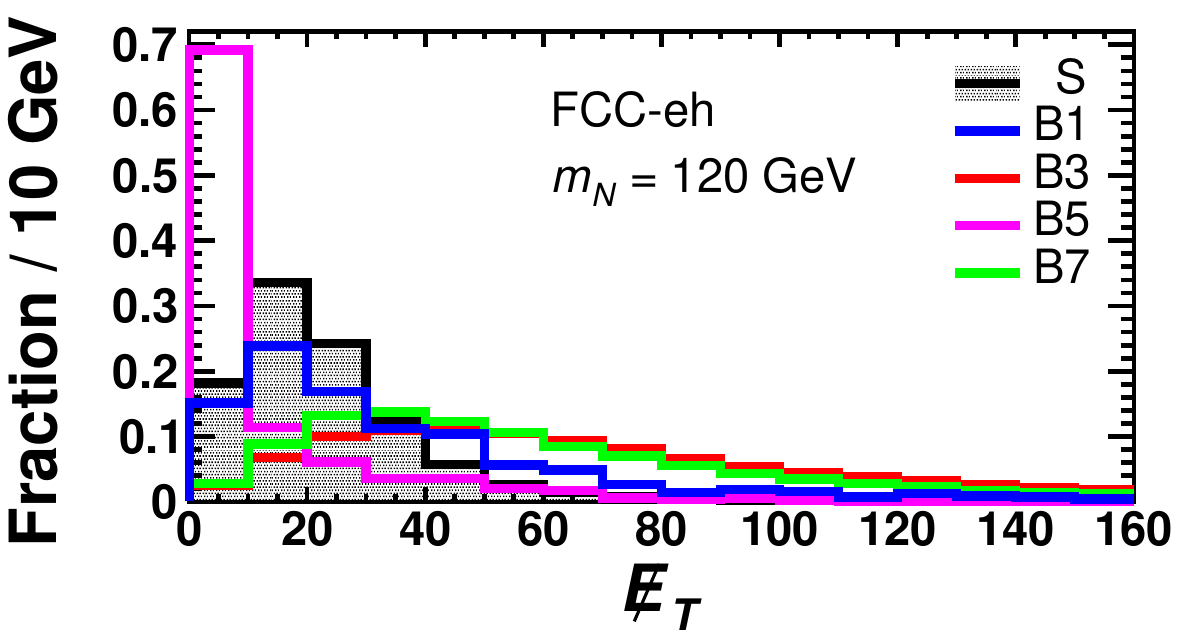}\,\,\,\,\,\,\,\,
\includegraphics[width=7cm,height=5cm]{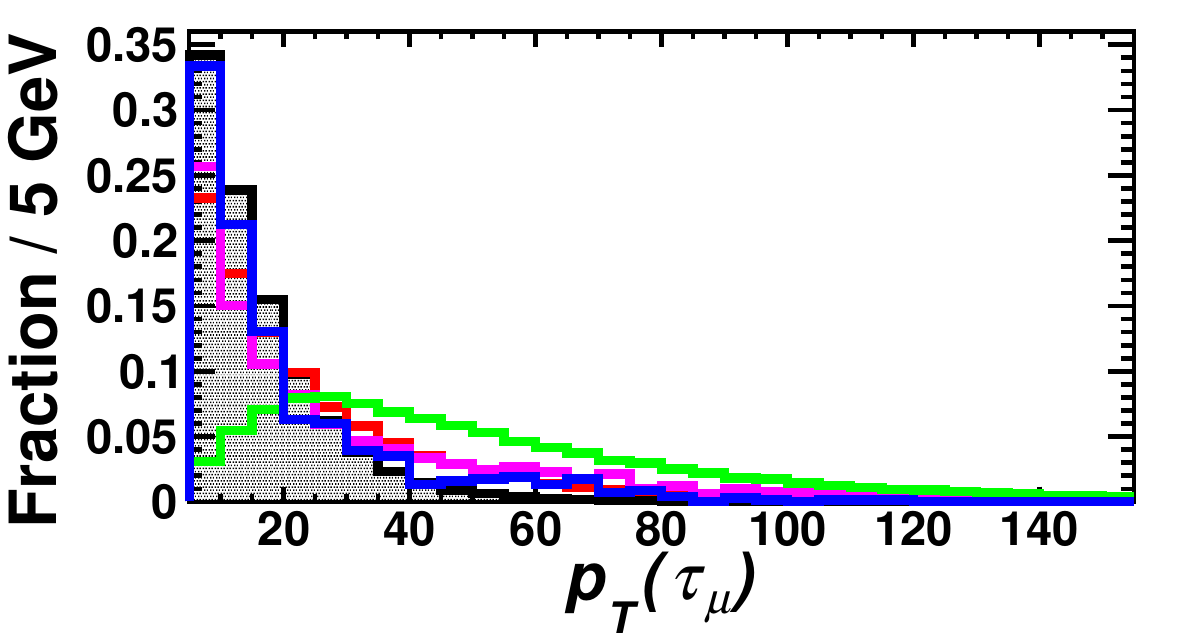}
\includegraphics[width=7cm,height=5cm]{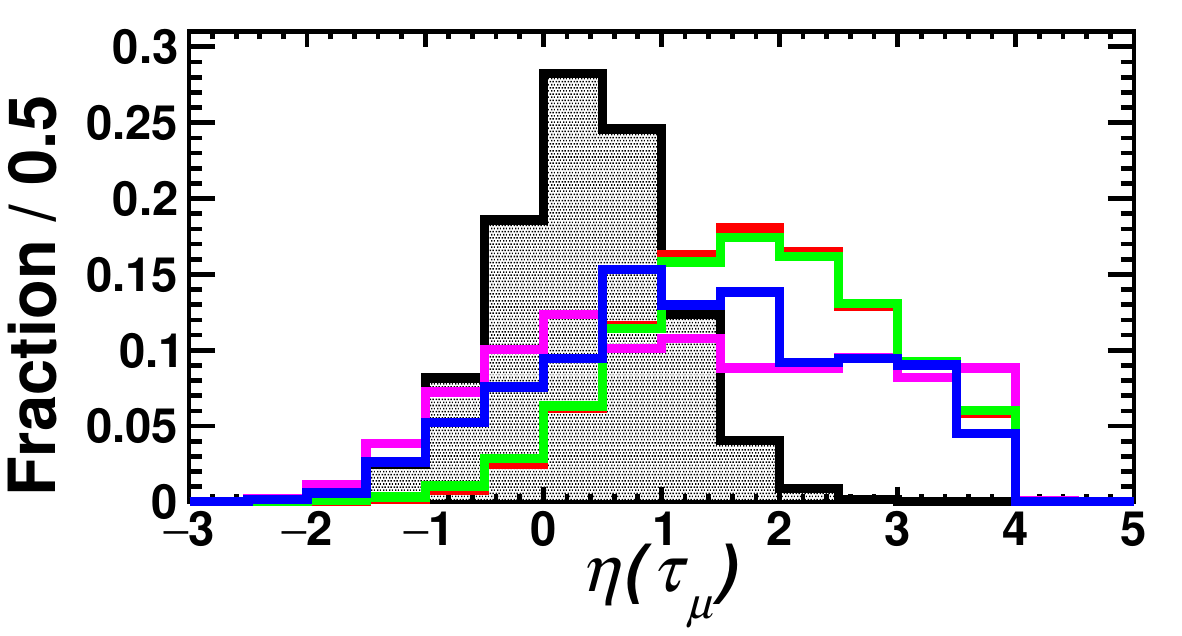}\,\,\,\,\,\,\,\,
\includegraphics[width=7cm,height=5cm]{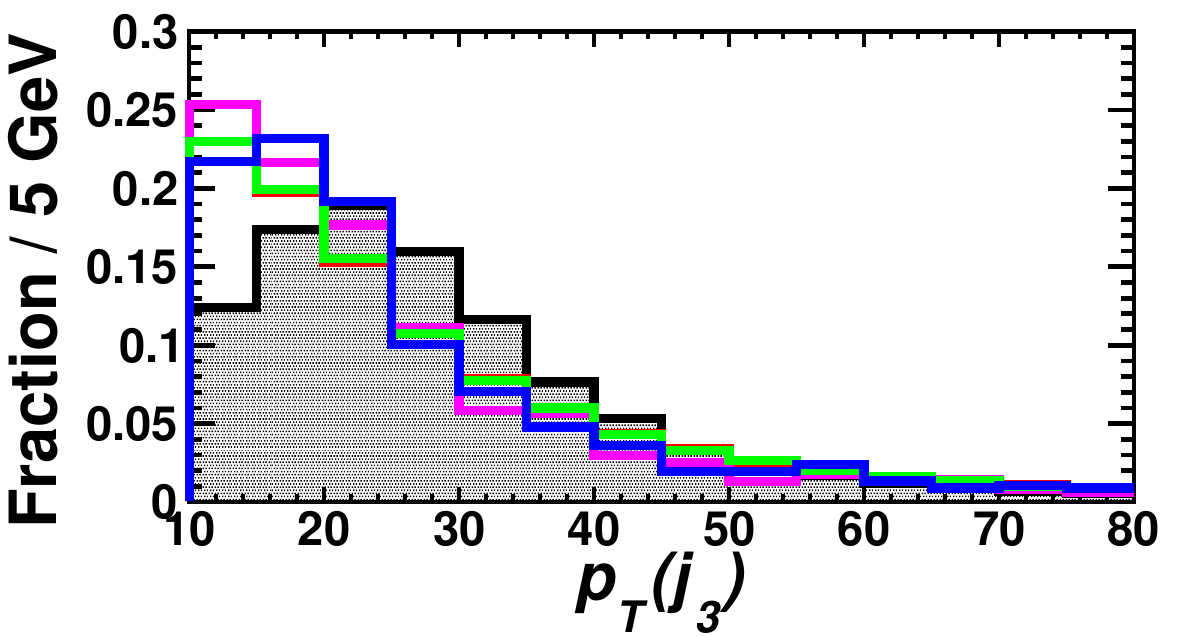}
\includegraphics[width=7cm,height=5cm]{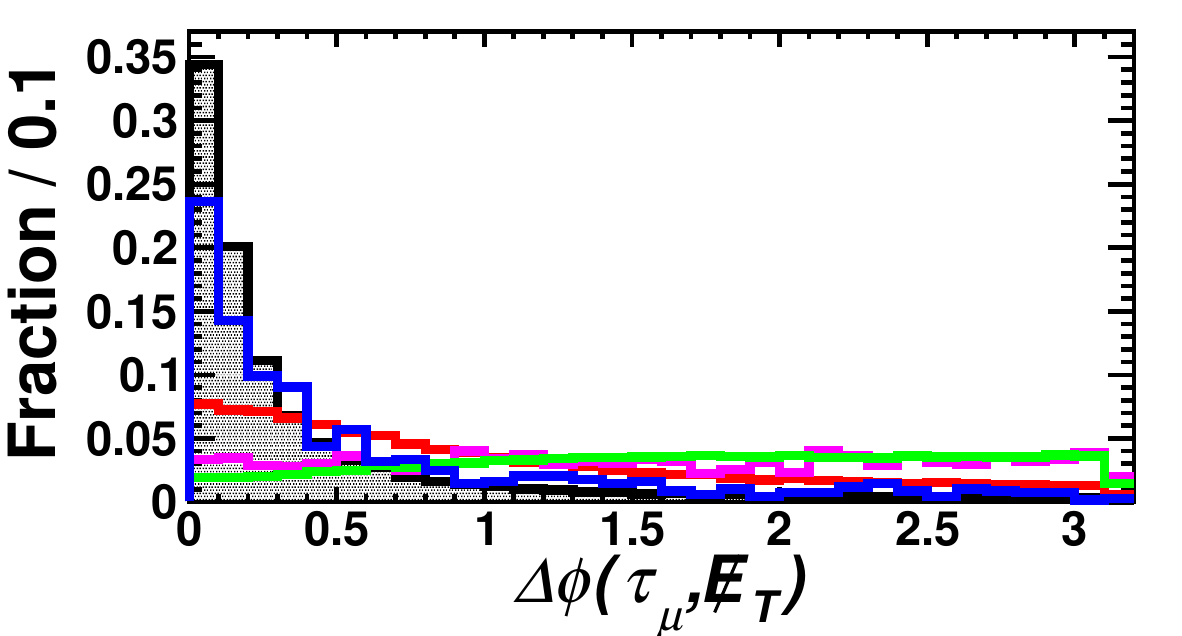} \,\,\,\,\,\,\,\,
\includegraphics[width=7cm,height=5cm]{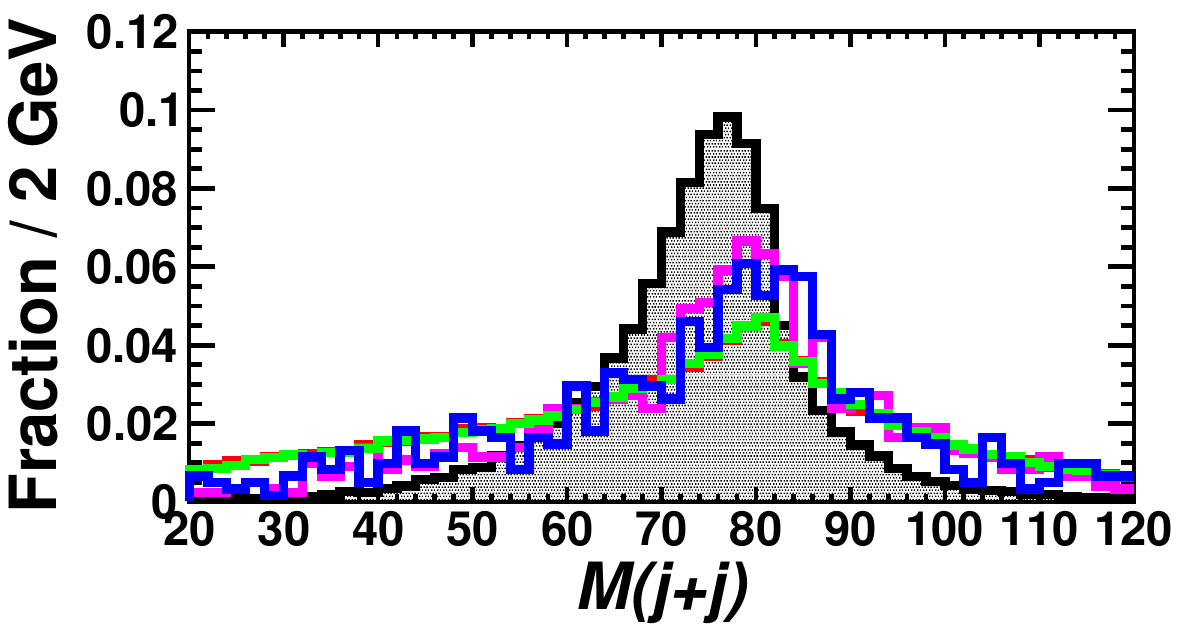}
\includegraphics[width=7cm,height=5cm]{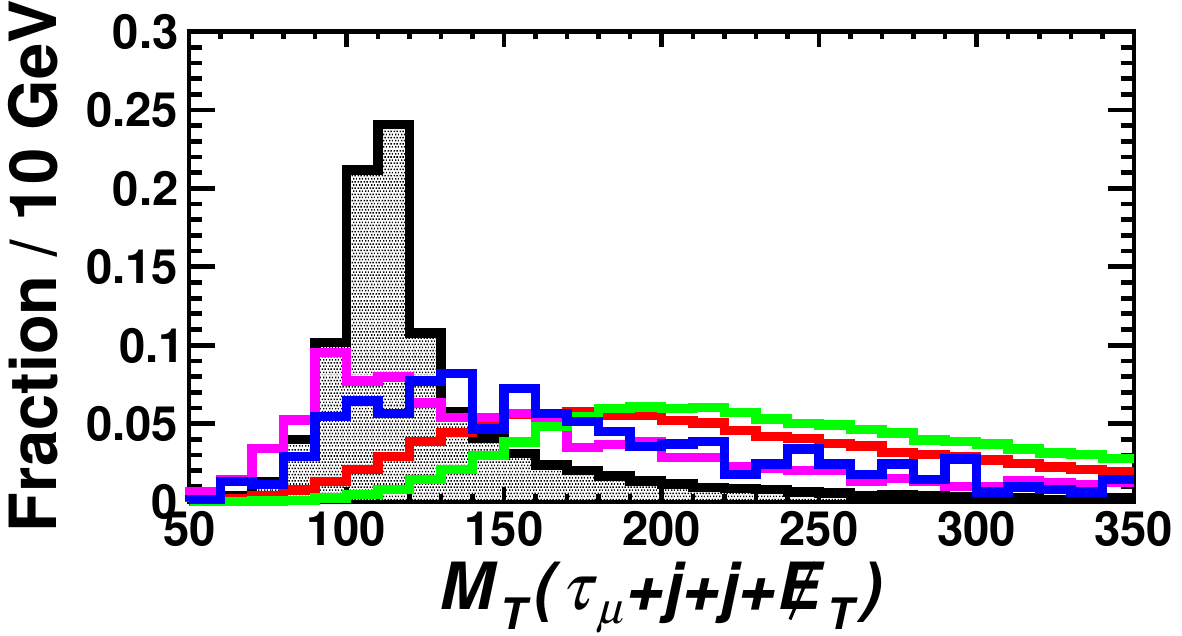}\,\,\,\,\,\,\,\,
\includegraphics[width=7cm,height=5cm]{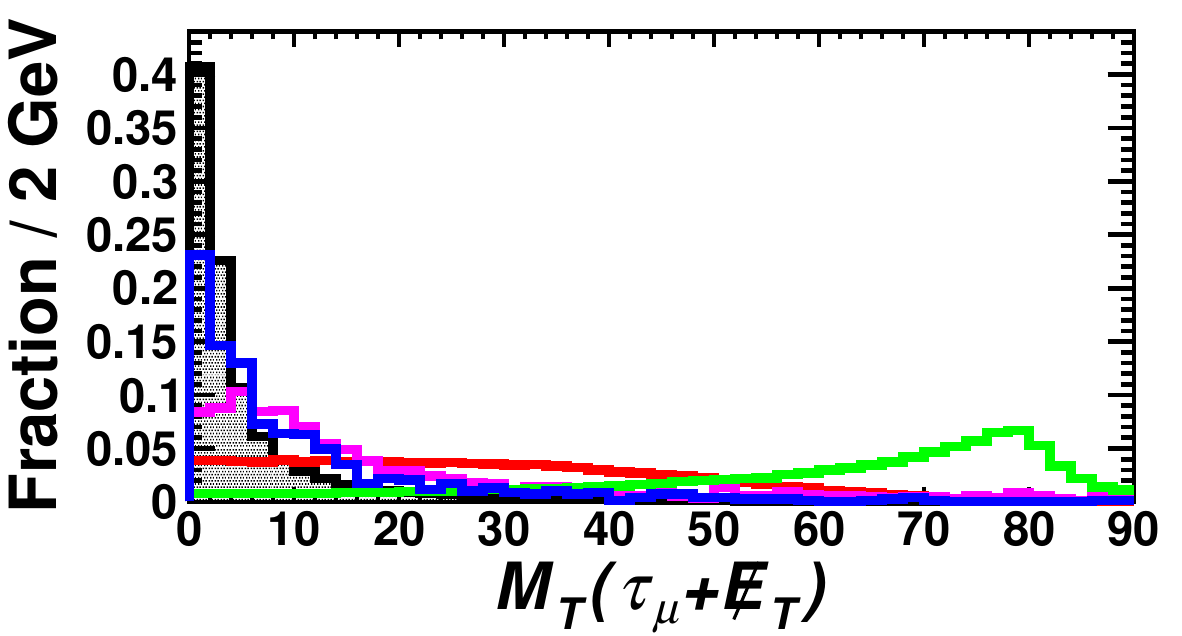}
\caption{
Similar as Fig.~\ref{fig:ObsLHeCLep}, but at the FCC-eh.
}
\label{fig:ObsFCCehLep}
\end{figure}

\section{The Selection efficiency tables}
\label{app:Eff}

\subsection{Hadronic $\tau_h$ final state }

\begin{table}[h]
\centering
\resizebox{\textwidth}{!}{
\begin{tabular}{ccccccccc}
\hline
\hline
		$m_N$ &  collider & selection  & signal  & $\tau^+ \tau^- e^- jjj$ & $\tau^+ \tau^- \nu_e jjj$  & $\tau^+ \nu_\tau  e^- jjj$  & $\tau^+ \nu_\tau  \nu_e jjj$ & $\nu_e jjjj$\\
		\hline
		\multirow{4}{*}{20 GeV} & \multirow{2}{*}{LHeC}    & preselection & $1.19\mltp10^{-2}$&$2.17\mltp10^{-3}$&$3.22\mltp10^{-2}$&$6.78\mltp10^{-3}$&$1.48\mltp10^{-1}$&$2.14\mltp10^{-4}$\\
		& & BDT$>$0.184  &$3.17\mltp10^{-1}$&$2.65\mltp10^{-2}$&$5.08\mltp10^{-5}$&$1.47\mltp10^{-4}$&$1.35\mltp10^{-5}$&$1.46\mltp10^{-4}$\\
		& \multirow{2}{*}{FCC-eh} &preselection&$1.89\mltp10^{-2}$&$2.00\mltp10^{-3}$&$4.25\mltp10^{-2}$&$7.30\mltp10^{-3}$&$1.64\mltp10^{-1}$&$4.16\mltp10^{-4}$\\
		&                                            & BDT$>$0.182 &$4.14\mltp10^{-1}$&$3.86\mltp10^{-2}$&$1.99\mltp10^{-4}$&$9.77\mltp10^{-4}$&$3.65\mltp10^{-5}$&$1.93\mltp10^{-4}$\\
		\hline
		\multirow{4}{*}{60 GeV} & \multirow{2}{*}{LHeC}    & preselection & $7.72\mltp10^{-2}$&$2.17\mltp10^{-3}$&$3.22\mltp10^{-2}$&$6.78\mltp10^{-3}$&$1.48\mltp10^{-1}$&$2.14\mltp10^{-4}$ \\
		& & BDT$>$0.196  &$3.05\mltp10^{-1}$&$2.93\mltp10^{-2}$&$8.25\mltp10^{-5}$&$1.26\mltp10^{-4}$&$2.70\mltp10^{-5}$&$1.46\mltp10^{-4}$\\
		& \multirow{2}{*}{FCC-eh} &preselection&$8.20\mltp10^{-2}$&$2.00\mltp10^{-3}$&$4.25\mltp10^{-2}$&$7.30\mltp10^{-3}$&$1.64\mltp10^{-1}$&$4.16\mltp10^{-4}$\\
		&                                            & BDT$>$0.196 &$3.94\mltp10^{-1}$&$4.59\mltp10^{-2}$&$8.76\mltp10^{-5}$&$4.14\mltp10^{-4}$&$3.65\mltp10^{-5}$&$9.64\mltp10^{-5}$\\
		\hline
		\multirow{4}{*}{120 GeV} & \multirow{2}{*}{LHeC}    & preselection & $1.48\mltp10^{-1}$&$1.56\mltp10^{-3}$&$2.76\mltp10^{-2}$&$5.66\mltp10^{-3}$&$1.33\mltp10^{-1}$&$1.49\mltp10^{-4}$\\
		& & BDT$>$0.164 &$3.79\mltp10^{-1}$&$4.14\mltp10^{-2}$&$2.15\mltp10^{-4}$&$8.41\mltp10^{-4}$&$5.40\mltp10^{-4}$&$1.26\mltp10^{-3}$\\
		& \multirow{2}{*}{FCC-eh} &preselection&$1.49\mltp10^{-1}$&$1.46\mltp10^{-3}$&$3.80\mltp10^{-2}$&$6.40\mltp10^{-3}$&$1.55\mltp10^{-1}$&$3.19\mltp10^{-4}$\\
		&                                            & BDT$>$0.206 &$2.52\mltp10^{-1}$&$1.30\mltp10^{-2}$&$5.35\mltp10^{-5}$&$3.42\mltp10^{-4}$&$6.44\mltp10^{-5}$&$2.51\mltp10^{-4}$\\
		\hline
		\multirow{4}{*}{200 GeV} & \multirow{2}{*}{LHeC}    & preselection & $1.94\mltp10^{-1}$&$1.56\mltp10^{-3}$&$2.76\mltp10^{-2}$&$5.66\mltp10^{-3}$&$1.33\mltp10^{-1}$&$1.49\mltp10^{-4}$\\
		& & BDT$>$0.170 &3.23$\mltp10^{-1}$&$1.92\mltp10^{-2}$&$5.70\mltp10^{-4}$&$1.90\mltp10^{-3}$&$9.16\mltp10^{-4}$&$2.10\mltp10^{-4}$\\
		& \multirow{2}{*}{FCC-eh} &preselection&$1.96\mltp10^{-1}$&$1.46\mltp10^{-3}$&$3.80\mltp10^{-2}$&$6.40\mltp10^{-3}$&$1.55\mltp10^{-1}$&$3.19\mltp10^{-4}$\\
		&                                            & BDT$>$0.188 &$3.19\mltp10^{-1}$&$3.20\mltp10^{-2}$&$8.29\mltp10^{-4}$&$2.46\mltp10^{-3}$&$2.58\mltp10^{-4}$&$1.25\mltp10^{-4}$\\
		\hline
		\multirow{4}{*}{600 GeV} & \multirow{2}{*}{LHeC}    & preselection & $1.98\mltp10^{-1}$&$1.56\mltp10^{-3}$&$2.76\mltp10^{-2}$&$5.66\mltp10^{-3}$&$1.33\mltp10^{-1}$&$1.49\mltp10^{-4}$\\
		& & BDT$>$0.196 &$6.42\mltp10^{-1}$&$3.60\mltp10^{-3}$&$4.81\mltp10^{-4}$&$2.29\mltp10^{-3}$&$9.76\mltp10^{-4}$&$2.10\mltp10^{-4}$\\
		& \multirow{2}{*}{FCC-eh} &preselection&$2.01\mltp10^{-1}$&$1.46\mltp10^{-3}$&$3.80\mltp10^{-2}$&$6.40\mltp10^{-3}$&$1.55\mltp10^{-1}$&$3.19\mltp10^{-4}$\\
		&                                            & BDT$>$0.196 &$5.52\mltp10^{-1}$&$2.96\mltp10^{-3}$&$2.49\mltp10^{-4}$&$2.73\mltp10^{-3}$&$5.80\mltp10^{-4}$&$1.25\mltp10^{-4}$\\
		\hline
		\multirow{4}{*}{1000 GeV} & \multirow{2}{*}{LHeC}    & preselection & $9.40\mltp10^{-2}$&$1.56\mltp10^{-3}$&$2.76\mltp10^{-2}$&$5.66\mltp10^{-3}$&$1.33\mltp10^{-1}$&$1.49\mltp10^{-4}$\\
		& & BDT$>$0.237 &$7.34\mltp10^{-1}$&$\cdots$&$1.11\mltp10^{-4}$&$2.19\mltp10^{-4}$&$1.35\mltp10^{-4}$&$2.10\mltp10^{-4}$\\
		& \multirow{2}{*}{FCC-eh} &preselection&$1.35\mltp10^{-1}$&$1.46\mltp10^{-3}$&$3.80\mltp10^{-2}$&$6.40\mltp10^{-3}$&$1.55\mltp10^{-1}$&$3.19\mltp10^{-4}$\\
		&                                            & BDT$>$0.202 &$6.17\mltp10^{-1}$&$2.37\mltp10^{-3}$&$2.05\mltp10^{-4}$&$2.55\mltp10^{-3}$&$7.08\mltp10^{-4}$&$1.25\mltp10^{-4}$\\
\hline
\hline
\end{tabular}
}
\caption{
Selection efficiencies of preselection and BDT 
requirements
for both the signal with representative $m_N$ assumptions
and fixing $ |V_{\tau N}|^2\, |V_{eN}|^2 / \left( |V_{\tau N}|^2 + |V_{eN}|^2 \right) = 5 \times 10^{-5}$,
and for
background processes at the LHeC and FCC-eh for the hadronic $\tau_h$ final state,
where ``$\cdots$'' means the number of events can be reduced to be negligible.
}
\label{tab:allEffHad}
\end{table}

\subsection{Leptonic $\tau_\mu$ final state }

\begin{table}[h]
\scalebox{0.62}{
\begin{tabular}{cccccccccccc}
\hline
\hline
$m_N$ &  collider & selection  & signal  & $\tau^+ \tau^- e^- jjj$ &  $\tau^+ \tau^- \nu_e jjj$ & $\tau^+ \nu_\tau  e^- jjj$ &  $\tau^+ \nu_\tau \nu_e jjj$ &  $\tau^+ \mu^- e^- jjj$& $\mu^+ \mu^- \nu_e jjj$&$\mu^+ \nu_\mu  e^- jjj$& $\mu^+ \nu_\mu \nu_e jjj$  \\
\hline
\multirow{4}{*}{20 GeV} & \multirow{2}{*}{LHeC}  & preselection & $5.27\mltp10^{-3}$&$8.70\mltp10^{-4}$&$1.77\mltp10^{-2}$&$3.65\mltp10^{-3}$&$7.30\mltp10^{-2}$&$1.83\mltp10^{-3}$&$1.67\mltp10^{-2}$&$2.80\mltp10^{-2}$&$5.93\mltp10^{-1}$\\
& & BDT$>$0.161  &$2.84\mltp10^{-1}$&$2.85\mltp10^{-2}$&$1.61\mltp10^{-4}$&$4.43\mltp10^{-4}$&$\cdots$&$1.42\mltp10^{-2}$&$1.75\mltp10^{-6}$&$2.22\mltp10^{-5}$&$3.37\mltp10^{-5}$\\
& \multirow{2}{*}{FCC-eh} &preselection&$9.46\mltp10^{-3}$&$8.12\mltp10^{-4}$&$2.44\mltp10^{-2}$&$3.94\mltp10^{-3}$&$4.60\mltp10^{-2}$&$8.59\mltp10^{-4}$&$2.22\mltp10^{-2}$&$2.87\mltp10^{-2}$&$3.74\mltp10^{-1}$\\
& & BDT$>$0.151 &$4.58\mltp10^{-1}$&$5.86\mltp10^{-2}$&$4.74\mltp10^{-6}$&$7.26\mltp10^{-4}$&$1.30\mltp10^{-4}$&$6.63\mltp10^{-2}$&$3.37\mltp10^{-4}$&$1.17\mltp10^{-4}$&$8.02\mltp10^{-5}$\\
\hline
\multirow{4}{*}{60 GeV} & \multirow{2}{*}{LHeC}    & preselection & $3.72\mltp10^{-2}$&$8.70\mltp10^{-4}$&$1.77\mltp10^{-2}$&$3.65\mltp10^{-3}$&$7.30\mltp10^{-2}$&$1.83\mltp10^{-3}$&$1.67\mltp10^{-2}$&$2.80\mltp10^{-2}$&$5.93\mltp10^{-1}$\\
& & BDT$>$0.139  &$3.07\mltp10^{-1}$&$2.42\mltp10^{-2}$&$1.73\mltp10^{-4}$&$6.13\mltp10^{-4}$&$2.74\mltp10^{-4}$&$9.29\mltp10^{-3}$&$2.45\mltp10^{-4}$&$4.43\mltp10^{-5}$&$5.06\mltp10^{-5}$\\
& \multirow{2}{*}{FCC-eh} &preselection&$4.37\mltp10^{-2}$&$8.12\mltp10^{-4}$&$2.44\mltp10^{-2}$&$3.94\mltp10^{-3}$&$4.60\mltp10^{-2}$&$8.59\mltp10^{-4}$&$2.22\mltp10^{-2}$&$2.87\mltp10^{-2}$&$3.74\mltp10^{-1}$\\
&                                            & BDT$>$0.153 &$4.19\mltp10^{-1}$&$4.80\mltp10^{-2}$&$2.78\mltp10^{-4}$&$7.79\mltp10^{-4}$&$5.22\mltp10^{-4}$&$4.79\mltp10^{-2}$&$4.18\mltp10^{-4}$&$9.11\mltp10^{-5}$&$9.35\mltp10^{-5}$\\		
\hline
\multirow{4}{*}{120 GeV} & \multirow{2}{*}{LHeC}    & preselection & $7.97\mltp10^{-2}$&$6.56\mltp10^{-4}$&$1.53\mltp10^{-2}$&$3.07\mltp10^{-3}$&$6.64\mltp10^{-2}$&$1.37\mltp10^{-3}$&$1.46\mltp10^{-2}$&$2.35\mltp10^{-2}$&$5.38\mltp10^{-1}$\\
& & BDT$>$0.120 &3.92$\mltp10^{-1}$&$3.85\mltp10^{-2}$&$6.54\mltp10^{-4}$&$2.14\mltp10^{-3}$&$2.05\mltp10^{-3}$&$5.17\mltp10^{-3}$&$5.99\mltp10^{-4}$&$1.05\mltp10^{-4}$&$8.00\mltp10^{-4}$\\
& \multirow{2}{*}{FCC-eh} &preselection&$8.68\mltp10^{-2}$&$5.94\mltp10^{-4}$&$2.19\mltp10^{-2}$&$3.47\mltp10^{-3}$&$4.39\mltp10^{-2}$&$6.57\mltp10^{-4}$&$1.99\mltp10^{-2}$&$2.53\mltp10^{-2}$&$3.56\mltp10^{-1}$\\
&                                            & BDT$>$0.139 &$5.40\mltp10^{-1}$&$9.48\mltp10^{-2}$&$1.25\mltp10^{-3}$&$3.90\mltp10^{-3}$&$2.37\mltp10^{-3}$&$7.32\mltp10^{-2}$&$1.25\mltp10^{-3}$&$3.84\mltp10^{-4}$&$5.90\mltp10^{-4}$\\
\hline
\multirow{4}{*}{200 GeV} & \multirow{2}{*}{LHeC}    & preselection & $1.18\mltp10^{-1}$&$6.56\mltp10^{-4}$&$1.53\mltp10^{-2}$&$3.07\mltp10^{-3}$&$6.64\mltp10^{-2}$&$1.37\mltp10^{-3}$&$1.46\mltp10^{-2}$&$2.35\mltp10^{-2}$&$5.38\mltp10^{-1}$\\
& & BDT$>$0.118 &3.48$\mltp10^{-1}$&$3.28\mltp10^{-2}$&$2.22\mltp10^{-3}$&$5.22\mltp10^{-3}$&$6.09\mltp10^{-3}$&$5.51\mltp10^{-3}$&$2.32\mltp10^{-3}$&$7.65\mltp10^{-4}$&$2.44\mltp10^{-3}$\\
& \multirow{2}{*}{FCC-eh} &preselection&$1.22\mltp10^{-1}$&$5.94\mltp10^{-4}$&$2.19\mltp10^{-2}$&$3.47\mltp10^{-3}$&$4.39\mltp10^{-2}$&$6.57\mltp10^{-4}$&$1.99\mltp10^{-2}$&$2.53\mltp10^{-2}$&$3.56\mltp10^{-1}$\\
&                                            & BDT$>$0.125 &$4.65\mltp10^{-1}$&$1.06\mltp10^{-1}$&$4.81\mltp10^{-3}$&$1.14\mltp10^{-2}$&$7.34\mltp10^{-3}$&$5.81\mltp10^{-2}$&$5.15\mltp10^{-3}$&$2.23\mltp10^{-3}$&$3.03\mltp10^{-3}$\\
\hline
\multirow{4}{*}{600 GeV} & \multirow{2}{*}{LHeC}    & preselection & $1.32\mltp10^{-1}$&$6.56\mltp10^{-4}$&$1.53\mltp10^{-2}$&$3.07\mltp10^{-3}$&$6.64\mltp10^{-2}$&$1.37\mltp10^{-3}$&$1.46\mltp10^{-2}$&$2.35\mltp10^{-2}$&$5.38\mltp10^{-1}$\\
& & BDT$>$0.252 &1.57$\mltp10^{-1}$&$\cdots$&$\cdots$&$\cdots$&$3.01\mltp10^{-5}$&$\cdots$&$4.00\mltp10^{-5}$&$\cdots$&$5.58\mltp10^{-5}$\\
& \multirow{2}{*}{FCC-eh} &preselection&$1.32\mltp10^{-1}$&$5.94\mltp10^{-4}$&$2.19\mltp10^{-2}$&$3.47\mltp10^{-3}$&$4.39\mltp10^{-2}$&$6.57\mltp10^{-4}$&$1.99\mltp10^{-2}$&$2.53\mltp10^{-2}$&$3.56\mltp10^{-1}$\\
&                                            & BDT$>$0.225 &$2.05\mltp10^{-1}$&$\cdots$&$4.64\mltp10^{-5}$&$2.10\mltp10^{-4}$&$2.28\mltp10^{-4}$&$\cdots$&$6.80\mltp10^{-5}$&$1.18\mltp10^{-4}$&$1.55\mltp10^{-4}$\\
\hline
\multirow{4}{*}{1000 GeV} & \multirow{2}{*}{LHeC}    & preselection & $6.23\mltp10^{-2}$&$6.56\mltp10^{-4}$&$1.53\mltp10^{-2}$&$3.07\mltp10^{-3}$&$6.64\mltp10^{-2}$&$1.37\mltp10^{-3}$&$1.46\mltp10^{-2}$&$2.35\mltp10^{-2}$&$5.38\mltp10^{-1}$\\
& & BDT$>$0.266 &3.04$\mltp10^{-1}$&$\cdots$&$2.67\mltp10^{-5}$&$\cdots$&$\cdots$&$\cdots$&$1.20\mltp10^{-4}$&$\cdots$&$1.86\mltp10^{-5}$\\
& \multirow{2}{*}{FCC-eh} &preselection&$8.80\mltp10^{-2}$&$5.94\mltp10^{-4}$&$2.19\mltp10^{-2}$&$3.47\mltp10^{-3}$&$4.39\mltp10^{-2}$&$6.57\mltp10^{-4}$&$1.99\mltp10^{-2}$&$2.53\mltp10^{-2}$&$3.56\mltp10^{-1}$\\
&                                            & BDT$>$0.226 &$2.34\mltp10^{-1}$&$\cdots$&$3.10\mltp10^{-5}$&$1.20\mltp10^{-4}$&$9.12\mltp10^{-5}$&$\cdots$&$9.07\mltp10^{-5}$&$2.51\mltp10^{-4}$&$9.83\mltp10^{-5}$\\
\hline
\hline
\end{tabular}	
}
\caption{
Similar as Table~\ref{tab:allEffHad}, but for the leptonic $\tau_\mu$ final state. 
}
\label{tab:allEffLep}
\end{table}

\acknowledgments

We thank Lingxiao Bai for analysing the background of misidentified taus and helpful comments.
We also thank 
Filmon Andom Ghebretinsae, Minglun Tian and Zeren Simon Wang for helpful discussions. 
H.G. and K.W. are supported by the National Natural Science Foundation of China under grant no.~11905162, 
the Excellent Young Talents Program of the Wuhan University of Technology under grant no.~40122102, and the research program of the Wuhan University of Technology under grant no.~2020IB024.
H.S. is supported by the National Natural Science Foundation of China under grants no.~12075043 and 12147205.
Y.N.M. is supported 
by 
the National Natural Science Foundation of China under grant no.~12205227 and the Fundamental Research Funds for the Central Universities (WUT:~2022IVA052).
The simulation and analysis work of this paper was completed with the computational cluster provided by the Theoretical Physics Group at the Department of Physics, School of Sciences, Wuhan University of Technology.




\bibliography{Refs}
\bibliographystyle{hep_v5}

\end{document}